\documentclass[floatfix,twocolumn,prd,showpacs,preprintnumbers,amsmath,nofootinbib,amssymb,noeprint,superscriptaddress]{revtex4-1}
\usepackage[utf8]{inputenc}
\usepackage{comment}
\usepackage[pdftex]{color}
\usepackage{graphicx,capt-of}
\usepackage{amsmath,amssymb,amsfonts,mathtools}
\usepackage[colorlinks=true,linkcolor=blue,filecolor=black,urlcolor=blue,citecolor=blue,pdftex,plainpages=false]{hyperref}
\usepackage[utf8]{inputenc}
\usepackage[normalem]{ulem}

\DeclareRobustCommand{\Eq}[1]{Eq.~\eqref{eq:#1}}

\DeclareRobustCommand{\fig}[1]{Fig.~\ref{fig:#1}}

\DeclareRobustCommand{\app}[1]{App.~\ref{app:#1}}
\DeclareRobustCommand{\sec}[1]{Sec.~\ref{sec:#1}}

\DeclareRobustCommand{\refcite}[1]{Ref.~\cite{#1}}
\newcommand\bets{\begin{table*}}
\newcommand\eets[1]{\label{tb:#1}\end{table*}}

\begin{document}
\preprint{MIT-CTP/6027}
\title{ 
{Proton isovector helicity PDF at NNLO and the twist-3 moment $\tilde{d}_2$ from lattice QCD at physical quark masses}}

\author{Xiang Gao}
\email{xgao@bnl.gov}
\affiliation{Physics Department, Brookhaven National Laboratory, Upton, New York 11973, USA}

\author{Andrew D. Hanlon}
\affiliation{Department of Physics, Kent State University, Kent, OH 44242, USA}

\author{Swagato Mukherjee}
\affiliation{Physics Department, Brookhaven National Laboratory, Upton, New York 11973, USA}

\author{Peter Petreczky}
\affiliation{Physics Department, Brookhaven National Laboratory, Upton, New York 11973, USA}

\author{Hai-Tao Shu}
\affiliation{Key Laboratory of Quark \& Lepton Physics (MOE) and Institute of Particle Physics, Central China Normal University, Wuhan 430079, China}

\author{Fei Yao}
\affiliation{Physics Department, Brookhaven National Laboratory, Upton, New York 11973, USA}

\author{Rui Zhang}
\email{rzhang93@mit.edu}
\affiliation{Center for Theoretical Physics - a Leinweber Institute, Massachusetts Institute of Technology, Cambridge, MA 02139, USA}

\author{Yong Zhao}
\affiliation{Physics Division, Argonne National Laboratory, Lemont, IL 60439, USA}

\begin{abstract}
We present a lattice quantum chromodynamics calculation of the $x$-dependent isovector quark helicity parton distribution function (PDF) of the proton in the large momentum effective theory (LaMET) framework. Through operator product expansion (OPE) we also extract the $\tilde{d}_2$ moment of the twist-3 PDF $g_T(x)$ for the first time in the $\overline{\rm MS}$ scheme, which is proportional to the average color Lorentz force experienced by the quark in the proton. This calculation is performed on a lattice of spacing $a$ = 0.076 fm at physical quark masses. The quasi-PDF matrix elements are measured in proton states boosted to momenta $P_z=\{0, 0.25, 1.02, 1.53\}$ GeV. We first extract the lowest few helicity PDF moments from the renormalization-group (RG) invariant ratios of the matrix elements with OPE. Combined with the matrix elements relevant for $g_T(x)$, we obtain $\tilde{d}_2^{u-d}(2\ {\rm GeV})=0.0024(46)$ at next-to-leading order in $\overline{\rm MS}$. 
Then, the helicity quasi-PDF matrix elements are renormalized in the hybrid scheme with linear renormalon resummation and Fourier transformed to the $x$-space after an asymptotic extrapolation. The quasi-PDF is perturbatively matched to the $\overline{\rm MS}$ PDF with RG and threshold resummations at next-to-leading power and next-to-next-to-leading logarithmic accuracies.
After resummations, we determine the PDF in the region $x\in[0.25,0.75]$. The end-point regions are then parameterized, combined with the LaMET prediction at moderate $x$, and fitted to the short-distance matrix elements in coordinate space.
\end{abstract}

\maketitle

\section{Introduction}

Parton distribution functions (PDFs) describe the one-dimensional internal structure of hadrons in terms
of partonic longitudinal momentum distributions. Specifically, they encode the nonperturbative probability
of finding a parton (quark or gluon) carrying a momentum fraction $x$ (Bjorken-$x$) in a fast-moving hadron.
At leading twist, there are three independent quark PDFs for the nucleon: the unpolarized distribution,
the transversity distribution, and the helicity distribution. The focus of this paper is the proton quark
helicity PDF, $g_1(x)$, which characterizes the difference between quark number densities with spin aligned
and antialigned with the longitudinally polarized proton.

The first moment of the quark helicity PDF gives the quark-spin contribution to the proton spin, which is common to both the Jaffe-Manohar decomposition\cite{Jaffe:1989jz} and Ji’s gauge-invariant decomposition\cite{Ji:1996ek}. The two decompositions differ, however, in how the remaining proton spin is partitioned among quark and gluon angular-momentum contributions, in particular in the treatment of orbital angular momentum and the gluon sector. The original EMC measurements indicated that the quark-spin contribution to the proton spin is much smaller than expected, giving rise to the well-known ``proton spin puzzle''~\cite{EuropeanMuon:1987isl}. Modern global analyses now suggest that quark spin accounts for roughly $30\%$ of the proton spin~\cite{Cocuzza:2022jye,Borsa:2024mss,Bertone:2024taw,Cruz-Martinez:2025ahf}.
A precise determination of $g_1(x)$
and its moments therefore provides an important first-principles test of this basic decomposition and helps
quantify how the remaining spin is distributed among gluons and orbital motion.

Experimentally, information on quark helicity PDFs is primarily obtained from polarized deep-inelastic
scattering (DIS) and semi-inclusive DIS (SIDIS). Key measurements include polarized $p$--$p$ collisions at
RHIC~\cite{STAR:2014wox,PHENIX:2014gbf,PHENIX:2015fxo}, open-charm muon production and muon--proton/deuteron
scattering at COMPASS~\cite{COMPASS:2012mpe,COMPASS:2009kiy,COMPASS:2010hwr}, and identified hadron production
at HERMES~\cite{HERMES:2004zsh}. Future programs such as Jefferson Lab 12~GeV~\cite{Dudek:2012vr}, the EIC
\cite{Accardi:2012qut,AbdulKhalek:2021gbh}, and the EicC~\cite{Anderle:2021dpv} are expected to significantly improve constraints,
especially in the small-$x$ region. 

Apart from experimental probes, first-principle lattice QCD calculations are also important for a quantitative understanding of the proton helicity structure. Because helicity PDFs are defined by nonlocal quark bilinear operators separated along the light-cone direction, they cannot be computed directly in Euclidean lattice QCD. Ji’s large-momentum effective theory (LaMET) shows that a class of quasidistributions defined from equal-time correlators in boosted hadrons can be matched to the corresponding light-cone distributions through a systematic large-momentum expansion~\cite{Ji:2013dva,Ji:2014gla,Ji:2020ect}. The universality class shares the same infrared physics as the light-cone PDF, thus can be matched to the latter perturbatively in the Bjorken-$x$ space, up to power corrections suppressed by the hadron momentum.  The simplest member of this universality class is the quasi-PDF that describes the momentum density distribution of quarks and gluons in a moving hadron. 
Since its proposal, substantial progress has been made in renormalization, matching, power corrections, and resummation~\cite{Ji:2017oey,Green:2017xeu,Ishikawa:2017faj,Alexandrou:2017huk,Chen:2017mzz,Constantinou:2017sej,Stewart:2017tvs,Ji:2020brr,LatticePartonLPC:2021gpi,Izubuchi:2018srq,Li:2020xml,Chen:2020ody,Cheng:2024wyu,Braun:2018brg,Zhang:2023bxs,Holligan:2023rex,Gao:2021hxl,Su:2022fiu,Ji:2023pba,Baker:2024zcd,Ji:2024hit,Holligan:2025baj}. More recently, Coulomb-gauge quasi-PDFs have also been proposed as a member of the same universality class and offers advantages for transverse-momentum-dependent observables~\cite{Gao:2023lny,Zhao:2023ptv,Bollweg:2024zet}.
Alternative approaches to extract the information of helicity PDFs include short-distance factorization in coordinate- or Ioffe-time space~\cite{Braun:2007wv,Radyushkin:2017cyf,Orginos:2017kos,Ma:2017pxb}, as well as the hadronic tensor~\cite{Liu:1993cv,Liang:2019frk} and Compton amplitudes~\cite{Detmold:2005gg,Detmold:2021uru,Chambers:2017dov}. Smearing and gradient-flowed based approaches have also been proposed to directly calculate the higher Mellin moments of the PDFs~\cite{Davoudi:2012ya,Shindler:2023xpd,Francis:2025rya}.

Based on these developments, lattice QCD computations have achieved substantial progress in recent years for the unpolarized~\cite{Alexandrou:2017huk,LatticeParton:2018gjr,Lin:2017ani,Alexandrou:2018pbm,Chen:2018xof, Alexandrou:2019lfo,Joo:2019jct,Joo:2020spy,Gao:2022uhg}, transversity~\cite{Alexandrou:2017huk,Liu:2018hxv, Alexandrou:2019lfo,Alexandrou:2018eet,LatticeParton:2022xsd,Gao:2023ktu}, and helicity~\cite{Chen:2016utp,Alexandrou:2017huk,Lin:2017ani,Lin:2018pvv,Alexandrou:2018pbm,Alexandrou:2019lfo, Lin:2019ocg,Alexandrou:2020qtt,Alexandrou:2021oih,Alexandrou:2020uyt,Fan:2020nzz,HadStruc:2022nay, Holligan:2024wpv} PDFs; see Refs.~\cite{Zhao:2018fyu,Cichy:2018mum,Monahan:2018euv,Ji:2020ect,Constantinou:2020pek} for recent reviews. To date, some helicity calculations have reached the continuum limit~\cite{Alexandrou:2020qtt, Holligan:2024wpv} and some are performed at the physical quark masses~\cite{Lin:2017ani,Lin:2018pvv, Alexandrou:2018pbm,Alexandrou:2019lfo}, but achieving both simultaneously remains challenging. In this work, we present results computed at the physical point on a fine lattice, and include the recent theory developments for a more accurate determination.

Beyond leading twist, the transverse spin structure function $g_T(x)=g_1(x)+g_2(x)$ and its twist-3 content
provide a direct window into quark-gluon correlations in the nucleon. A widely used benchmark is the reduced
twist-3 moment $\tilde d_2\equiv \int_{-1}^{1} dx\,x^2\,[3g_T(x)-g_1(x)]$ (or $d_2$ in an alternative normalization), which admits a semiclassical interpretation in terms of an average transverse color
Lorentz force on the struck quark~\cite{Burkardt:2008ps,SANE:2018pwx,Deur:2018roz,HERMES:2011xgd}. The
experimental determination of $g_2$ and $\tilde  d_2$ is demanding and remains less precise than leading-twist
observables, though recent Jefferson Lab measurements have extended the coverage and improved constraints
\cite{JeffersonLabHallA:2016neg,Cocuzza:2025qvf}.

The LaMET framework has also been used to compute the twist-3 PDFs on the lattice~\cite{Bhattacharya:2020cen,Bhattacharya:2020xlt,Bhattacharya:2021moj}.
The first lattice determination of $g_T(x)$~\cite{Bhattacharya:2020cen} employed the twist-2 matching kernel and
found that the Wandzura-Wilczek (WW) approximation~\cite{Wandzura:1977qf} describes the data well within uncertainties, consistent with a
small genuine twist-3 contribution. A complete NLO treatment, however, requires
including genuine three-parton (quark-gluon) contributions in the matching. Such a systematic twist-3
factorization framework, including the NLO violation of the WW relation and the
explicit appearance of the reduced moment $\tilde d_2$ in the OPE, was developed in Ref.~\cite{Braun:2021aon} for both LaMET and SDF. Building on
this framework, we extract $\tilde d_2$ in the $\overline{\rm MS}$ scheme from the short-distance matrix elements, providing a complementary first-principles constraint on the genuine twist-3 moment alongside our leading-twist helicity PDF.

On the lattice, $\tilde{d}_2$ has been computed directly from local twist-3 operator matrix elements, and the
results typically indicate a strong suppression of this lowest chiral-even twist-3 moment
\cite{Gockeler:2005vw,Burger:2021knd,Crawford:2024wzx}. A practical challenge is that twist-3 operators can mix with
lower-dimensional operators with coefficients proportional to $1/a$, requiring nonperturbative subtractions
and complicating scheme conversions. In some cases, results are therefore reported in
intermediate momentum subtraction schemes (and evolved to a reference scale) rather than converted fully to
$\overline{\rm MS}$. Complementary approaches that access twist-3 information while
reducing sensitivity to power-divergent mixing are thus valuable.

The remainder of this manuscript is organized as follows. In Sec.~\ref{sec:lattice} we provide the lattice
details used in this calculation, including the lattice setup, the definition of the matrix element, and
the relevant measurements. We then describe the extraction of the matrix elements corresponding to the quark
helicity from lattice data. Section~\ref{sec:mellin-moments} presents Mellin moments extracted using OPE.
In Sec.~\ref{sec:g2} we extract the $\tilde{d}_2$ moment for the twist-3 axial-vector PDF $g_T$. In
Sec.~\ref{sec:helicity-lamet} we discuss light-cone PDFs extracted in the LaMET framework and provide a
complementary analysis with SDF to constrain the 
end point regions. Section~\ref{sec:conclusion} summarizes our
findings and outlines prospects. Details of the three-point/two-point ratio fits, the axial charge, and the
momentum-space matching are provided in the Appendix.

\section{Lattice details and theoretical construction}
\label{sec:lattice}
\subsection{Lattice setup}
The ensemble used in this calculation is generated using the $n_f=2+1$ highly improved staggered quark (HISQ) action~\cite{Follana:2006rc} at physical light and strange quark masses by the HotQCD collaboration~\cite{Bazavov:2019www}. The coupling constant is $\beta=7.13$, which corresponds to lattice spacing $a = 0.076$ fm, determined using the $f_K$ scale~\cite{HotQCD:2014kol}. The lattice size is $N_{\sigma}^3\times N_{\tau} = 64^3\times 64$. The valence quark is of the tree-level tadpole-improved Clover-Wilson type~\cite{Sheikholeslami:1985ij}. We use $c_{\mathrm{sw}}=u_0^{-3/4}=1.0372$, determined by the plaquette expectation value $u_0$ measured on the one-step hypercubic (HYP)~\cite{Hasenfratz:2001hp} smeared gauge configurations. The valence light quark is tuned to correspond to a pion of nearly its physical mass, 140 MeV. The hadron is boosted to three momenta $P_z=\frac{2\pi}{aN_{\sigma}}\times\{1, 4, 6\}=\{0.25, 1.02, 1.53\}$ GeV, plus the one at rest. The source-sink separation in the temporal direction includes $t_{\mathrm{sep}}/a=6,8,10,12$. The $t_{\mathrm{sep}}/a=12$ data is only used in the calculation of axial vector charge which requires merely the zero momentum data where $t_{\mathrm{sep}}/a=12$ can have an acceptable signal. The other three separations are used in all calculations. There are 350 configurations in total for this ensemble. At each momentum and $t_{\mathrm{sep}}$, we generate multiple sources, including both exact and sloppy ones, to improve the signal-to-noise ratio. Since solving propagators at physical quark mass is computationally demanding, we use the multigrid solver in QUDA~\cite{Clark:2009wm,Babich:2011np} to speed up convergence. For the same reason, only a small portion of the propagators is calculated using a very small residual ($10^{-10}$) to guarantee the convergence. These are the ``exact solves''. We then calculate many more propagators using a larger residual ($10^{-4}$), called ``sloppy" solves. The exact and sloppy solves calculated from the same sources are used to estimate the differences between them. Adding the differences to the other sloppy solves, we reproduce correct results with significantly improved efficiently. This technique is called all mode averaging (AMA)~\cite{Shintani:2014vja}. In addition, we use Coulomb-gauge momentum smearing of the quark fields~\cite{Bali:2016lva} to improve the signal-to-noise ratio of boosted hadrons. The lattice setup is summarized in Table~\ref{tab:param-zeroT}. More details can be found in our previous works~\cite{Gao:2022uhg,Gao:2023ktu}.
\begin{table}[htb]
\centering
\begin{tabular}{cccccc}
\hline \hline
$\beta$ & $a$ [fm] & $m_\pi$ [GeV] & $N_{\sigma}$ & $N_{\tau}$ & \# conf. \\
\hline
7.13	& 0.076 & 0.14 & 64 & 64 & 350 \\
\hline \hline
\end{tabular}
\caption{Parameters of the lattice ensembles used in this work, including the inverse gauge coupling $\beta$, the lattice spacing, the valence pion mass, the spatial and temporal lattice extent $N_\sigma$ and $N_\tau$, and the number of configurations.}
\label{tab:param-zeroT}
\end{table}

\begin{figure*}[tbh]
\centerline{
\includegraphics[width=0.4\textwidth]{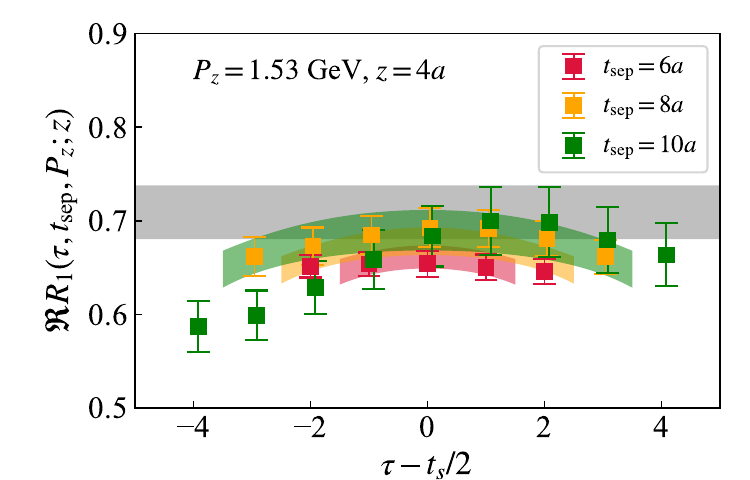}
\includegraphics[width=0.4\textwidth]{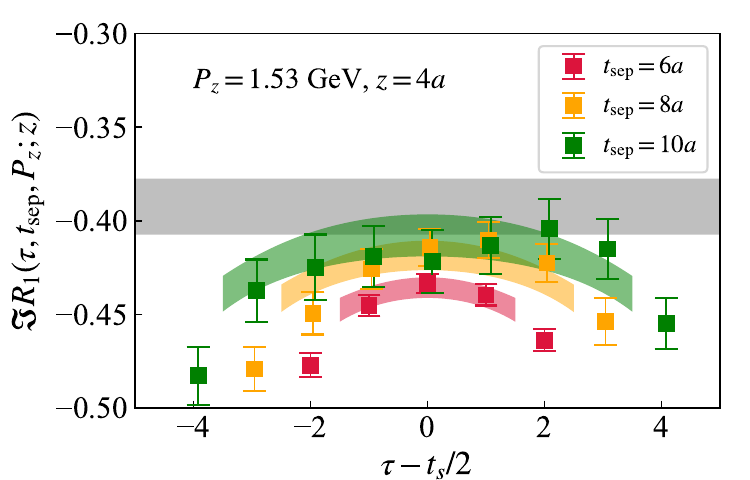}
}
\centerline{
\includegraphics[width=0.4\textwidth]{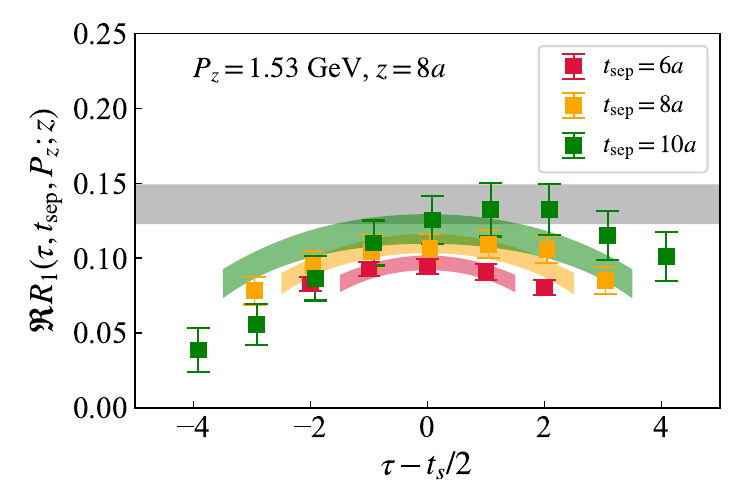}
\includegraphics[width=0.4\textwidth]{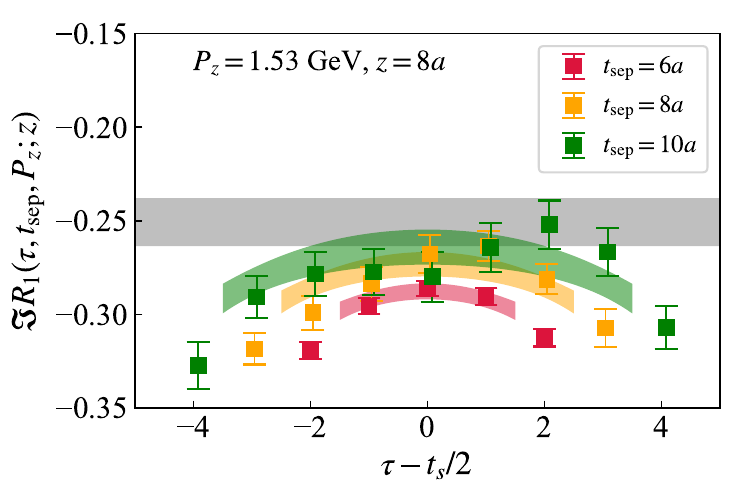}
}
\caption{The fit of the three-point function to two-point function ratio $R_1(\tau,t_{\mathrm{sep}},P_z;z)$ at the largest momentum $P_z=1.53$ GeV with $z=4a$ (upper panels) and $z=8a$ (lower panels). Left: the real part. Right: the imaginary part. The colored bands are reconstructed from two-state fit while the gray band denotes the extracted bare matrix element. }
\label{fig:fit-ratio}
\end{figure*}

\begin{figure*}[tbh]
\centerline{
\includegraphics[width=0.4\textwidth]{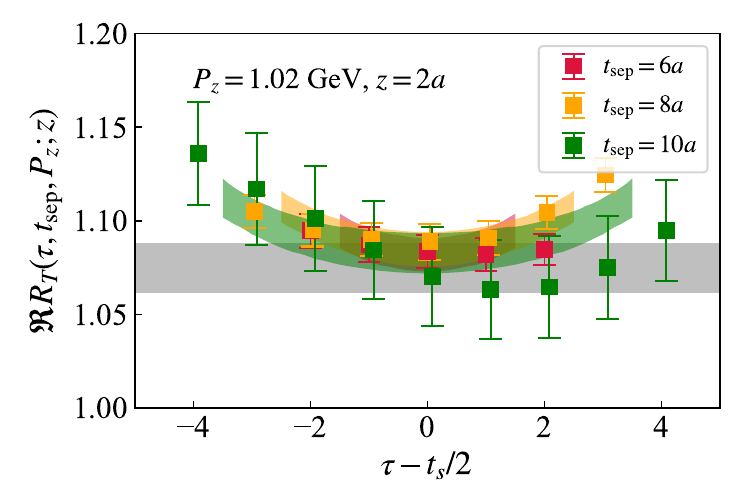}
\includegraphics[width=0.4\textwidth]{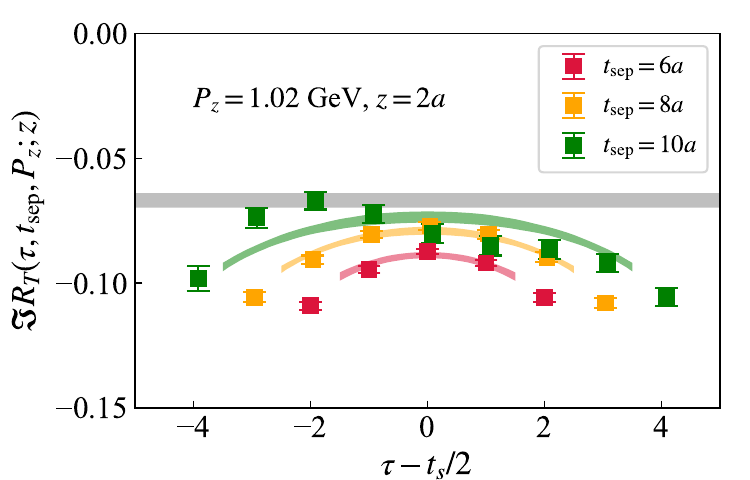}
}
\centerline{
\includegraphics[width=0.4\textwidth]{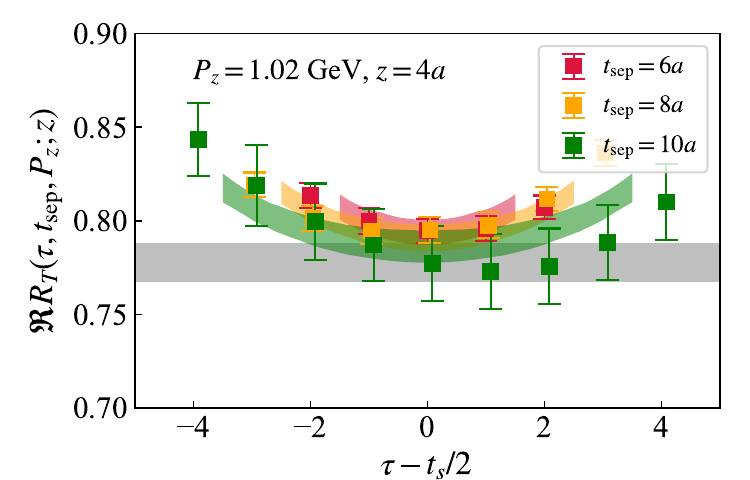}
\includegraphics[width=0.4\textwidth]{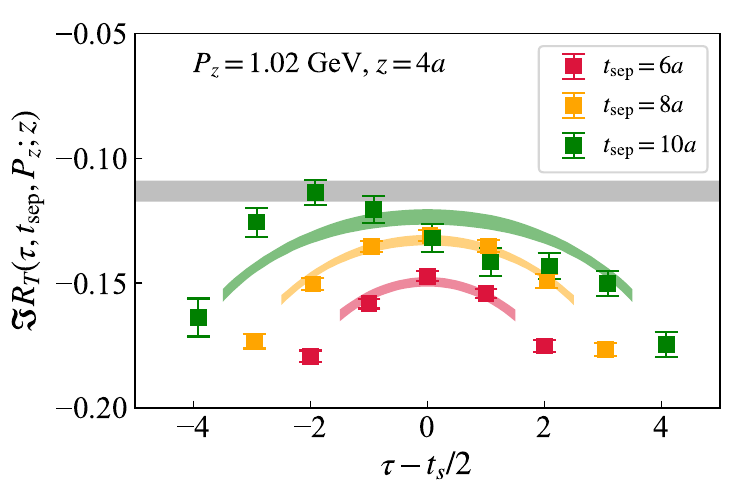}
}
\caption{Similar to \fig{fit-ratio} but for $R_T(\tau,t_{\mathrm{sep}},P_z;z)$ at momentum $P_z=1.02$ GeV with $z=2a$ (upper panels) and $z=4a$ (lower panels). }
\label{fig:fit-ratio-gT}
\end{figure*}

\subsection{Extraction of the bare matrix elements}
\label{sec:extract-mat}

Spin-dependent quark quasi-PDFs are accessed through equal-time, spatially separated nonlocal quark bilinears connected by a straight Wilson line. On the lattice, we compute nucleon matrix elements of
\begin{equation}
\mathcal{O}_\Gamma(z)\equiv \bar{\psi}(z\hat{z})\,\Gamma\,W(z\hat{z},0)\,\psi(0),
\end{equation}
where $z$ is the spatial separation along the $\hat{z}$ direction and $W(z\hat{z},0)$ is a straight Wilson line (implemented as a product of gauge links) that renders the bilinear gauge invariant.

For the twist-2 helicity channel, we take $\Gamma=i\gamma_5\gamma_z$ and define the bare matrix element
\begin{equation}
\begin{split}
\tilde{h}^{B}_1(z,P_z,a)
\equiv \frac{1}{2E}\,
\langle P,S \vert \bar{\psi}(z)\, \Gamma\, W(z,0)\, \psi(0) \vert P,S \rangle,
\end{split}
\label{eq:def-mat-1}
\end{equation}
where $P_\mu=(E,0,0,P_z)$ is the nucleon four-momentum and $S_\mu$ denotes the corresponding spin polarization.

The normalization factor $1/(2E)$ is fixed by Lorentz symmetry and the standard covariant decomposition of the axial-vector matrix element.
Using relativistically normalized nucleon states and Dirac spinors, one has
\begin{equation}
\bar u(P,S)\,\gamma^\mu\gamma_5\,u(P,S)=2m_N\,s^\mu,
\label{eq:spinor_bilinear}
\end{equation}
where $s^\mu$ is the covariant spin four-vector satisfying $s\cdot P=0$ and $s^2=-1$.
For a nucleon boosted along $\hat z$, $P^\mu=(E,0,0,P_z)$, the explicit form of $s^\mu$ depends on the polarization choice.
For longitudinal polarization (helicity) along the momentum direction, $s^\mu_{\parallel}=(\frac{P_z}{m_N},\,0,\,0,\,\frac{E}{m_N})$, we have,
\begin{equation}
\bar u(P,S)\,\gamma^z\gamma_5\,u(P,S)=2m_N s^z_{\parallel} \sim 2E.
\label{eq:s_longitudinal}
\end{equation}
Therefore the factor $1/(2E)$ in Eq.~(\ref{eq:def-mat-1}) removes the trivial boost factor carried by the external nucleon spinors.

For the transverse twist-3 channel relevant to $\tilde g_T(x)$, we use
\begin{equation}
\begin{split}
\tilde{h}^{B}_T(z,P_z,a)
\equiv \frac{1}{2m_N}\,
\langle P,S \vert \bar{\psi}(z)\, \Gamma\, W(z,0)\, \psi(0) \vert P,S \rangle,
\end{split}
\label{eq:def-mat-T}
\end{equation}
with $\Gamma=i\gamma_5\gamma_x$.
For a nucleon moving along $\hat z$ but polarized transversely along $\hat x$, a convenient covariant choice is $s^\mu_{\perp}=(0,\,1,\,0,\,0)$ and gives
\begin{equation}
\bar u(P,S)\,\gamma^x\gamma_5\,u(P,S)=2m_N s^x_{\perp}\sim 2m_N,
\label{eq:s_transverse}
\end{equation}
which again follows from Eq.~(\ref{eq:spinor_bilinear}) and the orthogonality condition $s\cdot P=0$.
With this polarization, dividing by $2m_N$ removes the trivial kinematic factor from the external spinors.

These matrix elements are obtained from nucleon three-point functions,
\begin{align}
C_{\mathrm{3pt}}^{\Gamma}&(\vec{P}, t_{\mathrm{sep}}, \tau, z)=
\sum_{\vec{y}, \vec{z}_0}e^{-i\vec{P}\cdot(\vec{y}-\vec{x}_0)}\,
\mathcal{P}_{\rm 3pt}\\
&\times \left\langle
N(\vec{y}, t_{\mathrm{sep}}+t_0)\,
\mathcal{O}_\Gamma(\vec{z}_0+z\hat{z}, \tau+t_0)\,
\bar{N}(\vec{x}_0, t_0)
\right\rangle \nonumber
\label{eq:3pt}
\end{align}
where $(\vec{x}_0,t_0)$, $(\vec{y},t_{\mathrm{sep}}+t_0)$, and $(\vec{z}_0+z\hat{z},\tau+t_0)$ denote the source, sink, and operator-insertion coordinates, respectively. Here $z\hat{z}$ is the spatial separation of the nonlocal operator along the $\hat{z}$ direction. The inserted operator has the same nonlocal structure as in Eqs.~(\ref{eq:def-mat-1}) and (\ref{eq:def-mat-T}). The nucleon interpolating operator $N$ is constructed from momentum-smeared quark fields with $\vec{k}\approx0.5\vec{P}$~\cite{Bali:2016lva},
\begin{align}
    N(x)=\epsilon_{abc} (d^T_{\vec{k}}(x)C\gamma_5u_{\vec{k}}(x))u_{\vec{k}}(x),
\end{align}
where the smeared quark fields $u_{\vec{k}}$ and $d_{\vec{k}}$ are obtained from the convolution with a Gaussian kernel $G(\vec{x},\vec{y})$ in the Coulomb gauge~\cite{Izubuchi:2019lyk},
\begin{align}
    q_k(\vec{x},t)=\sum_{\vec{y}}G(\vec{x},\vec{y})e^{i\vec{k}\cdot(\vec{y}-\vec{x})}q(\vec{y},t).
\end{align}
In this work we focus on the isovector combination to avoid disconnected contributions,
\begin{align}
\mathcal{O}_\Gamma&(\vec{z}_0+z\hat{z},\tau)=
\ \bar{u}(\vec{z}_0+z\hat{z},\tau)\, \Gamma\,
W(\vec{z}_0+z\hat{z},\tau;\vec{z}_0,\tau)\,
u(\vec{z}_0,\tau) \nonumber\\
&-\bar{d}(\vec{z}_0+z\hat{z},\tau)\, \Gamma\,
W(\vec{z}_0+z\hat{z},\tau;\vec{z}_0,\tau)\,
d(\vec{z}_0,\tau).
\label{eq:insert}
\end{align}
The projector $\mathcal{P}_{\rm 3pt}$ is chosen to isolate the desired parity and polarization structure. For the helicity channel we take
\begin{equation}
\mathcal{P}_{\rm 3pt}=\frac{1}{2}(1+\gamma_t)(i\gamma_5\gamma_z),
\qquad \Gamma=i\gamma_5\gamma_z,
\label{eq:proj3ptHeli}
\end{equation}
while for the twist-3 transverse channel we use
\begin{equation}
\mathcal{P}_{\rm 3pt}=\frac{1}{2}(1+\gamma_t)(i\gamma_5\gamma_x),
\qquad \Gamma=i\gamma_5\gamma_x.
\label{eq:proj3ptgT}
\end{equation}

The corresponding nucleon two-point correlator is
\begin{equation}
\begin{split}
C_{\mathrm{2pt}}(\vec{p}, &t_{\mathrm{sep}})=
\sum_{\vec{y}} e^{-i\vec{p}\cdot(\vec{y}-\vec{x}_0)}\,
\mathcal{P}_{\rm 2pt}\\
&\times\left\langle
N(\vec{y}, t_{\mathrm{sep}}+t_0)\,
\bar{N}(\vec{x}_0, t_0)
\right\rangle ,
\end{split}
\label{eq:2pt-polished}
\end{equation}
with the rest-frame positive-parity projector
\begin{equation}
\mathcal{P}_{\rm 2pt}=\frac{1}{2}(1+\gamma_t).
\label{eq:proj2pt-polished}
\end{equation}

\begin{figure*}[tbh]
\centerline{
\includegraphics[width=0.4\textwidth]{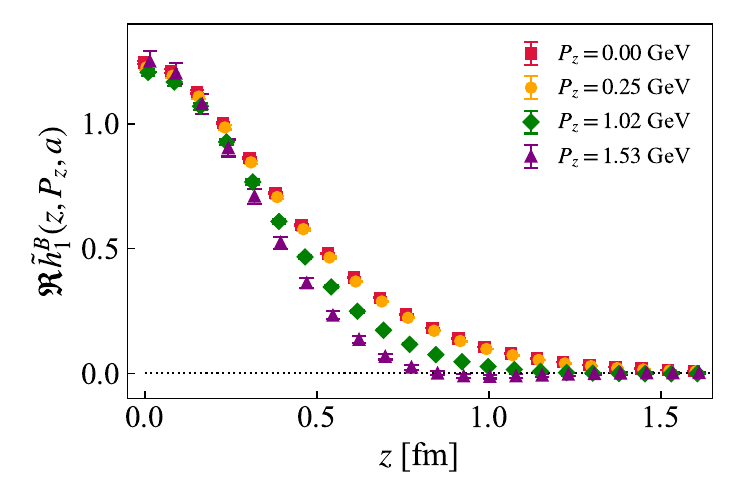}
\includegraphics[width=0.4\textwidth]{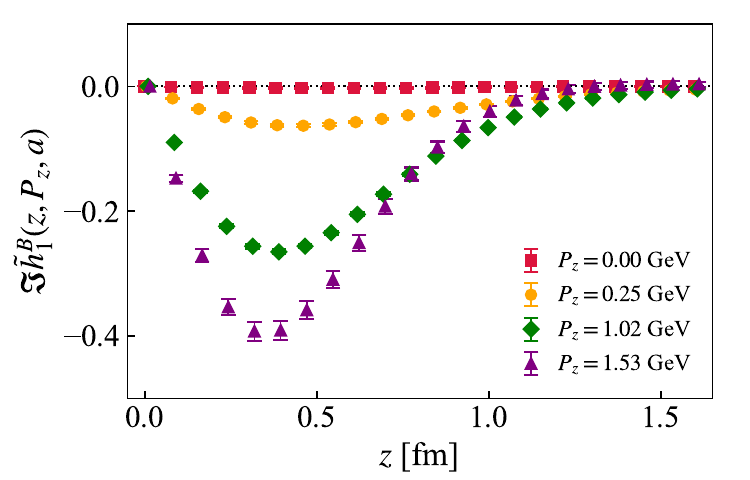}
}
\centerline{
\includegraphics[width=0.4\textwidth]{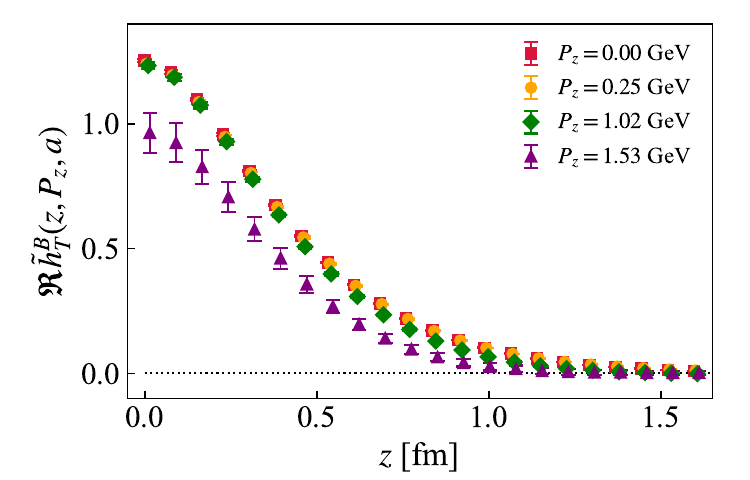}
\includegraphics[width=0.4\textwidth]{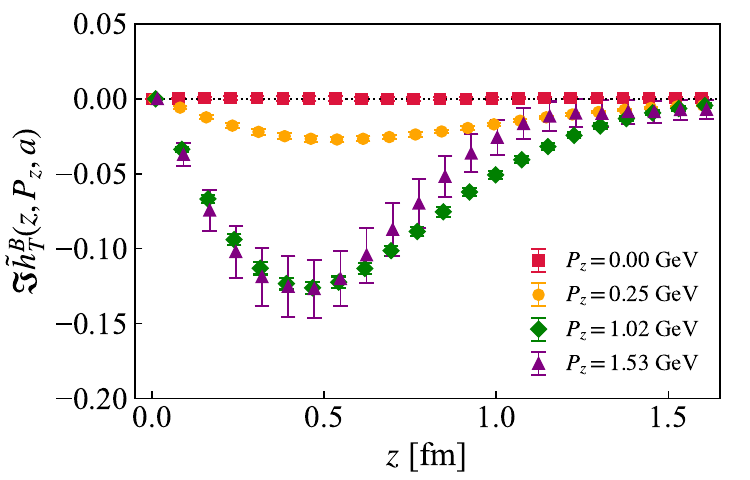}
}
\caption{
Isovector bare matrix elements as functions of $z$ for the helicity channel (top row, $\tilde h^{B}_1$) and the transverse twist-3 channel (bottom row, $\tilde h^{B}_T$), shown at several nucleon boosts $P_z$ (see legends).
Left (right) panels show the real (imaginary) parts.
}
\label{fig:bare-matrix}
\end{figure*}

The correlation functions admit the spectral decomposition
\begin{align}
C_{\mathrm{3pt}}^{\Gamma}(\vec{p},t_{\mathrm{sep}},\tau,z)
=&\sum_{m,n}
\langle \Omega \vert N \vert n\rangle\,
\langle n \vert \mathcal{O}_\Gamma(z) \vert m\rangle\,
\langle m \vert \bar{N} \vert \Omega \rangle\nonumber\\
&\times e^{-E_m \tau}\, e^{-E_n (t_{\mathrm{sep}}-\tau)},\\
C_{\mathrm{2pt}}(\vec{p},t_{\mathrm{sep}})
=&\sum_{n} \left|\langle \Omega \vert N \vert n\rangle\right|^2 e^{-E_n t_{\mathrm{sep}}},
\label{eq:joint_fit}
\end{align}
where $|\Omega\rangle$ denote the vacuum state and $n=0,1,...$ denote the ground and excited states. The desired ground-state matrix element is $\langle 0 \vert \mathcal{O}_\Gamma(z) \vert 0\rangle$.

To exploit correlations, we form the ratio of three-point to two-point correlators,
\begin{equation}
R(\tau,t_{\mathrm{sep}},P_z;z)
\equiv
\frac{C_{\mathrm{3pt}}^{\Gamma}(\vec{p},t_{\mathrm{sep}},\tau,z)}
{C_{\mathrm{2pt}}(\vec{p},t_{\mathrm{sep}})}\, .
\label{eq:ratio_def}
\end{equation}
We denote $R_1(\tau,t_{\mathrm{sep}},P_z;z)$ and $R_T(\tau,t_{\mathrm{sep}},P_z;z)$ for the helicity and twist-3 channels, respectively. In the asymptotic limit $t_{\mathrm{sep}}\gg \tau \gg 0$, one has
\begin{equation}
\begin{split}
&\lim_{t_{\mathrm{sep}}\gg \tau \gg 0} R_1(\tau,t_{\mathrm{sep}},P_z;z)= \tilde{h}^{B}_1(z,P_z,a),\\
&\lim_{t_{\mathrm{sep}}\gg \tau \gg 0} R_T(\tau,t_{\mathrm{sep}},P_z;z)= \frac{m_N}{E}\,\tilde{h}^{B}_T(z,P_z,a),
\end{split}
\label{eq:ratio_limit}
\end{equation}
where the factor $m_N/E$ reflects the $1/(2m_N)$ normalization used in Eq.~(\ref{eq:def-mat-T}) which suggests that the matrix elements of higher momentum are suppressed by the nucleon energy.

To determine the excited-state contributions, we first extract the energies $E_n$ and overlap amplitudes $A_n=\left|\langle \Omega \vert N \vert n\rangle\right|^2$ from two-state fits to the two-point correlator, and then use them as inputs in the three-point analysis, following the strategy of Refs.~\cite{Gao:2023ktu,Gao:2022uhg}. In our calculations we consider Wilson-line separations in the range $-32a \le z \le 32a$. Exploiting the symmetry properties under $z\to -z$ (real part even, imaginary part odd), we average data at $\pm z$ and restrict the analysis to $z\ge 0$, which reduces statistical fluctuations and suppresses potential systematic effects.

We find that data at different $\tau$ and $t_{\mathrm{sep}}$ are strongly correlated, leading to large condition numbers for the estimated covariance matrices and numerically unstable fully correlated fits. We therefore perform uncorrelated fits, with statistical uncertainties estimated using bootstrap resampling.

As an illustration, Fig.~\ref{fig:fit-ratio} shows representative two-state fits to
$R_1(\tau,t_{\mathrm{sep}},P_z;z)$ at our largest boost, $P_z=1.53~\mathrm{GeV}$, for
$z=4a$ and $8a$. Figure~\ref{fig:fit-ratio-gT} shows analogous examples for
$R_T(\tau,t_{\mathrm{sep}},P_z;z)$ at $P_z=1.02~\mathrm{GeV}$ with $z=2a$ and $4a$. When the operator insertion is close to the source or sink, the ratio exhibits sizable excited-state contamination. To mitigate this effect, we omit two time slices near each end (i.e., near $\tau\simeq 0$ and $\tau\simeq t_{\mathrm{sep}}$) from the fit range. The fit results shown as the colored bands are consistent with the data across the available $t_{\mathrm{sep}}$, indicating that the data can be well described by the two-state ansatz. The gray bands are our extracted bare matrix elements.

The extracted bare matrix elements $\tilde{h}^{B}_1$ (upper panels) and $\tilde{h}^{B}_T$ (lower panels) are summarized in Fig.~\ref{fig:bare-matrix}. At $z=0$, the matrix elements from different boosts and from both channels are consistent, as the nonlocal operator reduces to the local axial current; after renormalization this yields the axial charge,
$g_A = Z_A\, \tilde{h}^{B}_1(0,P_z,a)= Z_A\, \tilde{h}^{B}_T(0,P_z,a)$.
Details of our $g_A$ determination are presented in Appendix~\ref{app:av-charge}. For the twist-3 channel, the signal quality typically degrades at larger boosts. This is consistent with an additional kinematic suppression of order $m_N/E$ in the reduced transverse amplitudes. We also observe a clear momentum dependence in the $z$ behavior: the magnitude of $\tilde{h}^{B}_{1,T}(z,P_z,a)$ decreases more rapidly with $z$ as $P_z$ increases, as expected for boosted nonlocal matrix elements.

\section{
Moments of the helicity PDF}
\label{sec:mellin-moments}

\begin{figure*}[tbh]
\includegraphics[width=0.48\textwidth]{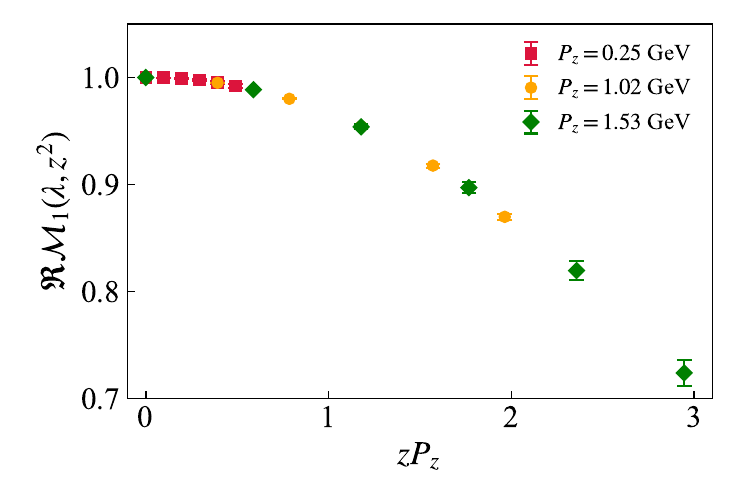}
\includegraphics[width=0.48\textwidth]{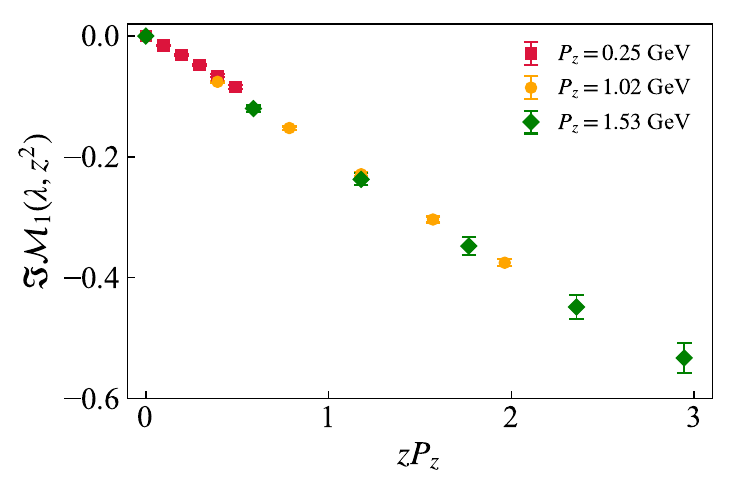}
\caption{The real (left panel) and imaginary (right panel) part of the RG-invariant ratios $\mathcal{M}_1(\lambda,z^2)$ are shown for $z\in[a,5a]$. }
\label{fig:itdg1}
\end{figure*}

Although the bare matrix elements contain linear and logarithmic UV divergences, they are proven to be multiplicatively renormalizable~\cite{Ji:2017oey,Green:2017xeu,Ishikawa:2017faj},
with renormalization constant $Z^R(z,a)$ independent of the external state. Thus one can form the following RG-invariant ratio~\cite{Orginos:2017kos,Fan:2020nzz} to cancel the UV divergences,
\begin{equation}
\mathcal{M}_1(\lambda,z^2 ) \equiv \frac{\tilde{h}^B_1(z, P_z)}{\tilde{h}^B_1(z, P_z^0)} \frac{\tilde{h}^B_1(0, P_z^0)}{\tilde{h}^B_1(0, P_z)}= \frac{\tilde{h}^{\overline{\rm MS}}_1(\lambda,z^2\mu^2)}{\tilde{h}^{\overline{\rm MS}}_1(\lambda^0, z^2\mu^2)},
\label{eq:itd}
\end{equation}
where the matrix elements $\tilde{h}^B_1(0, P_z)$ at $z=0$ are included for normalization $\mathcal{M}_1(\lambda, z^2)=1$. In this work, we choose $P_z^0=0$, and the data are shown in \fig{itdg1}.
\begin{figure*}[thbp!]
    \centering
    \includegraphics[width=0.48\linewidth]{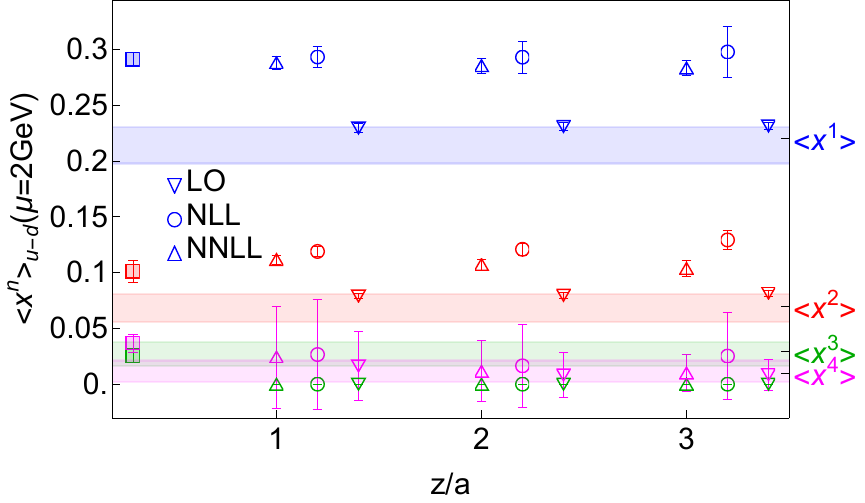}
    \includegraphics[width=0.48\linewidth]{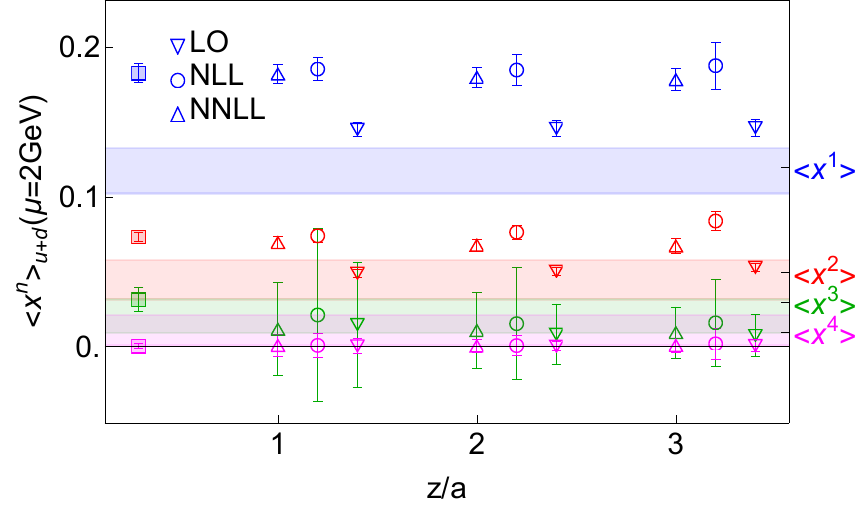}
    \caption{Helicity PDF moments $\langle x^n\rangle(\mu=2\rm{~GeV})$ fitted from the OPE of isovector (left) and isoscalar (right) 
    RG-invariant ratios. The consistency between NLL and NNLL accuracy suggest good perturbative convergence. The leading order (LO) results with $C_n=1$ have been included to demonstrate the importance of perturbative corrections. The combined correlated fit among different $z$ values are shown separately as filled squares on the left, which are consistent with individual fits for lower moments. For comparison, we integrate the quark helicity PDFs from JAM22~\cite{Buckley:2014ana,Cocuzza:2022jye}  to get the moments and plot as bands. 
    }
    \label{fig:moment_fit}
\end{figure*}

The ratio $\mathcal{M}_1(\lambda,z^2)$ is renormalization scheme independent, thus can be expressed in the $\overline{\rm MS}$ scheme with OPE,
\begin{align}
\mathcal{M}_1(\lambda,z^2) = &\frac{\sum_{n=0}C_n(\mu^2z^2) \frac{(-izP_z)^n}{n!} \langle x^n\rangle (\mu)}{\sum_{n=0}C_n(\mu^2z^2) \frac{(-izP_z^0)^n}{n!} \langle x^n\rangle (\mu) }\nonumber\\
& + \mathcal{O}(\Lambda_{{\mathrm{QCD}}}^2z^2),
\label{eq:ope}
\end{align}
where $\langle x^n\rangle$ are the moments of the helicity PDF, and in particular, $\langle x^0\rangle=g_A$ is the axial charge, $C_n(\mu^2z^2)$ are the Wilson coefficients for the helicity quasi-PDF. At NLO we have~\cite{Izubuchi:2018srq}, 
\begin{align}
&C_n(\mu^2z^2) = 1+\frac{\alpha_s C_F}{2\pi}\bigg{[}\left(\frac{3+2n}{2+3n+n^2}+2H_n\right)L_z\nonumber\\
& +\frac{7+2n}{2+3n+n^2}+2(1-H_n)H_n-2H_n^{(2)}\bigg{]}+\mathcal{O}(\alpha_s^2)
\label{eq:matching-kernel-nlo}
\end{align}
where $L_z=\ln\frac{\mu^2z^2e^{2\gamma_E}}{4}$, $H_n=\sum_{i=1}^n1/i$ and $H_n^{(2)}=\sum_{i=1}^n1/i^2$. At NNLO we can use the results for unpolarized case for $\gamma_z$ \cite{Li:2020xml} to obtain the corresponding Wilson coefficient.
Since the Wilson coefficients and the moments $\langle x^n\rangle$ are all real, when taking $P_z^0=0$, the real part of $\mathcal{M}_1(z^2,P)$ in Eq.~\eqref{eq:ope} only depends on the normalized even Mellin moments of the PDF $\langle x^{2k}\rangle/g_A$, while the imaginary parts only depends on the normalized odd Mellin moments $\langle x^{2k+1}\rangle/g_A$. The value of $g_A$ is obtained as $g_A^{u-d}=1.211(12)$ through independent analysis of the local matrix elements, as demonstrated in Appendix~\ref{app:av-charge}.

Although Eq.~\eqref{eq:ope} is RG invariant, the perturbative convergence of the Wilson coefficients in Eq.~\eqref{eq:matching-kernel-nlo} could still depend on the scale choice with fixed-order truncation, especially when the logarithm becomes large. It is thus more accurate to first extract the moments $\langle x^n(\mu_0)\rangle$ near the physical scale $\mu_0\sim\frac{2\kappa e^{-\gamma_E}}{z}$ with scale variation $\kappa\in[1/\sqrt{2},\sqrt{2}]$, which is perturbatively related to the commonly used fixed scale result at $\mu=2$~GeV  through the RG equations,
\begin{align}
    \frac{\partial}{\partial \ln\mu^2}\langle x^n(\mu)\rangle=\gamma_n(\alpha_s(\mu))\langle x^n(\mu)\rangle,
\end{align}
where the anomalous dimensions $\gamma_n(\alpha)$ of the helicity PDF moments are known up to 3-loop order~\cite{Moch:2014sna} and has been partially calculated at 4-loop order~\cite{Kniehl:2025ttz}. The solution to the RG equation is 
\begin{align}
    \langle x^n(\mu)\rangle=e^{\int_{\alpha_s(\mu_0)}^{\alpha_s(\mu)} \frac{\gamma_n(\alpha)}{\beta(\alpha)}d\alpha}\langle x^n(\mu_0)\rangle,
\end{align}
with the QCD beta functions $\beta(\alpha)$ known up to five loop order~\cite{Herzog:2017ohr}. For next-to-leading-log (NLL) perturbative accuracy, we use one-loop Wilson coefficient~\cite{Izubuchi:2018srq} with two-loop anomalous dimensions and two-loop QCD beta functions; for next-to-next-to-leading-log (NNLL) perturbative accuracy, we use two-loop Wilson coefficient~\cite{Li:2020xml} 
with three-loop anomalous dimensions~\cite{Moch:2014sna} and three-loop QCD beta functions.
Because $\alpha_s(\sqrt{2}e^{-\gamma_E}/z)\approx 1.04$ becomes too large at $z=4a$ in the NNLL Wilson coefficients, we perform independent fits at different $z$-values in the perturbative region $z\in[1,3]a$ which satisfies $z<0.3$~fm, as well as a correlated combined fit for all these $z$ values. Combined with the $g_A$ values above, we show the results in Fig.~\ref{fig:moment_fit}. The combined correlated fit among different $z$ values are consistent with individual fits, except for the higher moments that the data are insensitive to. The moments fitted from our short-distance matrix elements are $\langle x\rangle_{u-d}=0.291(7)$, $\langle x^2\rangle_{u-d}=0.101(10)$, $\langle x\rangle_{u+d}=0.183(6)$, $\langle x^2\rangle_{u+d}=0.073(3)$ at $\mu=2$~GeV, higher than those obtained from integrating the global fitting PDF by JAM22~\cite{Buckley:2014ana,Cocuzza:2022jye}. The uncertainties include both the statistical error and  the scale variations in the range $\kappa\in[1/\sqrt{2},\sqrt{2}]$.

\section{The $\tilde{d}_2$ moment}
\label{sec:g2}

\begin{figure*}[tbh]
\includegraphics[width=0.48\textwidth]{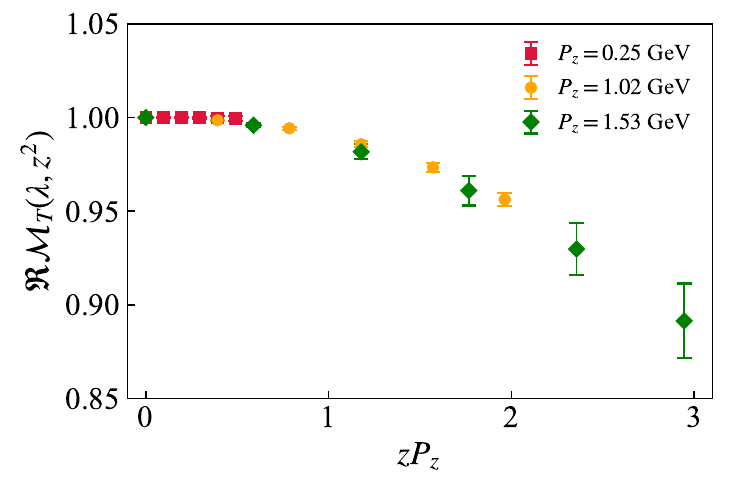}
\includegraphics[width=0.48\textwidth]{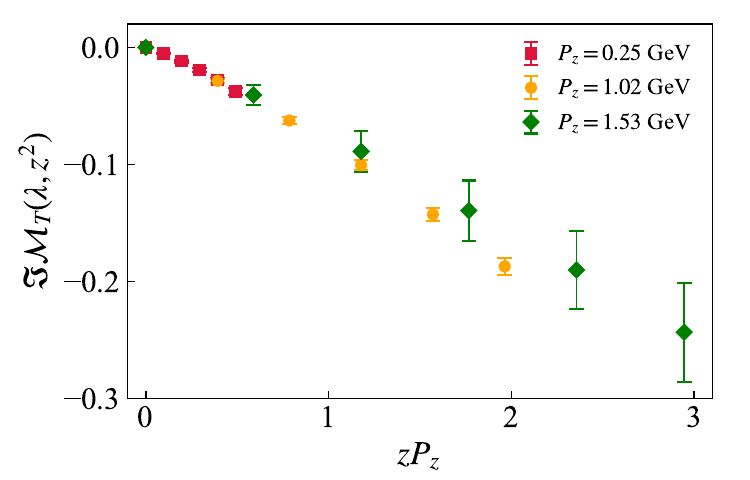}
\caption{Real (left) and imaginary (right) parts of the RG-invariant ratios $\mathcal{M}_T(\lambda,z^2)$
for $z\in[a,5a]$.}
\label{fig:itdg2}
\end{figure*}

In analogy with the RG invariant ratio introduced in
\Eq{itd}, we construct an ratio for the transverse correlator relevant for $g_T$,
\begin{equation}
\mathcal{M}_T(\lambda,z^2)\equiv
\frac{\tilde{h}^B_T(z,P_z)}{\tilde{h}^B_T(z,0)}\,
\frac{\tilde{h}^B_T(0,0)}{\tilde{h}^B_T(0,P_z)}
=\frac{\tilde{h}^{\overline{\rm MS}}_T(\lambda,z^2\mu^2)}{\tilde{h}^{\overline{\rm MS}}_T(0,z^2\mu^2)}.
\label{eq:RGIgT}
\end{equation}
By construction, $\mathcal{M}_T(\lambda,z^2)$ cancels the multiplicative renormalization associated
with the Wilson line as well as the leading UV-sensitive factors common to numerator and denominator,
so that the residual $\mu$ dependence is governed by short-distance logarithms in the factorization
kernels. In practice, $\mathcal{M}_T(\lambda,z^2)$ provides a stable observable for implementing the
short-distance expansion in $z^2$ and for isolating the twist-3 physics through its $\lambda$-dependence. Our numerical results on ${\cal M}(\lambda,z^2)$ as function of
$\lambda$ for different $P_z$ are shown in Fig.~\ref{fig:itdg2}.

We adopt the standard moment conventions
\begin{align}
  &\int_{-1}^{1} dx\,x^n\,g_1(x)=a_n,\nonumber\\
  &\int_{-1}^{1} dx\,x^n\,g_2(x)=\frac{n}{n+1}\big(d_n-a_n\big),
\end{align}
with $g_T(x)=g_1(x)+g_2(x)$. Following the convention used in Ref.~\cite{SANE:2018pwx}, we define
$\tilde d_n\equiv 2d_n$ and introduce
\begin{equation}
  \tilde d_2\equiv \int_{-1}^{1} dx\,x^2\,[3g_T(x)-g_1(x)]
  =\int[dx]\;S^-(x_1,x_2,x_3),
\end{equation}
where $\int[dx]=\int_{-1}^{1} dx_1\,dx_2\,dx_3\,\delta(x_1+x_2+x_3)$ and $S^-(x_1,x_2,x_3)$ is the
quark-antiquark-gluon correlator entering the genuine twist-3 contribution. This matrix element
is the lowest-dimension chiral-even twist-3 reduced matrix element and is the quantity we extract
from the short-distance behavior of $\mathcal{M}_T$.

The central ingredient is the short-distance factorization derived in Ref.~\cite{Braun:2021aon}, which
relates the transverse correlator to the longitudinal one and makes the appearance of $\tilde d_2$
at $\mathcal O(\lambda^2)$ explicit, with more details presented in Appendix~\app{gTSDF}. In the $\overline{\rm MS}$ scheme, we have the NLO matching~\cite{Braun:2021aon},
\begin{widetext}
\begin{align}\label{eq:hTh1}
\begin{split}
  \tilde{h}_T^{\overline{\rm MS}}(\lambda,z^2\mu^2)
  &
  =\int_0^1 d\alpha\;\tilde{h}_1^{\overline{\rm MS}}(\alpha\lambda,z^2\mu^2)-\frac{\alpha_s C_F}{\pi}\, a_0
  + i\,\lambda\,\frac{\alpha_s C_F}{3\pi}\, a_1 \\
  &\quad+\frac{\lambda^2}{3}\Bigg\{\tilde d_2
  +\frac{7\alpha_s C_F}{24\pi}\, a_2
  +\frac{\alpha_s}{4\pi}\tilde d_2\Big[\Big(\frac{43C_F}{12}+\frac{9}{4N_c}+\frac{3N_c}{4}\Big)L_z
  -\frac{13C_F}{4}+\frac{3}{4N_c}-\frac{3N_c}{4}\Big]\Bigg\}
  +\mathcal O(\lambda^3),
\end{split}
\end{align}

\end{widetext}
where $L_z=\ln\!\big(z^2\mu^2/(4e^{-2\gamma_E})\big)$.
Equation~\Eq{hTh1} exhibits two important features: (i) the naive WW subtraction does not
remove twist-2 effects at NLO, yielding the $\mathcal O(\alpha_s)\,a_n$ terms, and (ii) the genuine
twist-3 matrix element $\tilde d_2$ enters for the first time at $\mathcal O(\lambda^2)$ with a
perturbatively calculable coefficient. This provides a clean lever arm for extracting $\tilde d_2$
from the $\lambda$-dependence at fixed short distances. 

At zero momentum, \Eq{hTh1} reduces to,
\begin{align}
    \tilde{h}_T^{\overline{\rm MS}}(0,z^2\mu^2)
  =\tilde{h}_1^{\overline{\rm MS}}(0,z^2\mu^2)-\frac{\alpha_s C_F}{\pi}\, a_0,
\end{align}
where $\tilde{h}_1^{\overline{\rm MS}}(0,z^2\mu^2)=C_0(z^2\mu^2)a_0$ with $C_0$ from \Eq{matching-kernel-nlo} and $a_0=g_A$ derived in Appendix~\app{av-charge}. To implement the short-distance factorization above, we insert them into the RG-invariant ratios \Eq{itd} and \Eq{RGIgT}.
Working consistently to NLO, we define the 
subtracted combination
\begin{widetext}
\begin{align}\label{eq:MTratiotw3}
\begin{split}
  \mathcal{M}_T^{\rm tw3}(\lambda,z^2\mu^2)
  &\equiv
  \mathcal{M}_T(\lambda,z^2\mu^2)
  - \frac{\tilde{h}_1^{\overline{\rm MS}}(0,z^2\mu^2)}{\tilde{h}_T^{\overline{\rm MS}}(0,z^2\mu^2)}\int_0^1 d\alpha\,\mathcal{M}_1(\alpha\lambda,z^2\mu^2)-\frac{1}{\tilde{h}_T^{\overline{\rm MS}}(0,z^2\mu^2)}\left(-\frac{\alpha_s C_F}{\pi} a_0+i\,\lambda\,\frac{\alpha_s C_F}{3\pi}\, a_1 \right)\\
  &=\frac{\lambda^2}{3}\frac{\Bigg\{\tilde d_2
  +\frac{7\alpha_s C_F}{24\pi}\, a_2
  +\frac{\alpha_s}{4\pi}\tilde d_2\Big[\Big(\frac{43C_F}{12}+\frac{9}{4N_c}+\frac{3N_c}{4}\Big)L_z
  -\frac{13C_F}{4}+\frac{3}{4N_c}-\frac{3N_c}{4}\Big]\Bigg\}
  }{C_0(z^2\mu^2)a_0-\frac{\alpha_s C_F}{\pi}\, a_0}+\mathcal O(\lambda^3)
\end{split}
\end{align}
\end{widetext}
The formula makes explicit that, after the NLO-improved
twist-2 subtraction, the real part of $\mathcal{M}_T^{\rm tw3}$ scales as $\lambda^2$ at small $\lambda$, with the
leading coefficient controlled by $\tilde d_2$ (and a residual calculable twist-2 contamination
proportional to $a_2$). In the numerical analysis we use the helicity moments $a_n$ determined in
\sec{mellin-moments} as external inputs. For each fixed $z$, the $\alpha$ integral is evaluated numerically by interpolating $\mathcal{M}_1(\alpha\lambda,z^2\mu^2)$ as a function of $\alpha\lambda$.

\begin{figure}[tbh]
\includegraphics[width=0.45\textwidth]{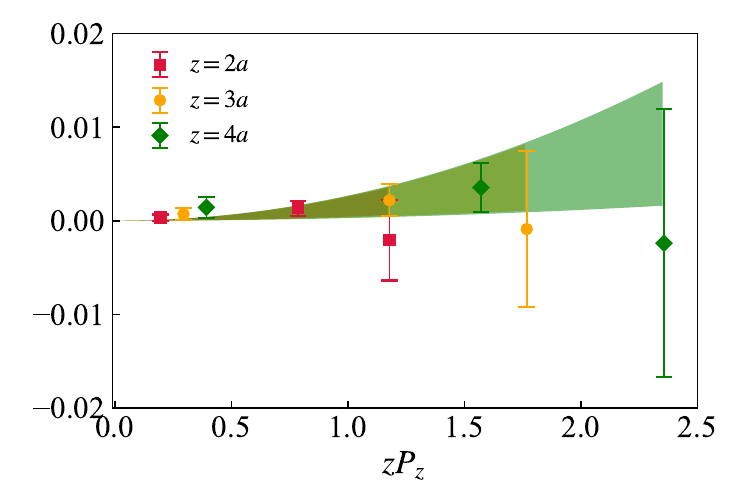}
\caption{The subtracted ratio $\mathcal{M}_T^{\rm tw3}(\lambda,z^2\mu^2)$ defined in 
\Eq{MTratiotw3}. The colored bands are reconstructed from the combined fit over $z\in[2a,4a]$.}
\label{fig:subtraction}
\end{figure}

\begin{figure}[tbh]
\includegraphics[width=0.45\textwidth]{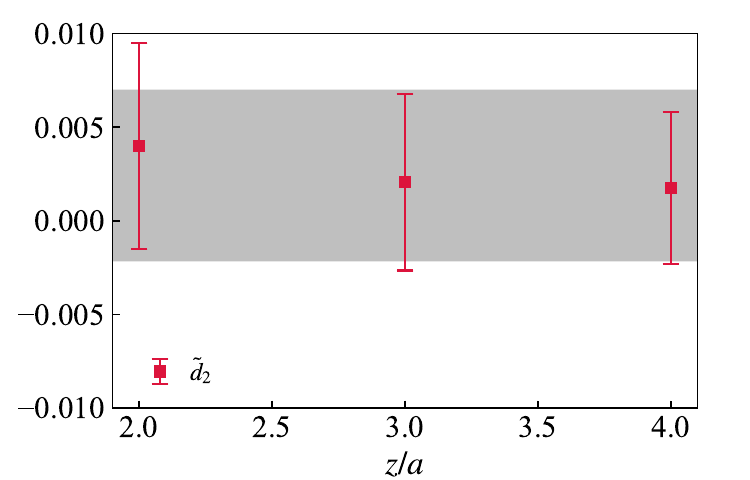}
\caption{The extracted NLO $\tilde d_2^{u-d}$ from fixed-$z$ fits as a function of $z$ for $z\in[2a,4a]$ at $\mu=2$ GeV. }
\label{fig:momentsd2}
\end{figure}

Our strategy is as follows. For each fixed $z$ in the short-distance window, we construct
$\mathcal{M}_T^{\rm tw3}(\lambda,z^2\mu^2)$ from the lattice data according to \Eq{MTratiotw3} and
fit its $\lambda$ dependence to the expected small-$\lambda$ behavior. This yields an estimate of
$\tilde d_2$ at each $z$. The resulting subtracted ratios are shown in \fig{subtraction}. We find
that $\mathcal{M}_T^{\rm tw3}(\lambda,z^2\mu^2)$ remains close to zero within uncertainties over
the range of $\lambda$ shown, already indicating a very small genuine twist--3 contribution and
hence a $\tilde d_2$ close to zero. Correspondingly, the extracted $\tilde d_2^{u-d}$ values from
fixed-$z$ fits, displayed in \fig{momentsd2}, are all statistically consistent with zero for
$z\in[2a,4a]$.

To obtain our final result, we perform a combined fit over $z\in[2a,4a]$, using the full NLO
kernel in \Eq{MTratiotw3} and a common $\tilde d_2$ parameter. We find
\begin{equation}
  \tilde d_2^{u-d}(\mu=2\ {\rm GeV})=0.0024(46),
\end{equation}
where the quoted uncertainty is statistical, and additional systematic effects remain to be studied. The result of the combined fit is shown as the gray band
in \fig{momentsd2}, while the corresponding colored bands in \fig{subtraction} are reconstructed
from the same fit. This combined-fit result is likewise consistent with
zero and thus supports the same conclusion suggested by the near-vanishing subtracted ratios. Physically, a small
$\tilde d_2^{u-d}$ implies that the lowest moment of the genuine chiral-even twist-3 quark-gluon
correlation in the isovector channel is strongly suppressed at the scale considered. Equivalently,
the $x^2$-weighted moment that isolates the genuine twist-3 part of $g_T(x)$ (or, in DIS language,
of $g_2(x)$ beyond the WW contribution) is small, indicating that quark-gluon correlation effects
in this particular moment are suppressed relative to the leading-twist structure. 

\section{Quark helicity PDFs from LaMET}
\label{sec:helicity-lamet}
The $x$-dependence of the helicity PDFs can be calculated 
through the LaMET approach~\cite{Ji:2013dva,Ji:2014gla,Ji:2020ect}, based on the expansion~\cite{Ji:2024oka},
\begin{align}\label{eq:factorization}
    g_1(x,\mu)=&\int \mathcal{C}(x,y,\mu,P_z)\tilde{g}_1(x,P_z,\mu)\nonumber\\
    &+\mathcal{O}\left(\frac{\Lambda_{\rm QCD}^2}{(2xP_z)^2},\frac{\Lambda_{\rm QCD}^2}{(2(1-x)P_z)^2}\right),
\end{align}
where $\mathcal{C}(x,y,\mu,P_z)$ is the perturbative matching known to 3-loop order~\cite{Li:2020xml,Chen:2020ody,Cheng:2024wyu}, $\tilde{g}_1$ is the $x$-dependent quasi-PDF obtained from the matrix elements $\tilde{h}_1^R(z, P_z, \mu)$,
\begin{align}
    \tilde{g}_1(x, P_z, \mu) = \int_{\infty}^{\infty}\frac{P_z\mathrm{d}z}{2\pi}e^{-ixP_z z}\ \tilde{h}_1^R(z, P_z, \mu),
\end{align}
renormalized in a scheme perturbatively convertible to $\overline{\rm MS}$.

The analysis thus includes the following steps: renormalization of the bare matrix elements in a scheme perturbatively convertible to $\overline{\rm MS}$; asymptotic extrapolation of the exponentially decaying long tail followed by a Fourier transformation to momentum space quasi-PDF; perturbative matching to light-cone PDF.

\subsection{Hybrid renormalization}
\label{sec:sub-hybrid}

Due to the self energy of the Wilson line and the vertices involving Wilson line and light quark, the bare matrix element calculated in the light-front frame on the lattice, $\tilde{h}_1^B(z, P_z)$, contains both linear and logarithmic ultraviolet (UV) divergences. These divergences can be regularized in various ways multiplicatively~\cite{Chen:2016fxx,Ji:2017oey,Ishikawa:2017faj,Green:2017xeu,Zhang:2018diq,Li:2018tpe}. Recently, it was proposed to renormalize the lattice data in the hybrid scheme~\cite{Ji:2020brr}. Structurally, the hybrid renormalization can be split into two parts
\begin{equation}
\begin{split}
& \tilde{h}_1^R(z,P_z)  =  \begin{cases} \frac{\tilde{h}_1^B(z, P_z, a)}{\tilde{h}_1^B(z,0,a)} & |z|\leq z_s \\
\frac{\tilde{h}_1^B(z, P_z, a)}{\tilde{h}_1^B(z_s,0,a)}e^{(\delta m+m_0)(z-z_s)} & |z|\geq z_s \end{cases}. \\
\end{split}
\label{eq:hyb}
\end{equation}
At small distances, it is can be renormalized in a ratio-type scheme~\cite{Orginos:2017kos,Chou:2022drv}. At large distances, it subtracts the linear divergence in the Wilson line as well as the logarithmic UV divergences.
The hybrid scheme renormalization constants include no nonperturbative scales at any $z$ value, thus is perturbatively convertible to the $\overline{\rm MS}$ scheme, where the factorization has been rigorously proven.  The overall logarithmic divergence in $a$ is canceled by the zero-momentum bare matrix elements, and the linear divergence is eliminated by a mass counterterm $\delta m$. For the ensemble used in this work, we have $a\delta m=0.1597(16)$ determined from the static quark-antiquark potential~\cite{HotQCD:2014kol,Bazavov:2016uvm,Bazavov:2017dsy,Bazavov:2018wmo}. Note, however, that $\delta m$ is scheme dependent and contains an infrared ambiguity of $\sim\mathcal{O}(\Lambda_{\rm QCD})$ known as the renormalon~\cite{Ji:2020brr}. To avoid introducing such ambiguity into the final light-cone PDF, it is proposed to define a renormalon regularization scheme $\tau$ and apply the coordinate-space renormalization and momentum-space perturbative matching consistently in the same $\tau$ scheme~\cite{Zhang:2023bxs,Holligan:2023rex}. In this approach, one defines a reasonable regularization for the renormalon in the perturbative coefficients as scheme $\tau$, then introduce a nonperturbative parameter $m_0(\tau)$ to convert $\delta m$ into the same $\tau$-scheme. Since such scheme definition works for all hadron momentum, it can also be applied to the renormalized $P=0$ matrix elements which should be identical to the perturbative Wilson coefficient $C_0$ in the same scheme. Once regularized, $m_0$ is determined by~\cite{Zhang:2023bxs},
\begin{align}
\label{eq:master_2}
    \frac{\tilde{h}_1^B(z,0,a)}{\tilde{h}_1^B(z',0,a)}e^{|z-z'|(\delta m(a)+m_0)}=\frac{C_0(z,\mu_0)e^{\mathcal{-I}(\mu_0)}}{C_0(z',\mu_0')e^{\mathcal{-I}(\mu_0')}},
\end{align}
where $\mu_0=2\kappa e^{-\gamma_E}/z$ and $\mu_0'=2\kappa e^{-\gamma_E}/z'$ are the physical scales of the system,  $\mathcal{I}(\mu)=\int d\alpha \frac{\gamma(\alpha)}{\beta(\alpha)}|_{\alpha=\alpha_s(\mu)}$ is the renormalization group evolution factor of $C_0$ that relates the quantity at two different physical scales, and $\kappa$ can be varied to examine the scale variations uncertainty. It is suggested in Ref.~\cite{Zhang:2023bxs} to define the regularization scheme of the renormalon series through a leading renormalon resummation (LRR) of its asymptotic form with principal value (PV) prescription, which improves the perturbative convergence of the Wilson coefficients. In this scheme, the Wilson coefficients are corrected as
\begin{widetext}
\begin{equation}
\begin{split}
C'_{0,\mathrm{PV}}\left(\alpha_s(\mu_0)\right) &= C^{\mathrm{LRR}}_{0,\mathrm{PV}}\left(\alpha_s(\mu_0)\right) +\left[ C_0\left(\alpha_s(\mu_0)\right) - C^{\mathrm{LRR,NNLO}}_{0,\mathrm{PV}}\left(\alpha_s(\mu_0)\right)\right]\\
&=C_0\left(\alpha_s(\mu_0)\right) +2N_m\frac{-2\pi e^{-\frac{2\pi}{\alpha_s\beta_0}}}{\beta_0}\Re\left[E_{b+1}\left(\frac{2\pi}{\beta_0}\right)+c_1E_{b}\left(\frac{2\pi}{\beta_0}\right)+c_2E_{b-1}\left(\frac{2\pi}{\beta_0}\right)\right]\\
&\ \ \ -2\alpha_s\kappa e^{-\gamma_E}N_m(1+c_1+c_2) - 2\alpha_s^2\kappa e^{-\gamma_E}N_m\frac{2\beta_0^2(1-c_2)+\beta_1(1+c_1+c_2)}{4\pi\beta_0},
\end{split}
\label{eq:lrr}
\end{equation}
\end{widetext}
where $\beta_0=9$, $\beta_1=64$, $\beta_2=643.83$ are coefficients of the QCD beta functions, the constants are numerically $b=0.395$, $c_1=-0.164$, $c_2=0.237$ and $N_m=0.575$ for $n_f=3$~\cite{Pineda:2001zq}, with more detailed definition in Appendix~\ref{app:matching}. $E_n(z)=\int_1^\infty e^{-zt}/t^n$ is the generalized exponential integral function~\cite{Zhang:2023bxs,Ding:2024saz}. Note that the above equations is defined at the physical scale $\mu_0$ because the renormalon lives in the constant terms of the perturbative series. It should be regularized at the system's physical scale to avoid being introduced into the logarithms through the RG evolution.

\begin{figure}
    \centering
    \includegraphics[width=0.9\linewidth]{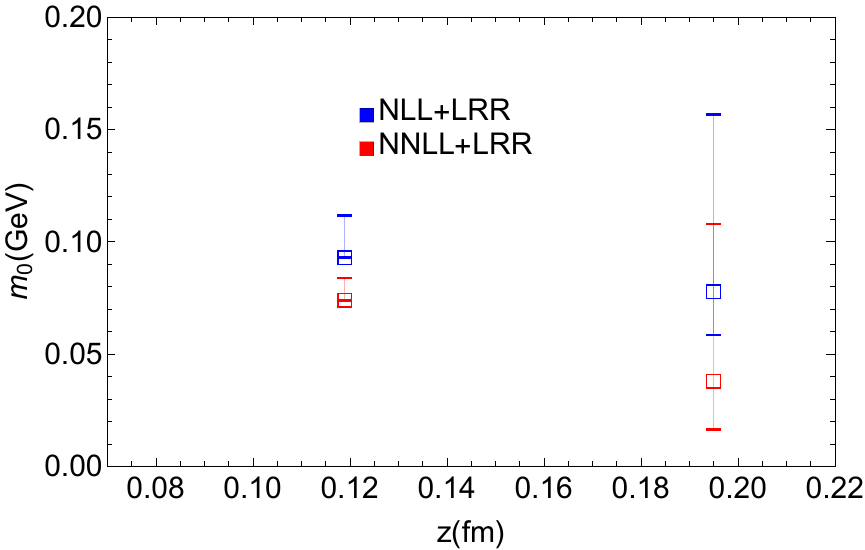}
    \caption{The $m_0$ parameter in the renormalization extracted at different $z$ and perturbative order. The error bar is the systematic uncertainty by varying $\kappa\in[\frac{1}{\sqrt{2}},\sqrt{2}]$.}
    \label{fig:m0_extraction}
\end{figure}

We extract the $m_0$ from the $\frac{\tilde{h}_1^B(z,0,a)}{\tilde{h}_1^B(z',0,a)}$ ratio with $\{z,z'\}=\{a,2a\}$ and $\{z,z'\}=\{2a,3a\}$, as shown in Fig. ~\ref{fig:m0_extraction}. The size of $z$-dependence and perturbative convergence is small, consistent with the observation in  previous works~\cite{Zhang:2023bxs}. The statistical error in the fitting is negligible, and the error bar shown is the systematic uncertainty by varying $\kappa\in[\frac{1}{\sqrt{2}},\sqrt{2}]$.

\subsection{Asymptotic extrapolation and Fourier transformation}
The $x$-dependent quasi PDFs are obtained from a Fourier transform of the renormalized matrix elements $\tilde{h}^R_1(\lambda, P_z)$, 
which requires its knowledge  in the entire $z$ space. However, on the lattice the $z$-range
is limited and the statistical errors on the matrix element grow exponentially at large $z$. 
The unphysical fluctuations at large $z$ could introduce unnecessary noise to the $x$-dependence, which has raised the question of how the Fourier transform uncertainty can be quantified~\cite{Dutrieux:2025jed}.
Since the spatial correlation functions decay exponentially at large $|z|$, an extrapolation based on their asymptotic expansion can be performed to enable the Fourier transform and estimate the associated uncertainties~\cite{Ji:2020brr,Gao:2021dbh,Chen:2025cxr}. 
The uncertainty from such an extrapolation is proven to be bounded from above, with the bound inversely proportional to $x$~\cite{Gao:2021dbh} and systematically improvable. The first proposed
functional form is inspired by the dispersive analysis and Regge behavior~\cite{Ji:2020brr,Gao:2021dbh} (labeled ``form 1''),
\begin{equation}
\tilde{h}^{\mathrm{asy}}_1(\lambda, P_z)= \frac{A e^{-m_{\mathrm{eff}}\lambda/P_z}}{|\lambda|^d},
\label{eq:extrp-model}
\end{equation}
where $A$, $d$ and $m_{\mathrm{eff}}$ are fit parameters, determined from joint fits of the real and imaginary part of $\tilde{h}_1^R(\lambda, P_z)$. The fit results suggest $m_{\mathrm{eff}}\approx 0.77(60)$ GeV with  $\chi^2/\rm{d.o.f}=1.47(1.39)$, corresponding to the effective mass related to the intermediate state. 

Recently, there has been a rigorous derivation of large-$z$ behavior, suggesting the following form up to next-to-leading asymptotics (labeled ``Form 2'')~\cite{Ji:2026vir},
\begin{equation}
\tilde{h}^{\mathrm{asy}}_1(z, P_z)= \left(Ae^{i\phi{\rm sgn}(z)}+\frac{A'e^{i\phi'{\rm sgn}(z)}}{|z|}\right)e^{-|z|\Lambda},
\label{eq:extrp-model}
\end{equation}
where $A, A',\phi,\phi',\Lambda$ are fit parameters.  The leading term proportional to $A$ is equivalent to ``Form 1''.
``Form 2'' has been demonstrated to describe the lattice data well beyond $z>\Lambda^{-1}$, and $\Lambda\sim0.5$~GeV estimated from the binding energy of a heavy-light pseudoscalar meson~\cite{Ji:2026vir}. Our fit results suggest $\Lambda\approx 0.47(11)$~GeV with $\chi^2/\rm{d.o.f}=0.78(0.68)$, consistent with the estimation.

\begin{figure*}[tbh]
\centerline{
\includegraphics[width=0.5\textwidth]{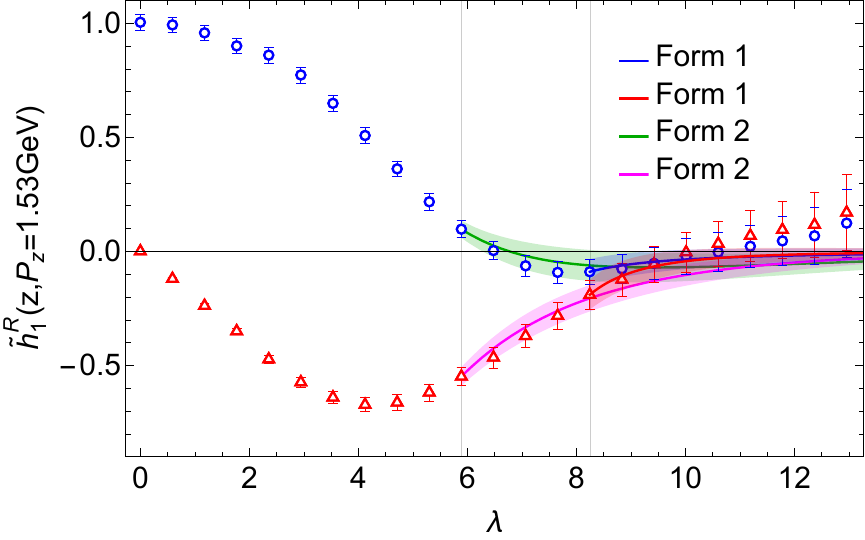}
\includegraphics[width=0.5\textwidth]{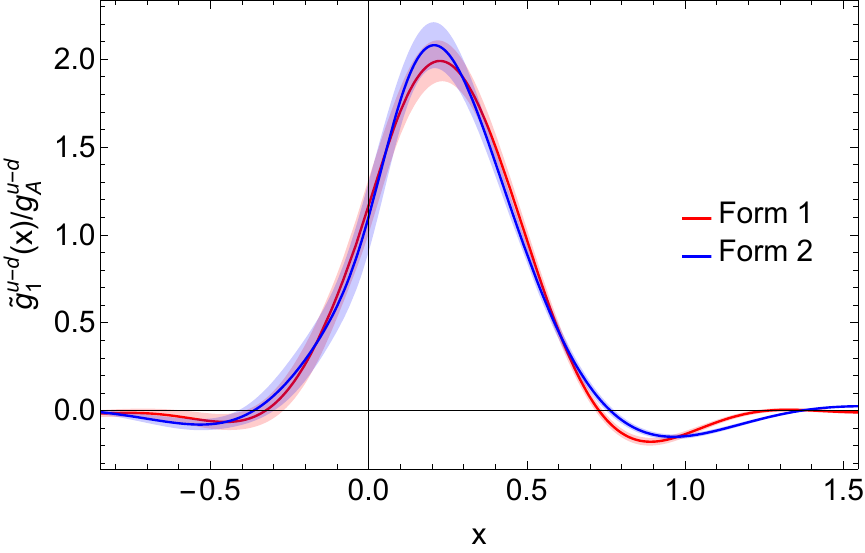}
}
\caption{Left: The asymptotic extrapolation for the real (blue) and imaginary (red) part of the isovector matrix elements obtained from the $P_z=1.53$ GeV data. The gray vertical grid lines labels where the extrapolation starts. Right: The corresponding Fourier transform to $x$-space quasi-PDFs. There is very little difference between renormalized quasi-PDF with NLL and NNLL accuracy from the slightly different $m_0$.  }
\label{fig:ft}
\end{figure*}

With these extrapolations, we then obtain the $x$ dependence by including the Fourier transformation of interpolated lattice data in $[-\lambda_0,\lambda_0]$ plus the long-tail contribution from the extrapolation.
An example of the extrapolation and the corresponding Fourier transformation is shown in Fig.~\ref{fig:ft}. In most regions of interest, the two extrapolations are consistent. We will include the discrepancy between the two extrapolations as a systematic error into the final results.

\subsection{Perturbative matching}

The perturbative matching Eq.~\eqref{eq:factorization} is applied to extract the light-cone PDF from the quasidistributions. The matching kernel can be expanded perturbatively in $\alpha_s$,
\begin{equation}
\mathcal{C}\left(\frac{x}{y}, \mu, z_s, P_z\right) = \delta \left(\frac{x}{y}-1 \right)+\sum_{n=1}^\infty \alpha_s^n\mathcal{C}^{(n)}\left(\frac{x}{y}, \mu, z_s, P_z \right).
\end{equation}
which has been calculated to NNLO in $\overline{\rm MS}$ scheme~\cite{Li:2020xml,Chen:2020ody}. The NNLO hybrid scheme correction has been derived for the unpolarized iso-vector PDF in Ref.~\cite{Su:2022fiu}, thus we follow a similar procedure for the helicity PDF, with more details provided in Appendix~\ref{app:matching}.

The higher-order logarithms in the matching kernel are non-negligible when any of the physical scales in the system becomes small, thus require a resummation. The RGR is needed for the region $x\to0$ when the physical scale $2xP_z$ related to the parton's momentum becomes soft~\cite{Su:2022fiu}, and the threshold resummation (TR) is needed for the region $x\to1$ when the physical scale $2(1-x)P_z$ related to the spectator's momentum becomes small~\cite{Ji:2023pba,Ji:2024hit}. 

To resum these logarithms, the matching kernel is evolved from the initial scale $\mu_h=2xP_z$~\cite{Su:2022fiu},
\begin{align}
    \mathcal{C}(x,y,\mu,P_z)= P_{\mathcal{C}}(\mu,\mu_h)\otimes\mathcal{C}(\mu_h,P_z),
\end{align}
where $P_{\mathcal{C}^{-1}}(\mu_h,\mu)$ is the same as the evolution kernel of the helicity PDF~\cite{Moch:2014sna}, and a further threshold factorization is applied to the matching kernel in the threshold limit $x\to y$~\cite{Ji:2024hit},
\begin{align}
    C(x,y,\mu_h,P_z)\xrightarrow{x\to y}S(\mu_h)\otimes H(\mu_h),
\end{align}
with resummed the soft function $S(\mu_h)$ (also used as the jet function $J$) from the semihard scale $\mu_i=2(1-x)P_z$,
\begin{align}
    S(\mu_h)= P_S(\mu_h,\mu_i)S(\mu_i),
\end{align}
where $P_S(\mu_h,\mu_i)$ is the evolution kernel of the soft function $S(\mu)$~\cite{Becher:2006mr}.

\begin{figure}
    \centering
    \includegraphics[width=0.99\linewidth]{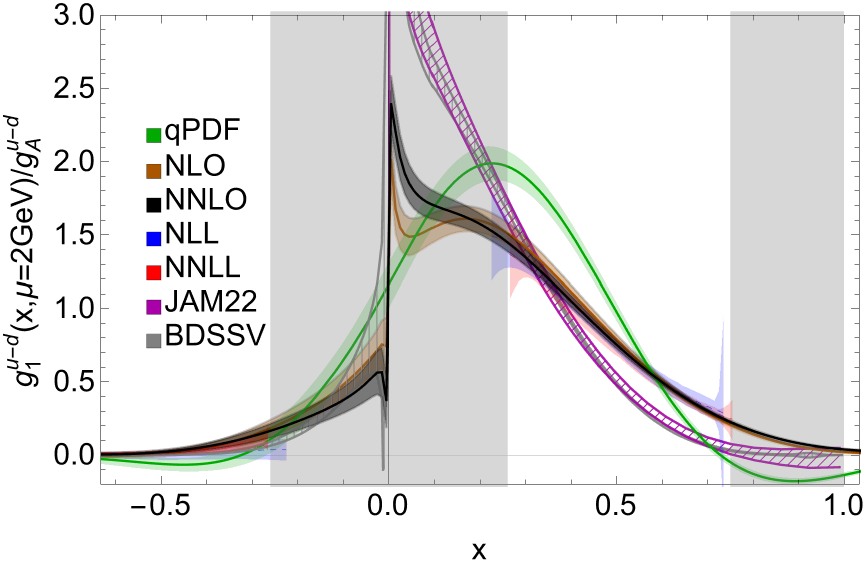}
    \caption{The isovector helicity PDFs at $\mu=2$~GeV, matched at different perturbative orders, compared with the global fits (hatched bands) from JAM22~\cite{Cocuzza:2022jye} and BDSSV24~\cite{Borsa:2024mss}.}
    \label{fig:matching_full}
\end{figure}

\begin{figure}
    \centering
    \includegraphics[width=0.99\linewidth]{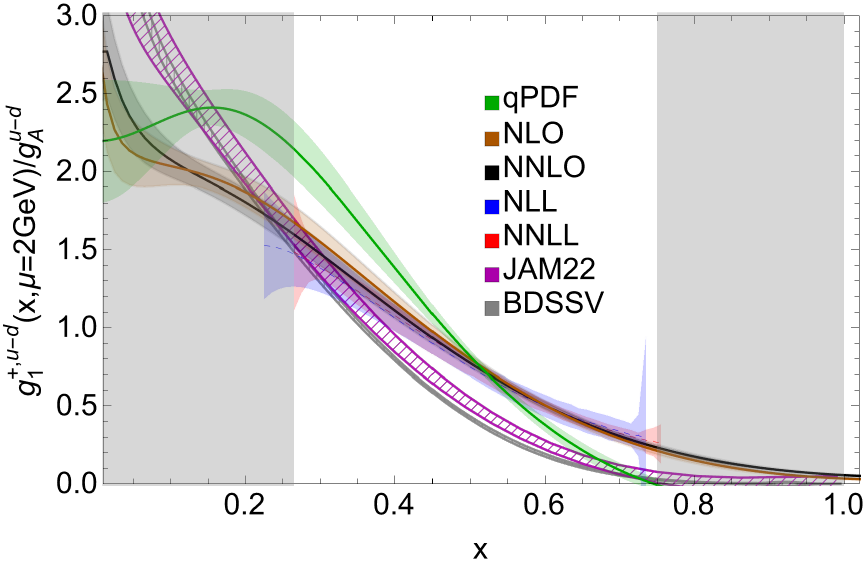}
    \caption{The $g_1+\bar{g}_1$ isovector helicity PDFs at $\mu=2\rm{~GeV}$, normalized by $g_A$ and matched at different perturbative order, compared with the global fitting PDF from JAM22~\cite{Cocuzza:2022jye} and BDSSV24~\cite{Borsa:2024mss}.}
    \label{fig:matching_real}
\end{figure}

In this work, we include both resummations to improve the matching kernel, with the LRR implicitly applied to cancel the linear power correction. The $x$-dependence before and after matching are shown in Fig.~\ref{fig:matching_full}. Meanwhile, we also show the $g_1^{+}(x)\equiv g_1(x)+g_1(-x)$ results from the real part of the matrix elements in Fig.~\ref{fig:matching_real}. The error band in the NLL and NNLL matching includes both statistical error and theoretical uncertainties, the latter is estimated from scale variation of the initial scale of resummation by a factor of $\kappa=\{\frac{1}{\sqrt{2}},\sqrt{2}\}$ and the discrepancy between the two different asymptotic extrapolation. {The results with different perturbative accuracy are consistent, but the NNLO matching has reduced scale variation in the perturbative region, which is expected because the uncertainty comes from higher order terms.}  
After resummation, we obtain the $x$-dependence for a region around $x\in[0.25,0.75]$ where the theoretical uncertainty from scale variations remains small. This range is mostly determined by where the physical scales $2xP_z$ and $2(1-x)P_z$ are both nonperturbative, such that the perturbative matching is still reliable, so it is inverse proportional to the hadron's momentum $P_z$.  {Our results suggest a more elevated distribution in the mid-$x$ region, $x\in[0.4,0.7]$, compared to the global fitting results. To understand the discrepancy requires more future studies on the lattice systematics and with higher hadron momentum.}

\begin{figure}
    \centering
    \includegraphics[width=0.99\linewidth]{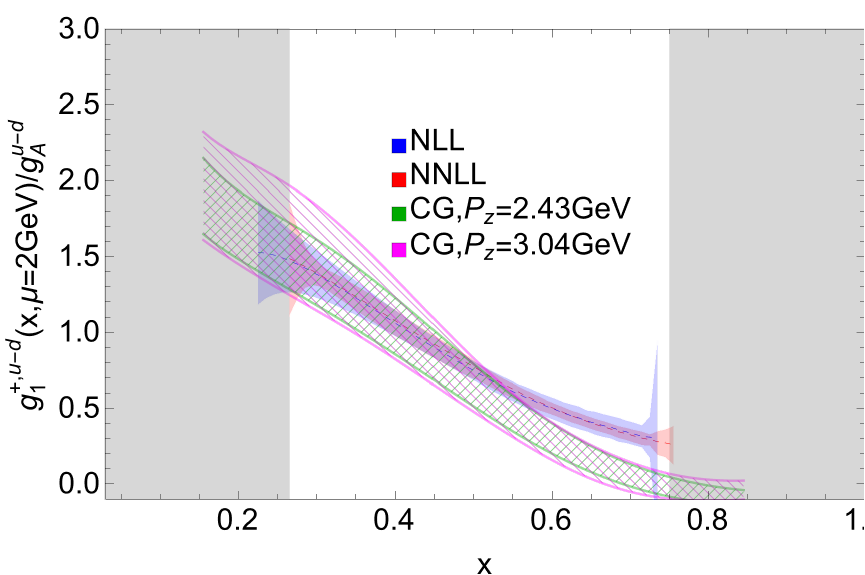}
    \caption{Comparison with the helicity PDF calculation from the Coulomb gauge method~\cite{Gao:2026hix}.}
    \label{fig:compare_cg}
\end{figure}

Comparing our results with the recent helicity PDF calculation in Coulomb gauge without a Wilson link at $3.04$~GeV~\cite{Gao:2026hix}, we observe a good consistency for $x<0.6$, but noticeable discrepancies for $x>0.6$, as shown in Fig.~\ref{fig:compare_cg}. Although there are many other lattice artifacts not being well controlled, the  discrepancy for $x>0.6$ could potentially come from the power corrections as we calculate at a much lower momentum $P_z\approx1.5$~GeV. 

To extend the LaMET's applicable range and reduce the power corrections, we need to access higher $P_z$ in future calculations, which could be improved with the help of the kinematically enhanced interpolators~\cite{Zhang:2025hyo}.

\subsection{Constraining the end point regions using SDF}
Since the LaMET expansion does not work for the end point regions $x\to\{0,1\}$, it has been proposed to compliment the $x$-dependence through model fitting of the end point regions with SDF or OPE ~\cite{Ji:2022ezo,Holligan:2023rex}.
In SDF analysis, the $x$-dependence of the PDFs are usually modeled by a functional form,
\begin{align}
    g_1(x)=x^a(1-x)^b\times({\rm  some \ continuous \ functions}),
\end{align}
where the $x^a$ and $(1-x)^b$ components come from the Regge behavior {and power counting rules}, while it is challenging to make the continuous function flexible enough to cover the actual $x$-dependence in the mid-$x$ region. The parameters $\{a,b\}$ and the continuous function model are all fitted to coordinate space lattice data $h(z^2,z\cdot P)$ at $z\ll \Lambda^{-1}_{\rm QCD}$, which is related to the $x$-dependent PDF through~\cite{Radyushkin:2017cyf,Ji:2017rah,Ma:2017pxb}, 
\begin{align}
    h(z^2,z\cdot P)=\int_{-1}^{1} K(x\omega,z^2,\mu)g_1(x,\mu) dx,
\end{align}
where the matching kernel $K(x\omega,z^2,\mu)$ is calculated to three loops~\cite{Li:2020xml,Chen:2020ody,Cheng:2024wyu} in $\overline{\rm MS}$, and are numerically evaluated up to NNLO for our work.
Since we are fitting to the short-distance data, it is convenient to actually fit to the RG-invariant ratio $\mathcal{M}_1(z^2,P,P^0)$, which is obtained by converting 
\begin{align}
    K(xw,z^2,\mu)\to \frac{K(xw,z^2,\mu=\frac{2e^{-\gamma_E}}{z})}{C_0(z^2,\mu=\frac{2e^{-\gamma_E}}{z})},
\end{align}
both at NNLO accuracy.

\begin{figure}
    \centering
    \includegraphics[width=0.99\linewidth]{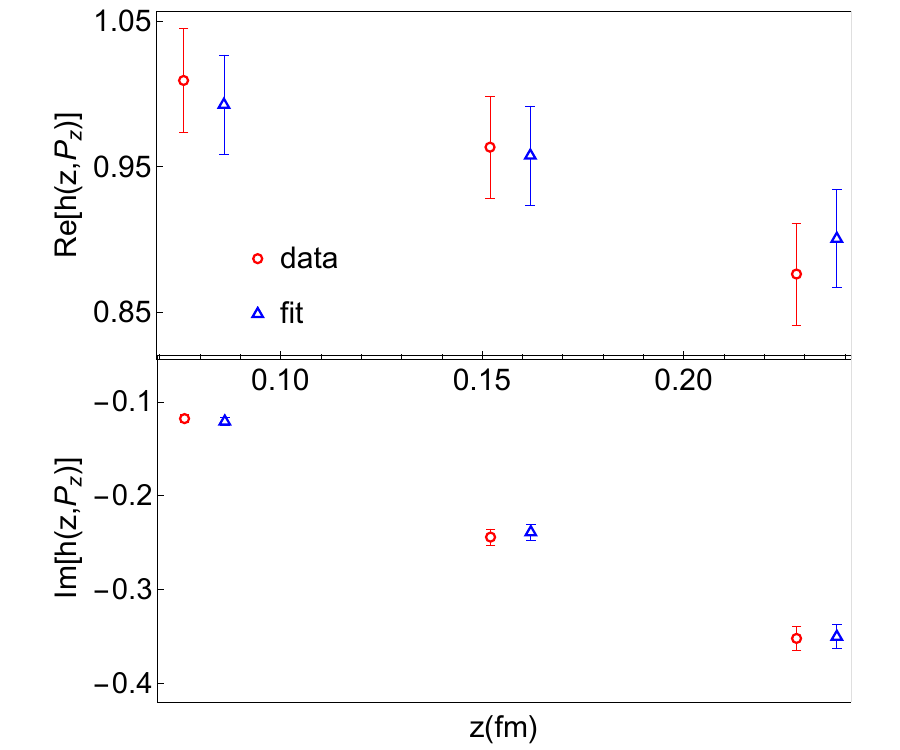}
    \caption{Fits of the PDF model to the lattice matrix elements using SDF.}
    \label{fig:sdf_fit}
\end{figure}

Although the behaviors of PDF in the end point regions {are constrained by physics},
it is difficult to allow enough flexibility for the mid-$x$ regions with only a few degree of freedom constrained by the lattice data in the perturbative region. However, as the LaMET calculations already provide local predictions of the mid-$x$ region with well-understood systematic uncertainty, the modeling of the PDF can now be simplified to the end point regions only, where simple power laws $x^a$ and $(1-x)^b$ are 
physically motivated approximations. Labeling the PDFs we calculated from LaMET as $f(x)$ for $x\in[x_0,1-x_0]$, we can model the PDF as
\begin{align}\label{eq:sdf_model}
    g_1(x)=\begin{cases} 
    f(x_0) \frac{x^a}{x_0^a}, & x<x_0 \\
    f(x), &x_0<x<1-x_0 \\
    f(1-x_0) \frac{(1-x)^b}{(1-x_0)^b},  & x>1-x_0
    \end{cases},
\end{align}
which can have more parameters 
if more short-distance lattice data are available. The model is fitted to the RG-invariant ratios at $z=\{0.076,0.152,0.228\}$ fm with $x_0=0.3$ to constrain the values of $a$ and $b$. For simplicity, we fit the following two degrees of freedom independently to the real and imaginary part of the matrix elements,
\begin{align}
    \Re[h(z,P_z)]=\int K(xw,z^2,\mu) {g_1^+}(x,\mu),\nonumber\\
    \Im[h(z,P_z)]=\int K(xw,z^2,\mu) {g_1^-}(x,\mu),
\end{align}
with four parameters $a_\pm$ and $b_\pm$ in total, where
\begin{align}
    {g_1^+}(x)={g_1}(x)+{g_1}(-x),\nonumber\\
    {g_1^-}(x)={g_1}(x)-{g_1}(-x),
\end{align}
representing the full and valence distributions.
The model fits to our data well, as shown in Fig.~\ref{fig:sdf_fit}.  The fit results suggest $a_-=-0.38(9)$ and $b_-=1.68(87)$ for $g_1^+$ with $\chi^2/\rm{d.o.f}=0.16(11)$, and  $a_-=-0.10(83)$ and $b_-=0.15(22)$ for $g_1^-$ with $\chi^2/\rm{d.o.f}=1.21(31)$. 
Notably, there are large uncertainties in the 
fitted parameters for the $g_1^-$ to the imaginary part.

\begin{figure}
    \centering
    \includegraphics[width=0.99\linewidth]{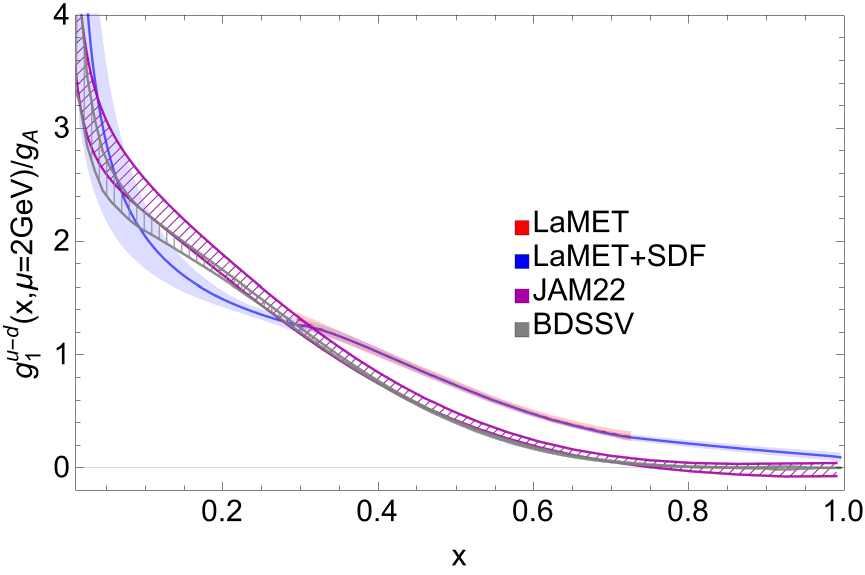}
    \includegraphics[width=0.99\linewidth]{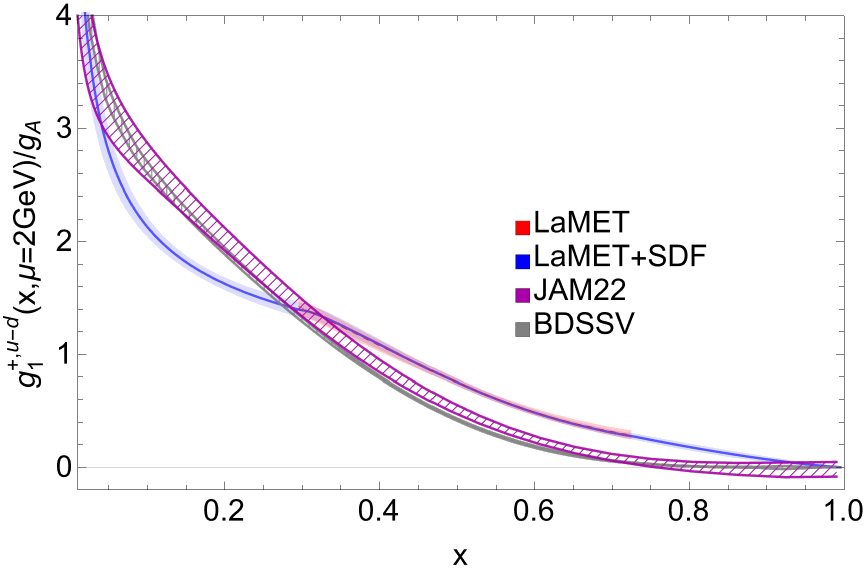}
    \caption{The LaMET calculation in mid-$x$ region complimented with end point modeling from SDF, compared with global fits.}
    \label{fig:complementarity}
\end{figure}

Combining the fitted end point regions with the LaMET calculation, we provide a full $x$-dependent helicity PDF in Fig.~\ref{fig:complementarity}. The disagreement with global fitting PDF in mid-$x$ region remains the same as Fig.~\ref{fig:matching_full}, thus the resulting $b$ parameter is much smaller than global fitting results.
Note that the end point regions in blue are still subject to model uncertainties due to the finite cutoff $x_0$ and the nonsmoothness there, which have not been quantified in this work. Also, because of the limited amount of data points at short distance, we can only fit to the simplest power-law form, and the connection between LaMET calculation and SDF fits are naively enforced by the continuity condition in Eq.~\eqref{eq:sdf_model} at the connecting points $x=x_0$ and $x=1-x_0$. In principle, these artifacts can be improved in the future at finer lattice spacing, and the model dependence can be examine by comparing the fits to various more flexible functional forms.

\section{Conclusion}
\label{sec:conclusion}

In this work we present a physical-point lattice-QCD study of the proton spin-dependent quark structure on a fine
lattice, combining short-distance information from coordinate-space matrix elements with an $x$-dependent
analysis in the LaMET framework. Our results provide a consistent picture for the leading-twist helicity
PDF at NNLO and, in addition, a direct constraint on the lowest chiral-even twist-3 moment $\tilde{d}_2$ in the $\overline{\rm MS}$ scheme for the first time.

We first extract Mellin moments of the helicity PDF using the OPE of RG-invariant ratios of the quasi-PDF matrix elements.
In this approach, the small-$z$ expansion relates Euclidean equal-time quark bilinear matrix elements to the PDF
moments in the $\overline{\rm MS}$ scheme, allowing a controlled determination of the lowest ones. Extending the same strategy beyond leading twist, we implement the NLO OPE for the transverse axial-vector correlator and constructed an improved subtraction that
isolates the genuine twist-3 contribution. This enabled us to extract the twist-3 moment $\tilde d_2$ directly in
$\overline{\rm MS}$, from the $\lambda$ dependence of the short-distance matrix elements.
Our result, $\tilde d_2^{u-d}=0.0024(46)$ at $\mu=2~\mathrm{GeV}$, is numerical small.
This behavior is compatible with phenomenological constraints and polarized-DIS extractions of $\tilde d_2$
\cite{Cocuzza:2025qvf,SANE:2018pwx,JeffersonLabHallA:2016neg}, and recent model-based studies~\cite{zhang2026transverseforcetomographyinside}. While a full quantification of systematic uncertainties remains for future work, the present analysis demonstrates the feasibility of the SDF approach that isolates a genuine twist-3 moment from nonlocal correlators directly in $\overline{\rm MS}$,
thereby complementing earlier lattice determinations based on local twist-3 operator matrix elements that are
often quoted in momentum subtraction schemes~\cite{Gockeler:2005vw,Burger:2021knd,Crawford:2024wzx}.

We then determine the $x$-dependence of the isovector helicity PDF using LaMET together with a hybrid renormalization scheme. The renormlization of linear divergence in the Wilson line self-energy is done with leading renormalon resummation, which regularizes the leading renormalon ambiguity, thus ensuring the leading power accuracy as the perturbative matching kernel is regularized in the same way. We perform an asymptotic expansion of the large $z$ data by implementing physical constraints of an exponential decaying mode at long distance, and Fourier transform the quasi-PDF matrix elements to momentum space. The $x$-dependent helicity PDFs are then obtained from the quasi-PDFs by a perturbative matching. To improve perturbative accuracy and better control the theoretical uncertainties, we improve the NNLO fixed-order matching with renormalization group and threshold resummations, corresponding to the large logarithm contributions when the parton's and spectator's momenta become small. The resummations manifest the applicable region of perturbative matching in the large momentum expansion, which only works when both physical scales remain perturbative. With this condition, we obtain the LaMET prediction of $x$-dependent helicity PDF in a range of $[x_0,1-x_0]$, where $x_0\propto \frac{\Lambda_{\rm QCD}}{2P_z} $ around $0.25$ in this calculation, which can be improved with larger hadron momentum in the future. Although the end point regions $x\to\{0,1\}$ are not directly calculable, we model them with the asymptotic behaviors and fit the parameters with the SDF of coordinate space matrix elements, complementing the LaMET calculation. 

Further progress will require tighter control of lattice systematics. Simulations at multiple lattice spacings, together with increased statistics and improved excited-state control, will sharpen the $\tilde d_2$ extraction by tightening the short-distance window and reducing residual contamination. It will also expand the reliable LaMET window
in $x$ by enabling higher $P_z$ with reduced power corrections and cutoff effects.

\section*{ACKNOWLEDGMENTS}

We thank Christina Cocuzza for sharing the data of the global helicity PDF from  JAM22. We thank Jinchen He for sharing the results in CG method for comparison. 

This material is based upon work supported by The U.S. Department of Energy, Office of Science, Office of Nuclear Physics through Contract No.~DE-SC0012704 and Contract No.~DE-AC02-06CH11357, and within the frameworks of Scientific Discovery through Advanced Computing (SciDAC) award Fundamental Nuclear Physics at the Exascale and Beyond and the Topical Collaboration in Nuclear Theory 3D quark-gluon structure of hadrons: mass, spin, and tomography. Y.Z. is support by the U.S. Department of Energy, Office of Science, Office of Nuclear Physics, Early Career Award through Contract No.~DE-SCL0000017. R.Z. is supported by the U.S. Department of Energy, Office of Science, Office of Nuclear Physics under Contract No.~DE-SC0011090 and  DOE Quark-Gluon Tomography (QGT) Topical Collaboration under award No.~DE-SC0023646.

This research used awards of computer time provided by: The INCITE program at Argonne Leadership Computing Facility, a DOE Office of Science User Facility operated under Contract No.~DE-AC02-06CH11357, the INCITE and ALCC program at the Oak Ridge Leadership Computing Facility, which is a DOE Office of Science User Facility supported under Contract DE-AC05-00OR22725, and the National Energy Research Scientific Computing Center, a DOE Office of Science User Facility supported by the Office of Science of the U.S.~Department of Energy under Contract DE-AC02-05CH11231 using NERSC award NP-ERCAP0035739. Computations for this work were carried out in part on facilities of the USQCD
Collaboration, which are funded by the Office of Science of the
U.S. Department of Energy. Part of the data analysis are carried out on Swing, a high-performance computing cluster operated by the Laboratory Computing Resource Center at Argonne National Laboratory.

The computation of the correlators was carried out with the \texttt{Qlua} software suite~\cite{qlua}, which utilized the multigrid solver in \texttt{QUDA}~\cite{Clark:2009wm,Babich:2011np}. %

\begin{widetext}

\appendix

\section{Axial vector charge}
\label{app:av-charge}

The renormalized axial vector charge reads
\begin{equation}
g^R_A = Z_A g^B_A = \frac{Z_A}{Z_V} \frac{g_A^B}{g_V^B},
\end{equation}
where we have used the charge conservation $Z_V g_V^B = 1$. The subscripts $R$ and $B$ represent ``renormalized" and ``bare", respectively. $Z_A/Z_V$ has been determined in the RI-MOM scheme for this ensemble, and converted  to the $\overline{\mathrm{MS}}$ scheme at $\mu=2$ GeV in two different ways~\cite{Gao:2022vyh}. The statistical uncertainty of each way is much smaller than the systematic uncertainty between different methods, thus we safely ignore the statistical errors and use the mean of two results as the central value and half the difference as uncertainty, which yields $Z_A/Z_V=1.0160(8)$. With this, we generate gaussian bootstrap samples for $Z_A/Z_V$. Together with the bootstrap samples for $\frac{g_A^{\mathrm{bare}}}{g_V^{\mathrm{bare}}}$, we determine the renormalized axial vector charge.

The bare axial vector charge $g_A^B$ is calculated using the bare matrix element Eq.~(\ref{eq:joint_fit}) obtained at $P=0$ and $z=0$. The fit results is shown in Fig.\ref{fig:gAgV}, which gives $g_A^B=1.260(12)$. $g_V^B$ was determined in \refcite{Gao:2022uhg} using bare matrix elements of unpolarized quasi-PDF at $P=0$ and $z=0$, which gives $g_V^B=1.057(1)$. Together, we get the ratio $\frac{g_A^{\mathrm{bare}}}{g_V^{\mathrm{bare}}}=1.192(12)$, Our final results for the renormalized axial vector charge is $g^{u-d}_A=1.211(12)$.

\begin{figure*}[tbh]
\includegraphics[width=0.5\textwidth]{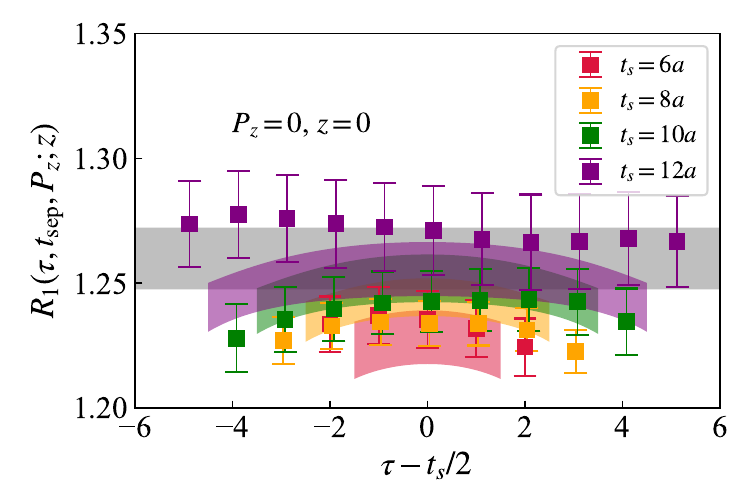}
\caption{The fit of the three-point function to two-point function ratio in the isovector case at the zero momentum with $z=0$. Left: the real part. The colored bands are reconstructed from two-state fit while the gray band denotes the extracted bare matrix element.}
\label{fig:gAgV}
\end{figure*}

\section{SDF for the twist-three quasi-PDF matrix elements}
\label{app:gTSDF}

This appendix summarizes the short-distance OPE for the
equal-time axial-vector bilocal matrix elements that enter the lattice extraction of the
chiral-even twist-3 distribution $g_T(x)$.
Throughout, we work at spacelike separations $z^2<0$ relevant for Euclidean correlators
and consider the short-distance limit $z^2\to 0$ at fixed Ioffe time
\begin{equation}
  \lambda \equiv P\!\cdot\! z ,
\end{equation}
so that power corrections are organized in $z^2\Lambda_{\rm QCD}^2$.
In this regime, QCD factorization follows directly from the existence of the OPE, and the
bilocal matrix elements admit a systematic matching onto light-cone correlation functions
with perturbatively calculable coefficient functions; see Ref.~\cite{Braun:2021aon} for a
complete derivation at twist-3 accuracy and one-loop order. In particular, the transverse
axial-vector correlator contains both twist-2 and twist-3 collinear structures already at
leading power in $z^2$~\cite{Braun:2021aon}.

\subsection{Twist decomposition and the Wandzura--Wilczek (WW) approximation}

We denote by $\tilde h_1(\lambda,z^2\mu^2)$ and $\tilde h_T(\lambda,z^2\mu^2)$ the quasi-PDF
matrix elements corresponding to the longitudinal ($\gamma_z\gamma_5$) and transverse
($\gamma_\perp\gamma_5$) projections, respectively. At short distances, they can be factorized into $h_1(\lambda,\mu)$ and $h_T(\lambda,\mu)$ which are the light-cone helicity (twist-2) correlation function and the light-cone transverse axial-vector correlator that defines $g_T(x)$.
The latter can be decomposed into twist-2 and twist-3 parts,
\begin{equation}
  h_T(\lambda)=h_T^{\rm tw2}(\lambda)+h_T^{\rm tw3}(\lambda).
\end{equation}

The WW approximation refers to retaining only the kinematic (twist-2) contribution to
$g_T$ that is completely determined by the helicity distribution $g_1(x)$, while
neglecting the genuine quark-gluon correlation encoded in twist-3 three-parton
operators. On the light cone, the WW relation is an operator identity for the twist-2
projection, implying in position space~\cite{Braun:2021aon}
\begin{equation}
  h_T^{\rm tw2}(\lambda)=\int_0^1 d\xi\, h_1(\xi\lambda),
\end{equation}
equivalently in $x$-space $g_T^{\rm tw2}(x)=\int_x^1 \frac{dy}{y}\,g_1(y)$ (with the
standard extension to $x<0$ for antiquarks)~\cite{Braun:2021aon}.
Physically, WW isolates the part of $g_T$ that arises from quark intrinsic transverse
motion and equations of motion without explicit quark-gluon interactions; the omitted
piece probes genuine quark-gluon correlations.

A crucial point for lattice quasi-/pseudo-observables is that the WW relation is not
preserved by the perturbative matching at NLO: the twist-2 coefficient functions for the
longitudinal and transverse correlators are different,
$C_T^{(1)}\neq C_1^{(1)}$, so that a residual twist-2 contamination remains after the
naive WW subtraction at finite $z^2$~\cite{Braun:2021aon}. In the language of SDF this
appears explicitly as~\cite{Braun:2021aon}
\begin{equation}
  \tilde{h}_T^{\rm tw2}(\lambda,z^2)-\int_0^1 d\alpha\, \tilde{h}_1(\alpha\lambda,z^2)\neq 0 ,
\end{equation}
with the NLO remainder given in Eq.~(4.29) of Ref.~\cite{Braun:2021aon}.
We note that Ref.~\cite{Braun:2021aon} formulates the matching in coordinate space with the
natural rescaling $(\lambda,z^2)\!\to\!(\alpha\lambda,\alpha^2 z^2)$, which is convenient for the
quasi-PDF (LaMET) representation where $z$ is the Fourier variable.
In our short-distance-factorization setup we keep $z^2$ fixed in the small-$z$ regime and
analyze the $\lambda=P\!\cdot\! z$ dependence; accordingly we rewrite the NLO matching in a
fixed-$z$ form and rederive below the central relation from which $\tilde d_2$ enters at
$\mathcal O(\lambda^2)$.

At leading twist, the short-distance factorization can be written as
\begin{align}
\begin{split}
  \tilde{h}_1(\lambda,z^2\mu^2)&=\int_0^1 d\alpha\,C_1(\alpha,z^2\mu^2)\,h_1(\alpha\lambda;\mu),\\
  \tilde{h}_T^{\rm tw2}(\lambda,z^2\mu^2)&=\int_0^1 d\alpha\,C_T(\alpha,z^2\mu^2)\,h_T^{\rm tw2}(\alpha\lambda;\mu),
\end{split}
\end{align}
where $C_{1,T}=\delta(1-\alpha)+C_{1,T}^{(1)}+\cdots$.
We employ the explicit NLO kernels from Ref.~\cite{Braun:2021aon}:
\begin{align}
\begin{split}
  C_1^{(1)}(\alpha,z^2\mu^2)&=\frac{\alpha_s C_F}{2\pi}\Bigg[\Bigg(-\frac{1+\alpha^2}{1-\alpha}L_z-\frac{4\ln(1-\alpha)}{1-\alpha}+4(1-\alpha)-\frac{\alpha^2+1}{1-\alpha}\Bigg)_+  +\delta(1-\alpha)\Bigg(\frac{3}{2}L_z+\frac{7}{2}\Bigg)\Bigg],\\
  C_T^{(1)}(\alpha,z^2\mu^2)&=\frac{\alpha_s C_F}{2\pi}\Bigg[\Bigg(-\frac{1+\alpha^2}{1-\alpha}L_z-\frac{4\ln(1-\alpha)}{1-\alpha}+2(1-\alpha)-\frac{\alpha^2+1}{1-\alpha}\Bigg)_+  +\delta(1-\alpha)\Bigg(\frac{3}{2}L_z+\frac{5}{2}\Bigg)\Bigg],
\end{split}
\end{align}
with
\begin{equation}
  L_z \equiv \ln\!\left(\frac{-z^2\mu^2}{4e^{-2\gamma_E}}\right),
\end{equation}
and the plus distribution defined by
\begin{equation}
\int_0^1 d\alpha\,[g(\alpha)]_+\,\varphi(\alpha)
=\int_0^1 d\alpha\,g(\alpha)\big(\varphi(\alpha)-\varphi(1)\big).
\end{equation}

Using the WW representation for $h_T^{\rm tw2}$, one can form a combination that cancels
the LO twist-2 contribution and leaves a calculable NLO remainder:
\begin{align}
\begin{split}
  \tilde{h}_T^{\rm tw2}(\lambda,z^2\mu^2)-\int_0^1 d\beta\,\tilde{h}_1(\beta\lambda,z^2\mu^2)
  =\int_0^1 d\beta\int_0^1 d\alpha\;\Delta C^{(1)}(\alpha,z^2\mu^2)\,h_1(\alpha\beta\lambda;\mu),
\end{split}
\end{align}
where $\Delta C^{(1)} \equiv C_T^{(1)}-C_1^{(1)}$.
Expanding $h_1(\alpha\beta\lambda;\mu)=\sum_{n\ge 0}\frac{(-i\lambda)^n}{n!}(\alpha\beta)^n a_n(\mu)$
with $a_n(\mu)=\int_{-1}^1 dx\,x^n\,g_1(x)$ yields
\begin{align}\label{eq:hTtw2}
\begin{split}
  \tilde{h}_T^{\rm tw2}(\lambda,z^2\mu^2)-\int_0^1 d\beta\,\tilde{h}_1(\beta\lambda,z^2\mu^2)
  =\sum_{n=0}^\infty \frac{(-i\lambda)^n}{n!}\,\Delta C_n^{(1)}(z^2\mu^2)\,a_n(\mu),
\end{split}
\end{align}
with
\begin{equation}
  \Delta C_n^{(1)}(z^2\mu^2)=\frac{\alpha_s C_F}{2\pi}\frac{-n^2-3n-4}{(n+1)^2(n+2)}.
\end{equation}
This makes explicit that the WW subtraction at finite $z^2$ leaves a perturbatively
controlled, but non-vanishing, twist-2 remainder starting already at $\mathcal O(\alpha_s)$,
consistent with the general statement that WW is violated for quasi-/pseudo-observables at NLO
\cite{Braun:2021aon}.

\subsection{Twist-3 matching: Two-parton vs three-parton correlators and the $\tilde d_2$ moment}

Beyond the WW (twist-2) approximation, the transverse correlator $\tilde h_T$ necessarily
receives genuine twist-3 contributions from quark-gluon operators. At NLO, the
short-distance factorization takes the schematic form of a two-particle (``2p'') piece
multiplying the two-parton twist-3 light-cone correlator and a genuine three-particle
(``3p'') piece involving the quark-antiquark-gluon correlator $S^-$~\cite{Braun:2021aon}
\begin{align}\label{eq:hTtw3}
\begin{split}
  \tilde{h}_T^{\rm tw3}(\lambda,z^2\mu^2)
  &=h_T^{\rm tw3}(\lambda)
    +C_{\rm 2p}^{(1)}\circledast h_T^{\rm tw3}
    +2\lambda^2\, C_{\rm 3p}^{(1)}\circledast \widehat{S}^-(\lambda,\beta\lambda,\alpha\lambda),
\end{split}
\end{align}
where $\circledast$ denotes the convolution
\begin{equation}
  [C\circledast f](\lambda)=\int_0^1 d\alpha\, C(\alpha)\, f(\alpha\lambda),
\end{equation}
and $\widehat S^-$ is the position-space three-parton correlator whose Fourier transform
defines $S^-(x_1,x_2,x_3)$ as in Ref.~\cite{Braun:2021aon}:
\begin{align}
\begin{split}
  \widehat{S}^{\pm}(\lambda_1,\lambda_2,\lambda_3)
  &=\int[dx]\;e^{-i(\lambda_1x_1+\lambda_2x_2+\lambda_3x_3)}\,S^{\pm}(x_1,x_2,x_3),\\
  \int[dx]&\equiv\int_{-1}^1 dx_1\,dx_2\,dx_3\;\delta(x_1+x_2+x_3).
\end{split}
\end{align}
In $x$-space, the genuine twist-3 part of $g_T$ can be expressed through $S^-$,
\begin{align}
\begin{split}
  g_T^{\rm tw3}(x)=&\;2\int[dx]\int_{0}^{1}d\alpha\left(\frac{\delta(x+\alpha x_1)}{x_1x_3}
  +\frac{\delta(x+\alpha x_1+\alpha x_2)}{x_2x_3}
  +\frac{\delta(x+\alpha x_1)}{x_1x_2}\right)S^-(x_1,x_2,x_3),
\end{split}
\end{align}
and the corresponding light-cone Ioffe-time correlator reads~\cite{Braun:2021aon}
\begin{align}\label{eq:htw3_def}
  h_T^{\rm tw3}(\lambda)=2\lambda^2\int_0^1 d\alpha\int_\alpha^1 d\beta\,(1-\beta)\,
  \widehat{S}^-(\lambda,\beta\lambda,\alpha\lambda).
\end{align}

We adopt the standard moment conventions
\begin{align}
  &\int_{-1}^1 dx\,x^n\,g_1(x)=a_n,\nonumber\\
  &\int_{-1}^1 dx\,x^n\, g_2(x)=\frac{n}{n+1}\big(d_n-a_n\big),
\end{align}
and $g_T(x)=g_1(x)+g_2(x)$. Following the convention of Ref.~\cite{SANE:2018pwx}, we define
$d_n=\tilde d_n/2$ and introduce the twist-3 reduced matrix element
\begin{equation}\label{eq:d2_def}
  \tilde d_2=\int_{-1}^1 dx\,x^2\,[3g_T(x)-g_1(x)]
  \;=\;\int[dx]\,S^-(x_1,x_2,x_3),
\end{equation}
which is the lowest-dimension chiral-even twist-3 matrix element entering the first
nontrivial moment of $g_2$ and, equivalently, the lowest moment of the three-parton correlator
$S^-$~\cite{Braun:2021aon}.

The small-$\lambda$ power counting implied by \Eq{htw3_def} makes the appearance of
$\tilde d_2$ transparent. Expanding $\widehat S^-(\lambda,\beta\lambda,\alpha\lambda)$ around
$\lambda=0$ and keeping terms up to $\mathcal O(\lambda^0)$ inside the integral yields
\begin{align}
\begin{split}
  h_T^{\rm tw3}(\lambda)
  &=2\lambda^2\left[\int_0^1 d\alpha\int_\alpha^1 d\beta\,(1-\beta)\right]\widehat S^-(0,0,0)
  +\mathcal O(\lambda^3)
  =\frac{\lambda^2}{3}\,\widehat S^-(0,0,0)+\mathcal O(\lambda^3)\\
  &=\frac{\lambda^2}{3}\int[dx]\,S^-(x_1,x_2,x_3)+\mathcal O(\lambda^3)
  =\frac{\lambda^2}{3}\,\tilde d_2+\mathcal O(\lambda^3),
\end{split}
\end{align}
where we used $\int_0^1 d\alpha\int_\alpha^1 d\beta\,(1-\beta)=1/6$ and the overall prefactor $2$
in \Eq{htw3_def}.

We now evaluate the NLO corrections in \Eq{hTtw3} consistently up to $\mathcal O(\lambda^2)$.
For the two-particle part, one may write (using the notation and kernels of
Ref.~\cite{Braun:2021aon})
\begin{align}
\begin{split}
  C_{\rm 2p}^{(1)}\circledast h_T^{\rm tw3}
  &=\int_0^1 d\alpha\left[
  C_T^{(1)}(\alpha,z^2\mu^2)+N_c\Big(L_z(\delta(1-\alpha)-\alpha)+\alpha+2\delta(1-\alpha)\Big)
  \right]h_T^{\rm tw3}(\alpha\lambda;\mu)\\
  &=\frac{\lambda^2}{3}\,\tilde d_2\;C_{{\rm 2p},2}^{(1)}(z^2\mu^2)+\mathcal O(\lambda^3),
\end{split}
\end{align}
where in the last step we used $h_T^{\rm tw3}(\alpha\lambda)=\alpha^2\,\frac{\lambda^2}{3}\tilde d_2+\mathcal O(\lambda^3)$.
The moment-space coefficient can be expressed as
\begin{align}
\begin{split}
  C_{{\rm 2p},n}^{(1)}(z^2\mu^2)=&
  C_F\left[\left(\frac{3+2n}{2+3n+n^2}+2H_n\right)L_z
  +2(1-H_n)H_n-2H^{(2)}_n+\frac{5+2n}{2+3n+n^2}\right]\\
  &+N_c\left[\left(\frac{n+1}{n+2}\right)L_z+\frac{2n+5}{n+2}\right],
\end{split}
\end{align}
with $H_n=\sum_{k=1}^n \frac{1}{k}$ and $H_n^{(2)}=\sum_{k=1}^n \frac{1}{k^2}$.

For the genuine three-particle part, the NLO kernel contains a logarithmic piece proportional
to $L_z$ and a finite remainder. Using the representation in Ref.~\cite{Braun:2021aon}, we write
\begin{align}
\begin{split}
  C_{\rm 3p}^{(1)}\circledast \widehat{S}^-=&-L_z\,{\rm P_{tw3}}\circledast\widehat{S}^-\\
  &+\int_0^1d\alpha\bigg\{\int_\alpha^1d\beta\bigg(2N_c\ln\beta+\frac{1}{N_c}\frac{\alpha^2}{2}\bigg)\widehat{S}^-(\lambda,\beta\lambda,\alpha\lambda)\\
  &\qquad\qquad+\frac{1}{N_c}\int_0^\alpha d\beta\bigg[\frac{\beta^2}{2}\widehat{S}^-(\bar{\alpha}\lambda,\bar{\beta}\lambda,0)
  -\bigg(\frac{\beta(2+\beta)}{2}-\frac{2\beta}{\alpha}(1+\ln\alpha)\bigg)\widehat{S}^-(\lambda,\beta\lambda,\alpha\lambda)\bigg]\bigg\},
\end{split}
\end{align}
where the logarithmic convolution is
\begin{align}
\begin{split}
  {\rm P_{tw3}}\circledast\widehat{S}^-=&\frac{1}{N_c}\int_0^1d\alpha\bigg\{\int_\alpha^1d\beta\frac{\alpha(\alpha-2)}{2}\widehat{S}^-(\lambda,\beta\lambda,\alpha\lambda)\\
  &+\int_0^\alpha d\beta\bigg[\frac{\beta(\beta-2)}{2}\widehat{S}^-(\bar{\alpha}\lambda,\bar{\beta}\lambda,0)
  +\bigg(\frac{\beta(2-\beta)}{2}-\frac{\beta}{\alpha}\bigg)\widehat{S}^-(\lambda,\beta\lambda,\alpha\lambda)\bigg]\bigg\}.
\end{split}
\end{align}
Keeping only terms up to $\mathcal O(\lambda^0)$ inside $\widehat S^-$, the convolution reduces
to a contribution proportional to $\tilde d_2$,
\begin{align}
\begin{split}
  C_{\rm 3p}^{(1)}\circledast \widehat{S}^-
  =\tilde{d}_2\left[\frac{3}{8N_c}L_z+\frac{1-4N_c^2}{8N_c}\right]+\mathcal O(\lambda),
\end{split}
\end{align}
so that the three-particle NLO correction enters the $\mathcal O(\lambda^2)$ term in
\Eq{hTtw3} through the explicit prefactor $2\lambda^2$.

Finally, collecting the twist-2 contribution in \Eq{hTtw2} and the twist-3 contribution in
\Eq{hTtw3} and keeping terms up to $\mathcal O(\lambda^2)$, we arrive at
\begin{align}\label{eq:GTG1}
\begin{split}
  \tilde{h}_T(\lambda,z^2\mu^2)&-\int_0^1 d\alpha\,\tilde{h}_1(\alpha\lambda,z^2\mu^2)
  =-\frac{\alpha_s C_F}{\pi} a_0 + i\,\lambda\,\frac{\alpha_s C_F}{3\pi} a_1 \\
  &\quad+\frac{\lambda^2}{3}\Bigg\{\tilde d_2+\frac{7\alpha_s C_F}{24\pi} a_2
    +\frac{\alpha_s}{4\pi}\tilde d_2\Big[\Big(\frac{43C_F}{12}+\frac{9}{4N_c}+\frac{3N_c}{4}\Big)L_z
    -\frac{13C_F}{4}+\frac{3}{4N_c}-\frac{3N_c}{4}\Big]\Bigg\}
  +\mathcal O(\lambda^3),
\end{split}
\end{align}
which makes explicit that the genuine twist-3 matrix element $\tilde d_2$ enters at
$\mathcal O(\lambda^2)$, accompanied by a calculable twist-2 contamination (through $a_2$)
and the NLO logarithm $L_z$ dictated by SDF.

\section{Forward matching via matrix expansion in $\alpha_s$}
\label{app:matching}
Following the spirit of effective theory, the light cone PDF can be factorized into quasi-PDF with a power expansion~\cite{Ji:2024oka}. Thus it can be calculated from the $x$-space quasi-PDF through a forward matching. In the forward matching with leading renormalon resummation, it is more convenient to include the resummed renormalon contribution in its exponential form~\cite{Ji:2024hit}. The matching of the quasi helicity PDF $\tilde{g}_1(y)$ to the light cone helicity PDF $g_1(x,\mu)$ including RGR can be obtained by extending Eq.~(\ref{eq:factorization}) as follows

\begin{align}
g_1(x,\mu)=\int_{-\infty}^{+\infty}\frac{\mathrm{d}w}{|w|} \int_{-\infty}^{+\infty} \frac{\mathrm{d}y'}{|y'|}  \int_{-\infty}^{+\infty} \frac{\mathrm{d}y}{|y|} \ \mathcal{C}_{\mathrm{evo}}\left(\frac{x}{w},\mu,\mu_0\right) \ &\mathcal{C}_{\mathrm{hyb,LRR}}\left(\frac{w}{y'},\mu_0, z_s, y'P_z\right)M_{\rm LRR}\left(\frac{y'}{y},\mu_0,z_s,yP_z\right) \tilde{g}_1(y)\nonumber\\
&+\mathcal{O}\left(\frac{\Lambda^2_{\mathrm{QCD}}}{(xP_z)^2}, \frac{\Lambda^2_{\mathrm{QCD}}}{\big{(}(1-x)P_z\big{)}^2}\right)
\label{eq:matching-general}
\end{align}
where $\lambda_s=z_s P_z$, $\mathcal{C}_{\mathrm{evo}}$ is the DGLAP evolution of the light cone PDF from the initial scale $\mu_0$ to $\mu=2$ GeV, and $M_{\rm LRR}(\xi',\mu_0,z_s,P_z)$ is the resummed mass renormalon contribution to the matching kernel~\cite{Zhang:2023bxs,Ji:2024hit},
\begin{align}
    M_{\rm LRR}\left(\xi',\mu_0,z_s,p_z\right)&=\int_{-\infty}^{\infty}\frac{d\lambda}{2\pi}e^{i\lambda(\xi'-1)}\exp[{-\theta(|\lambda|/p_z-z_s) (|\lambda|/p_z-z_s)\mu\sum_{n=0}^{\infty} r_n \alpha^{n+1}_s(\mu_0)|_{\rm p.v.}] },\nonumber\\
    &= \frac{\sin(z_sp_z(1-{\xi}'))}{\pi(1-{\xi}')}+ \Re\left[\frac{e^{i z_sp_z(1-{\xi}')}}{(\mu/p_z)\sum_{n=0}^{\infty} r_n \alpha^{n+1}_s(\mu_0)|_{\rm p.v.}-i(1-{\xi}')}\right], \nonumber\\ 
    r_n=N_m\beta_0^n&\frac{\Gamma(n+1+b)}{\Gamma(1+b)}\left(1+\frac{b}{b+n}c_1+\frac{b(b-1)}{(n+b)(n+b-1)}c_2+\dots\right), \nonumber\\
    \sum r_n\alpha^{n+1}_s(\mu_0)|_{\rm p.v.}=&N_m\frac{-2\pi e^{-\frac{2\pi}{\alpha_s\beta_0}}}{\beta_0}\Re\left[E_{b+1}\left(\frac{2\pi}{\beta_0}\right)+c_1E_{b}\left(\frac{2\pi}{\beta_0}\right)+c_2E_{b-1}\left(\frac{2\pi}{\beta_0}\right)\right],
\end{align}
with $N_m=0.575$ for $n_f=3$, $E_n(z)\equiv\int_1^\infty e^{-zt}/t^n dt$ being the generalized exponential integral, and the constants are defined as
\begin{align}
    b&=\frac{\beta_1}{\beta_0^2},\\
    c_1&=\frac{\beta_1^2-\beta_0\beta_2}{4b\beta_0^4},\\
    c_2&=\frac{-\beta_3\beta_0^4+2\beta_1\beta_2\beta_0^3+(\beta_2^2-\beta_1^3)\beta_0^2-2\beta_1^2\beta_2\beta_0+\beta_1^4}{2(b-1)b\beta_0^8}.
\end{align}
The matching kernel \(\mathcal{C}_{\mathrm{hyb,LRR}}\left(\frac{w}{y},\mu_0,z_s,P_z\right)\) in the hybrid scheme with the renormalon contribution subtracted takes the following form up to NNLO
\begin{equation}
\mathcal{C}_{\mathrm{hyb,LRR}}\left(\xi=\frac{w}{y},\mu_0, z_s, P_z\right)= \delta(1-\xi)+ \delta\mathcal{C}_{\mathrm{ratio}}(\xi, \mu_0, P_z) + \delta\mathcal{C}_{\mathrm{hyb}}\left(\xi, \mu_0,  z_s, P_z\right) + \delta\mathcal{C}_{M}(\xi, \mu_0, z_s, P_z),
\end{equation}
The $\delta(1-\xi)$ term gives the LO contribution, while $\delta\mathcal{C}_{\mathrm{ratio}}$ and $\delta\mathcal{C}_{\mathrm{hyb}}$ denote the ratio- and hybrid-scheme corrections. The superscripts (1) and (2) indicate the NLO and NNLO contributions, respectively. The NLO correction for choosing the ratio scheme reads~\cite{Izubuchi:2018srq,Su:2022fiu}
\begin{equation}
\begin{split}
& \delta\mathcal{C}^{(1)}_{\mathrm{ratio}}(\xi, \mu_0, P_z)  = -\frac{\alpha_{s}(\mu_0) C_{F}}{2 \pi} \begin{cases}\left(\frac{1+\xi^{2}}{1-\xi} \ln \frac{\xi}{\xi-1}+1-\frac{3}{2(1-\xi)}\right)_{+(1)}^{[1, \infty]} & \xi>1 \\ \left(\frac{1+\xi^{2}}{1-\xi}\left[-\ln \frac{\mu_0^{2}}{4 {w}^{2} P_{z}^{2}}+\ln (\frac{1-\xi}{\xi})-1\right]+1+\frac{3}{2(1-\xi)}+2(1-\xi)\right)_{+(1)}^{[0,1]} & 0<\xi<1 \\ \left(-\frac{1+\xi^{2}}{1-\xi} \ln \frac{-\xi}{1-\xi}-1+\frac{3}{2(1-\xi)}\right)_{+(1)}^{[-\infty, 0]} & \xi<0 \end{cases}.
\end{split}
\label{eq:ratiokernelNLO}
\end{equation}
$\delta\mathcal{C}_{\mathrm{hyb}}^{(1)}\left(\xi, \mu_0  z_s, P_z\right)$ accounts for the correction for switching from the ratio scheme to the hybrid scheme~\cite{Su:2022fiu} 
\begin{equation}
\delta\mathcal{C}^{(1)}_{\mathrm{hyb}}\left(\xi, \mu_0, z_s, P_z\right) = -
\frac{\alpha_s C_{F}}{2\pi}\frac{3}{2}\left[\mathcal{R}(\xi,z_s,P_z)\,\right]_{+(1)}^{[-\infty,\infty]} 
\quad \text{with} \quad
\mathcal{R}(\xi, z_s, P_z)=-
\frac{1}{|1-\xi|} + \frac{2\,{\rm Si}[(1-\xi)|y|z_s P_z]}{\pi(1-\xi)},
\label{eq:hybridkernelNLOcorrection} 
\end{equation}
where Si is the sine integral function.
The NNLO ratio-scheme correction is
\begin{align}
\delta\mathcal{C}&^{(2)}_{\mathrm{ratio}}(\xi,\mu_0,P_z)
= -\delta\mathcal{C}^{(2)}_{\overline{\mathrm{MS}}}(\xi,\mu_0,P_z) 
 -\left(\frac{\alpha_s}{2\pi}\right)^{\!2}
 \Bigg\{
 \left(\frac{11 C_A C_F}{8}
      -\frac{C_F n_f T_F}{2}
      +\frac{9 C_F^2}{8}\right)
 \frac{2}{|1-\xi|}\ln\!\left(\dfrac{\mu_0^2 \xi^2}{4 {w
 }^2 P_z^2 (1-\xi)^2}\right)
 \notag \\[4pt]
&\quad
+ \left[\left(\frac{203}{24}-\frac{\pi^2}{6}\right) C_A C_F
      +\left(\frac{37}{8}+\frac{2\pi^2}{3}\right) C_F^2
      -\frac{19 C_F n_f T_F}{6}\right]
 \frac{1}{|1-\xi|} 
 \Bigg\}_{+(1)}^{[-\infty,\infty]}  .
\label{eq:ratiokernelNNLO}
\end{align}
Here $\delta\mathcal{C}^{(2)}_{\overline{\mathrm{MS}}}$ denotes the two-loop
matching kernel from light cone to quasi-PDF in the $\overline{\mathrm{MS}}$ scheme, as given in
Ref.~\cite{Li:2020xml}.  In deriving Eq.~\eqref{eq:ratiokernelNNLO}, all terms
proportional to $\delta(1-\xi)$ are omitted, and an overall plus prescription
is applied at the end to ensure the current conservation.
The corresponding NNLO hybrid-scheme correction reads
\begin{align}
\delta C^{(2)}_\mathrm{hyb}&\left(\xi, \mu_0,  z_s, P_z\right)\\
&=-\bigg\{\left(\frac{\alpha_s}{2\pi}\right)^2 
\left(\frac{11 C_A C_F}{8}-\frac{C_F n_f T_F}{2}+\frac{9 C_F^2}{8}\right)
\Bigg[
\frac{2}{|1-\xi|}\,
\ln\left(y^2 P_z^2 z_s^2 e^{2\gamma_{E}} (1-\xi)^2 \right)+2 \,\mathcal{R}(\xi,z_s)\,
\ln\left(\frac{z_s^{2}\mu_0^{2} e^{2\gamma_E}}{4}\right)
\notag\\
&\quad
-\frac{4|y|P_z z_s}{\pi}
\left(
{}_3 F_3\!\left(1,1,1;2,2,2;i(\xi-1)|y|P_z z_s\right)
+{}_3 F_3\!\left(1,1,1;2,2,2;-i(\xi-1)|y|P_z z_s\right)
\right)
\Bigg]+\left(\frac{\alpha_s}{2\pi}\right)^2 \mathcal{R}(\xi,z_s)
\notag\\[6pt]
&\quad\!\times\!
\left(
\left(\frac{203}{24}-\frac{\pi^2}{6}\right) C_A C_F
\!+\!
\left(\frac{37}{8}+\frac{2\pi^2}{3}\right) C_F^2
\!-\!
\frac{19 C_F n_f T_F}{6}
\right)
\!-\!\left(\frac{\alpha_s}{2\pi}\right)^2 C_F^2
\left(
\frac{3}{2}\ln\!\left(\frac{z_s^{2}\mu_0^{2} e^{2\gamma_E}}{4}\right)
\!+\!\frac{7}{2}
\right)
\frac{3}{2}
\mathcal{R}(\xi,z_s)\notag\\[6pt]
&\quad
-\int_{-\infty}^{\infty}\frac{d\nu}{2\pi} e^{i\nu\xi}
\int_{-1}^{1}du\, e^{-iu\nu}\,
Z_{\rm ratio}^{(1)}\!\left(u,\mu_0^{2}\left(\frac{\nu}{yP_{z}}\right)^{2}\right)\,
\frac{\alpha_s C_{F}}{2\pi}\frac{3}{2}
\ln\!\left[
\left(\frac{\nu}{yP_{z}z_s}\right)^{2}
\right]\,
\theta\left(z_s-\left|\frac{\nu}{yP_{z}}\right|\right)\Bigg\}_{+(1)}^{[-\infty,+\infty]}
\notag\\[6pt]
&\quad
-  \delta\mathcal{C}^{(1)}_{\mathrm{ratio}}(\xi, \mu_0, P_z)
\frac{\alpha_{s} C_{F}}{2\pi}
\left(
\frac{3}{2}\ln\!\left(\frac{z_s^{2}\mu_0^{2} e^{2\gamma_E}}{4}\right)
+\frac{7}{2}
\right) + \int\frac{d\nu}{|\nu|}\left(\delta\mathcal{C}^{(1)}_{\mathrm{ratio}}\left(\frac{\xi}{\nu}\right)+\delta\mathcal{C}^{(1)}_{\mathrm{hyb}}\left(\frac{\xi}{\nu}\right)\right)\left(\delta\mathcal{C}^{(1)}_{\mathrm{ratio}}(\nu)+\delta\mathcal{C}^{(1)}_{\mathrm{hyb}}(\nu)\right),
\notag
\end{align}
with the one-loop coordinate-space ratio-scheme matching coefficient
\begin{align}
   Z_{\rm ratio}^{(1)}\!\left(u,\mu_0^{2}z^{2}\right) =\frac{\alpha_{s} C_{F}}{2 \pi}\left[-\frac{1+u^{2}}{1-u} \ln\left( \frac{  \mu_0^2z^2 e^{2 \gamma_{E}} }{4} \right) +\frac{3-8 u+3u^{2}-4 \ln (1-u)}{1-u}\right]_{+(1)}^{[0,1]} \theta(u)\theta(1-u).
\end{align}
The renormalon subtraction term  $\delta\mathcal{C}_M$ is obtained from the fixed-order expansion of the resummed renormalon term $M_{\rm LRR}$. At NLO, it is~\cite{Zhang:2023bxs,Ji:2024hit},
\begin{align}
    \delta\mathcal{C}_M^{(1)}(\xi,\mu_0,z_s,yP_z)=r_0\alpha_s\frac{\mu_0}{p_z} \Big{[}- \frac{(\bar{\xi}^2-\epsilon^2)\cos(z_syP_z\bar{\xi})+2\bar{\xi}\epsilon\sin(z_syP_z\bar{\xi})}{\pi(\bar{\xi}^2+\epsilon^2)^2}\exp(-\epsilon z_sP_z) \Big{]}_{+(1)}^{[-\infty,\infty]} ,
\end{align}
where the variable $\bar{\xi}=1-\xi$, and $\epsilon$ is a small number close to zero to regularize the Fourier transformation of the linear $z$ term. The NNLO correction is,
\begin{align}
    \delta\mathcal{C}&_M^{(2)}(\xi,\mu_0,z_s,yP_z)=r_1\alpha^2_s\frac{\mu_0}{p_z} \Big{[}- \frac{(\bar{\xi}^2-\epsilon^2)\cos(z_syP_z\bar{\xi})+2\bar{\xi}\epsilon\sin(z_syP_z\bar{\xi})}{\pi(\bar{\xi}^2+\epsilon^2)^2}\exp(-\epsilon z_sP_z) \Big{]}_{+(1)}^{[-\infty,\infty]}  \nonumber \\
    &+ \int\frac{d\nu}{|\nu|}\left(\delta\mathcal{C}^{(1)}_{\mathrm{ratio}}\left(\frac{\xi}{\nu}\right)+\delta\mathcal{C}^{(1)}_{\mathrm{hyb}}\left(\frac{\xi}{\nu}\right)\right)\delta\mathcal{C}_M^{(1)}(\nu) + \frac{1}{2} \int\frac{d\nu}{|\nu|}\delta\mathcal{C}_M^{(1)}\left(\frac{\xi}{\nu}\right)\delta\mathcal{C}_M^{(1)}(\nu).
\end{align}

The RG resummation is implemented by setting $\mu_0=2\kappa xP_z$ for each $x$ in Eq.~\eqref{eq:matching-general}. The higher-order uncertainties can be estimated by varying $\kappa\in\{1/\sqrt{2},\sqrt{2}\}$ to change the initial scale, which is absorbed into our error band as systematics. The numerical implementation in the matrix form follows the same procedure as Refs.~\cite{Su:2022fiu,Ding:2024saz}.
\end{widetext}


\begin{thebibliography}{133}%
\makeatletter
\providecommand \@ifxundefined [1]{%
 \@ifx{#1\undefined}
}%
\providecommand \@ifnum [1]{%
 \ifnum #1\expandafter \@firstoftwo
 \else \expandafter \@secondoftwo
 \fi
}%
\providecommand \@ifx [1]{%
 \ifx #1\expandafter \@firstoftwo
 \else \expandafter \@secondoftwo
 \fi
}%
\providecommand \natexlab [1]{#1}%
\providecommand \enquote  [1]{``#1''}%
\providecommand \bibnamefont  [1]{#1}%
\providecommand \bibfnamefont [1]{#1}%
\providecommand \citenamefont [1]{#1}%
\providecommand \href@noop [0]{\@secondoftwo}%
\providecommand \href [0]{\begingroup \@sanitize@url \@href}%
\providecommand \@href[1]{\@@startlink{#1}\@@href}%
\providecommand \@@href[1]{\endgroup#1\@@endlink}%
\providecommand \@sanitize@url [0]{\catcode `\\12\catcode `\$12\catcode
  `\&12\catcode `\#12\catcode `\^12\catcode `\_12\catcode `\%12\relax}%
\providecommand \@@startlink[1]{}%
\providecommand \@@endlink[0]{}%
\providecommand \url  [0]{\begingroup\@sanitize@url \@url }%
\providecommand \@url [1]{\endgroup\@href {#1}{\urlprefix }}%
\providecommand \urlprefix  [0]{URL }%
\providecommand \Eprint [0]{\href }%
\providecommand \doibase [0]{http://dx.doi.org/}%
\providecommand \selectlanguage [0]{\@gobble}%
\providecommand \bibinfo  [0]{\@secondoftwo}%
\providecommand \bibfield  [0]{\@secondoftwo}%
\providecommand \translation [1]{[#1]}%
\providecommand \BibitemOpen [0]{}%
\providecommand \bibitemStop [0]{}%
\providecommand \bibitemNoStop [0]{.\EOS\space}%
\providecommand \EOS [0]{\spacefactor3000\relax}%
\providecommand \BibitemShut  [1]{\csname bibitem#1\endcsname}%
\let\auto@bib@innerbib\@empty
\bibitem [{\citenamefont {Jaffe}\ and\ \citenamefont
  {Manohar}(1990)}]{Jaffe:1989jz}%
  \BibitemOpen
  \bibfield  {author} {\bibinfo {author} {\bibfnamefont {R.~L.}\ \bibnamefont
  {Jaffe}}\ and\ \bibinfo {author} {\bibfnamefont {A.}~\bibnamefont
  {Manohar}},\ }\href {\doibase 10.1016/0550-3213(90)90506-9} {\bibfield
  {journal} {\bibinfo  {journal} {Nucl. Phys. B}\ }\textbf {\bibinfo {volume}
  {337}},\ \bibinfo {pages} {509} (\bibinfo {year} {1990})}\BibitemShut
  {NoStop}%
\bibitem [{\citenamefont {Ji}(1997)}]{Ji:1996ek}%
  \BibitemOpen
  \bibfield  {author} {\bibinfo {author} {\bibfnamefont {X.-D.}\ \bibnamefont
  {Ji}},\ }\href {\doibase 10.1103/PhysRevLett.78.610} {\bibfield  {journal}
  {\bibinfo  {journal} {Phys. Rev. Lett.}\ }\textbf {\bibinfo {volume} {78}},\
  \bibinfo {pages} {610} (\bibinfo {year} {1997})}\BibitemShut {NoStop}%
\bibitem [{\citenamefont {Ashman}\ \emph {et~al.}(1988)\citenamefont {Ashman}
  \emph {et~al.}}]{EuropeanMuon:1987isl}%
  \BibitemOpen
  \bibfield  {author} {\bibinfo {author} {\bibfnamefont {J.}~\bibnamefont
  {Ashman}} \emph {et~al.} (\bibinfo {collaboration} {European Muon}),\ }\href
  {\doibase 10.1016/0370-2693(88)91523-7} {\bibfield  {journal} {\bibinfo
  {journal} {Phys. Lett. B}\ }\textbf {\bibinfo {volume} {206}},\ \bibinfo
  {pages} {364} (\bibinfo {year} {1988})}\BibitemShut {NoStop}%
\bibitem [{\citenamefont {Cocuzza}\ \emph {et~al.}(2022)\citenamefont
  {Cocuzza}, \citenamefont {Melnitchouk}, \citenamefont {Metz},\ and\
  \citenamefont {Sato}}]{Cocuzza:2022jye}%
  \BibitemOpen
  \bibfield  {author} {\bibinfo {author} {\bibfnamefont {C.}~\bibnamefont
  {Cocuzza}}, \bibinfo {author} {\bibfnamefont {W.}~\bibnamefont
  {Melnitchouk}}, \bibinfo {author} {\bibfnamefont {A.}~\bibnamefont {Metz}}, \
  and\ \bibinfo {author} {\bibfnamefont {N.}~\bibnamefont {Sato}} (\bibinfo
  {collaboration} {Jefferson Lab Angular Momentum (JAM)}),\ }\href {\doibase
  10.1103/PhysRevD.106.L031502} {\bibfield  {journal} {\bibinfo  {journal}
  {Phys. Rev. D}\ }\textbf {\bibinfo {volume} {106}},\ \bibinfo {pages}
  {L031502} (\bibinfo {year} {2022})}\BibitemShut {NoStop}%
\bibitem [{\citenamefont {Borsa}\ \emph {et~al.}(2024)\citenamefont {Borsa},
  \citenamefont {Stratmann}, \citenamefont {Vogelsang}, \citenamefont
  {de~Florian},\ and\ \citenamefont {Sassot}}]{Borsa:2024mss}%
  \BibitemOpen
  \bibfield  {author} {\bibinfo {author} {\bibfnamefont {I.}~\bibnamefont
  {Borsa}}, \bibinfo {author} {\bibfnamefont {M.}~\bibnamefont {Stratmann}},
  \bibinfo {author} {\bibfnamefont {W.}~\bibnamefont {Vogelsang}}, \bibinfo
  {author} {\bibfnamefont {D.}~\bibnamefont {de~Florian}}, \ and\ \bibinfo
  {author} {\bibfnamefont {R.}~\bibnamefont {Sassot}},\ }\href {\doibase
  10.1103/PhysRevLett.133.151901} {\bibfield  {journal} {\bibinfo  {journal}
  {Phys. Rev. Lett.}\ }\textbf {\bibinfo {volume} {133}},\ \bibinfo {pages}
  {151901} (\bibinfo {year} {2024})}\BibitemShut {NoStop}%
\bibitem [{\citenamefont {Bertone}\ \emph {et~al.}(2025)\citenamefont
  {Bertone}, \citenamefont {Chiefa},\ and\ \citenamefont
  {Nocera}}]{Bertone:2024taw}%
  \BibitemOpen
  \bibfield  {author} {\bibinfo {author} {\bibfnamefont {V.}~\bibnamefont
  {Bertone}}, \bibinfo {author} {\bibfnamefont {A.}~\bibnamefont {Chiefa}}, \
  and\ \bibinfo {author} {\bibfnamefont {E.~R.}\ \bibnamefont {Nocera}}
  (\bibinfo {collaboration} {MAP (Multi-dimensional Analyses of Partonic
  distributions)}),\ }\href {\doibase 10.1016/j.physletb.2025.139497}
  {\bibfield  {journal} {\bibinfo  {journal} {Phys. Lett. B}\ }\textbf
  {\bibinfo {volume} {865}},\ \bibinfo {pages} {139497} (\bibinfo {year}
  {2025})}\BibitemShut {NoStop}%
\bibitem [{\citenamefont {Cruz-Martinez}\ \emph {et~al.}(2025)\citenamefont
  {Cruz-Martinez}, \citenamefont {Hasenack}, \citenamefont {Hekhorn},
  \citenamefont {Magni}, \citenamefont {Nocera}, \citenamefont {Rabemananjara},
  \citenamefont {Rojo}, \citenamefont {Sharma},\ and\ \citenamefont {van
  Seeventer}}]{Cruz-Martinez:2025ahf}%
  \BibitemOpen
  \bibfield  {author} {\bibinfo {author} {\bibfnamefont {J.}~\bibnamefont
  {Cruz-Martinez}}, \bibinfo {author} {\bibfnamefont {T.}~\bibnamefont
  {Hasenack}}, \bibinfo {author} {\bibfnamefont {F.}~\bibnamefont {Hekhorn}},
  \bibinfo {author} {\bibfnamefont {G.}~\bibnamefont {Magni}}, \bibinfo
  {author} {\bibfnamefont {E.~R.}\ \bibnamefont {Nocera}}, \bibinfo {author}
  {\bibfnamefont {T.~R.}\ \bibnamefont {Rabemananjara}}, \bibinfo {author}
  {\bibfnamefont {J.}~\bibnamefont {Rojo}}, \bibinfo {author} {\bibfnamefont
  {T.}~\bibnamefont {Sharma}}, \ and\ \bibinfo {author} {\bibfnamefont
  {G.}~\bibnamefont {van Seeventer}},\ }\href {\doibase
  10.1007/JHEP07(2025)168} {\bibfield  {journal} {\bibinfo  {journal} {JHEP}\
  }\textbf {\bibinfo {volume} {07}},\ \bibinfo {pages} {168} (\bibinfo {year}
  {2025})}\BibitemShut {NoStop}%
\bibitem [{\citenamefont {Adamczyk}\ \emph {et~al.}(2015)\citenamefont
  {Adamczyk} \emph {et~al.}}]{STAR:2014wox}%
  \BibitemOpen
  \bibfield  {author} {\bibinfo {author} {\bibfnamefont {L.}~\bibnamefont
  {Adamczyk}} \emph {et~al.} (\bibinfo {collaboration} {STAR}),\ }\href
  {\doibase 10.1103/PhysRevLett.115.092002} {\bibfield  {journal} {\bibinfo
  {journal} {Phys. Rev. Lett.}\ }\textbf {\bibinfo {volume} {115}},\ \bibinfo
  {pages} {092002} (\bibinfo {year} {2015})}\BibitemShut {NoStop}%
\bibitem [{\citenamefont {Adare}\ \emph {et~al.}(2014)\citenamefont {Adare}
  \emph {et~al.}}]{PHENIX:2014gbf}%
  \BibitemOpen
  \bibfield  {author} {\bibinfo {author} {\bibfnamefont {A.}~\bibnamefont
  {Adare}} \emph {et~al.} (\bibinfo {collaboration} {PHENIX}),\ }\href
  {\doibase 10.1103/PhysRevD.90.012007} {\bibfield  {journal} {\bibinfo
  {journal} {Phys. Rev. D}\ }\textbf {\bibinfo {volume} {90}},\ \bibinfo
  {pages} {012007} (\bibinfo {year} {2014})}\BibitemShut {NoStop}%
\bibitem [{\citenamefont {Adare}\ \emph {et~al.}(2016)\citenamefont {Adare}
  \emph {et~al.}}]{PHENIX:2015fxo}%
  \BibitemOpen
  \bibfield  {author} {\bibinfo {author} {\bibfnamefont {A.}~\bibnamefont
  {Adare}} \emph {et~al.} (\bibinfo {collaboration} {PHENIX}),\ }\href
  {\doibase 10.1103/PhysRevD.93.011501} {\bibfield  {journal} {\bibinfo
  {journal} {Phys. Rev. D}\ }\textbf {\bibinfo {volume} {93}},\ \bibinfo
  {pages} {011501} (\bibinfo {year} {2016})}\BibitemShut {NoStop}%
\bibitem [{\citenamefont {Adolph}\ \emph {et~al.}(2013)\citenamefont {Adolph}
  \emph {et~al.}}]{COMPASS:2012mpe}%
  \BibitemOpen
  \bibfield  {author} {\bibinfo {author} {\bibfnamefont {C.}~\bibnamefont
  {Adolph}} \emph {et~al.} (\bibinfo {collaboration} {COMPASS}),\ }\href
  {\doibase 10.1103/PhysRevD.87.052018} {\bibfield  {journal} {\bibinfo
  {journal} {Phys. Rev. D}\ }\textbf {\bibinfo {volume} {87}},\ \bibinfo
  {pages} {052018} (\bibinfo {year} {2013})}\BibitemShut {NoStop}%
\bibitem [{\citenamefont {Alekseev}\ \emph {et~al.}(2009)\citenamefont
  {Alekseev} \emph {et~al.}}]{COMPASS:2009kiy}%
  \BibitemOpen
  \bibfield  {author} {\bibinfo {author} {\bibfnamefont {M.}~\bibnamefont
  {Alekseev}} \emph {et~al.} (\bibinfo {collaboration} {COMPASS}),\ }\href
  {\doibase 10.1016/j.physletb.2009.08.065} {\bibfield  {journal} {\bibinfo
  {journal} {Phys. Lett. B}\ }\textbf {\bibinfo {volume} {680}},\ \bibinfo
  {pages} {217} (\bibinfo {year} {2009})}\BibitemShut {NoStop}%
\bibitem [{\citenamefont {Alekseev}\ \emph {et~al.}(2010)\citenamefont
  {Alekseev} \emph {et~al.}}]{COMPASS:2010hwr}%
  \BibitemOpen
  \bibfield  {author} {\bibinfo {author} {\bibfnamefont {M.~G.}\ \bibnamefont
  {Alekseev}} \emph {et~al.} (\bibinfo {collaboration} {COMPASS}),\ }\href
  {\doibase 10.1016/j.physletb.2010.08.034} {\bibfield  {journal} {\bibinfo
  {journal} {Phys. Lett. B}\ }\textbf {\bibinfo {volume} {693}},\ \bibinfo
  {pages} {227} (\bibinfo {year} {2010})}\BibitemShut {NoStop}%
\bibitem [{\citenamefont {Airapetian}\ \emph {et~al.}(2005)\citenamefont
  {Airapetian} \emph {et~al.}}]{HERMES:2004zsh}%
  \BibitemOpen
  \bibfield  {author} {\bibinfo {author} {\bibfnamefont {A.}~\bibnamefont
  {Airapetian}} \emph {et~al.} (\bibinfo {collaboration} {HERMES}),\ }\href
  {\doibase 10.1103/PhysRevD.71.012003} {\bibfield  {journal} {\bibinfo
  {journal} {Phys. Rev. D}\ }\textbf {\bibinfo {volume} {71}},\ \bibinfo
  {pages} {012003} (\bibinfo {year} {2005})}\BibitemShut {NoStop}%
\bibitem [{\citenamefont {Dudek}\ \emph {et~al.}(2012)\citenamefont {Dudek}
  \emph {et~al.}}]{Dudek:2012vr}%
  \BibitemOpen
  \bibfield  {author} {\bibinfo {author} {\bibfnamefont {J.}~\bibnamefont
  {Dudek}} \emph {et~al.},\ }\href {\doibase 10.1140/epja/i2012-12187-1}
  {\bibfield  {journal} {\bibinfo  {journal} {Eur. Phys. J. A}\ }\textbf
  {\bibinfo {volume} {48}},\ \bibinfo {pages} {187} (\bibinfo {year}
  {2012})}\BibitemShut {NoStop}%
\bibitem [{\citenamefont {Accardi}\ \emph {et~al.}(2016)\citenamefont {Accardi}
  \emph {et~al.}}]{Accardi:2012qut}%
  \BibitemOpen
  \bibfield  {author} {\bibinfo {author} {\bibfnamefont {A.}~\bibnamefont
  {Accardi}} \emph {et~al.},\ }\href {\doibase 10.1140/epja/i2016-16268-9}
  {\bibfield  {journal} {\bibinfo  {journal} {Eur. Phys. J. A}\ }\textbf
  {\bibinfo {volume} {52}},\ \bibinfo {pages} {268} (\bibinfo {year}
  {2016})}\BibitemShut {NoStop}%
\bibitem [{\citenamefont {Abdul~Khalek}\ \emph {et~al.}(2022)\citenamefont
  {Abdul~Khalek} \emph {et~al.}}]{AbdulKhalek:2021gbh}%
  \BibitemOpen
  \bibfield  {author} {\bibinfo {author} {\bibfnamefont {R.}~\bibnamefont
  {Abdul~Khalek}} \emph {et~al.},\ }\href {\doibase
  10.1016/j.nuclphysa.2022.122447} {\bibfield  {journal} {\bibinfo  {journal}
  {Nucl. Phys. A}\ }\textbf {\bibinfo {volume} {1026}},\ \bibinfo {pages}
  {122447} (\bibinfo {year} {2022})}\BibitemShut {NoStop}%
\bibitem [{\citenamefont {Anderle}\ \emph {et~al.}(2021)\citenamefont
  {Anderle}, \citenamefont {Hou}, \citenamefont {Xing}, \citenamefont {Yan},
  \citenamefont {Yuan},\ and\ \citenamefont {Zhao}}]{Anderle:2021dpv}%
  \BibitemOpen
  \bibfield  {author} {\bibinfo {author} {\bibfnamefont {D.~P.}\ \bibnamefont
  {Anderle}}, \bibinfo {author} {\bibfnamefont {T.-J.}\ \bibnamefont {Hou}},
  \bibinfo {author} {\bibfnamefont {H.}~\bibnamefont {Xing}}, \bibinfo {author}
  {\bibfnamefont {M.}~\bibnamefont {Yan}}, \bibinfo {author} {\bibfnamefont
  {C.~P.}\ \bibnamefont {Yuan}}, \ and\ \bibinfo {author} {\bibfnamefont
  {Y.}~\bibnamefont {Zhao}},\ }\href {\doibase 10.1007/JHEP08(2021)034}
  {\bibfield  {journal} {\bibinfo  {journal} {JHEP}\ }\textbf {\bibinfo
  {volume} {08}},\ \bibinfo {pages} {034} (\bibinfo {year} {2021})}\BibitemShut
  {NoStop}%
\bibitem [{\citenamefont {Ji}(2013)}]{Ji:2013dva}%
  \BibitemOpen
  \bibfield  {author} {\bibinfo {author} {\bibfnamefont {X.}~\bibnamefont
  {Ji}},\ }\href {\doibase 10.1103/PhysRevLett.110.262002} {\bibfield
  {journal} {\bibinfo  {journal} {Phys. Rev. Lett.}\ }\textbf {\bibinfo
  {volume} {110}},\ \bibinfo {pages} {262002} (\bibinfo {year}
  {2013})}\BibitemShut {NoStop}%
\bibitem [{\citenamefont {Ji}(2014)}]{Ji:2014gla}%
  \BibitemOpen
  \bibfield  {author} {\bibinfo {author} {\bibfnamefont {X.}~\bibnamefont
  {Ji}},\ }\href {\doibase 10.1007/s11433-014-5492-3} {\bibfield  {journal}
  {\bibinfo  {journal} {Sci. China Phys. Mech. Astron.}\ }\textbf {\bibinfo
  {volume} {57}},\ \bibinfo {pages} {1407} (\bibinfo {year}
  {2014})}\BibitemShut {NoStop}%
\bibitem [{\citenamefont {Ji}\ \emph {et~al.}(2021{\natexlab{a}})\citenamefont
  {Ji}, \citenamefont {Liu}, \citenamefont {Liu}, \citenamefont {Zhang},\ and\
  \citenamefont {Zhao}}]{Ji:2020ect}%
  \BibitemOpen
  \bibfield  {author} {\bibinfo {author} {\bibfnamefont {X.}~\bibnamefont
  {Ji}}, \bibinfo {author} {\bibfnamefont {Y.-S.}\ \bibnamefont {Liu}},
  \bibinfo {author} {\bibfnamefont {Y.}~\bibnamefont {Liu}}, \bibinfo {author}
  {\bibfnamefont {J.-H.}\ \bibnamefont {Zhang}}, \ and\ \bibinfo {author}
  {\bibfnamefont {Y.}~\bibnamefont {Zhao}},\ }\href {\doibase
  10.1103/RevModPhys.93.035005} {\bibfield  {journal} {\bibinfo  {journal}
  {Rev. Mod. Phys.}\ }\textbf {\bibinfo {volume} {93}},\ \bibinfo {pages}
  {035005} (\bibinfo {year} {2021}{\natexlab{a}})}\BibitemShut {NoStop}%
\bibitem [{\citenamefont {Ji}\ \emph {et~al.}(2018)\citenamefont {Ji},
  \citenamefont {Zhang},\ and\ \citenamefont {Zhao}}]{Ji:2017oey}%
  \BibitemOpen
  \bibfield  {author} {\bibinfo {author} {\bibfnamefont {X.}~\bibnamefont
  {Ji}}, \bibinfo {author} {\bibfnamefont {J.-H.}\ \bibnamefont {Zhang}}, \
  and\ \bibinfo {author} {\bibfnamefont {Y.}~\bibnamefont {Zhao}},\ }\href
  {\doibase 10.1103/PhysRevLett.120.112001} {\bibfield  {journal} {\bibinfo
  {journal} {Phys. Rev. Lett.}\ }\textbf {\bibinfo {volume} {120}},\ \bibinfo
  {pages} {112001} (\bibinfo {year} {2018})}\BibitemShut {NoStop}%
\bibitem [{\citenamefont {Green}\ \emph {et~al.}(2018)\citenamefont {Green},
  \citenamefont {Jansen},\ and\ \citenamefont {Steffens}}]{Green:2017xeu}%
  \BibitemOpen
  \bibfield  {author} {\bibinfo {author} {\bibfnamefont {J.}~\bibnamefont
  {Green}}, \bibinfo {author} {\bibfnamefont {K.}~\bibnamefont {Jansen}}, \
  and\ \bibinfo {author} {\bibfnamefont {F.}~\bibnamefont {Steffens}},\ }\href
  {\doibase 10.1103/PhysRevLett.121.022004} {\bibfield  {journal} {\bibinfo
  {journal} {Phys. Rev. Lett.}\ }\textbf {\bibinfo {volume} {121}},\ \bibinfo
  {pages} {022004} (\bibinfo {year} {2018})}\BibitemShut {NoStop}%
\bibitem [{\citenamefont {Ishikawa}\ \emph {et~al.}(2017)\citenamefont
  {Ishikawa}, \citenamefont {Ma}, \citenamefont {Qiu},\ and\ \citenamefont
  {Yoshida}}]{Ishikawa:2017faj}%
  \BibitemOpen
  \bibfield  {author} {\bibinfo {author} {\bibfnamefont {T.}~\bibnamefont
  {Ishikawa}}, \bibinfo {author} {\bibfnamefont {Y.-Q.}\ \bibnamefont {Ma}},
  \bibinfo {author} {\bibfnamefont {J.-W.}\ \bibnamefont {Qiu}}, \ and\
  \bibinfo {author} {\bibfnamefont {S.}~\bibnamefont {Yoshida}},\ }\href
  {\doibase 10.1103/PhysRevD.96.094019} {\bibfield  {journal} {\bibinfo
  {journal} {Phys. Rev. D}\ }\textbf {\bibinfo {volume} {96}},\ \bibinfo
  {pages} {094019} (\bibinfo {year} {2017})}\BibitemShut {NoStop}%
\bibitem [{\citenamefont {Alexandrou}\ \emph {et~al.}(2017)\citenamefont
  {Alexandrou}, \citenamefont {Cichy}, \citenamefont {Constantinou},
  \citenamefont {Hadjiyiannakou}, \citenamefont {Jansen}, \citenamefont
  {Panagopoulos},\ and\ \citenamefont {Steffens}}]{Alexandrou:2017huk}%
  \BibitemOpen
  \bibfield  {author} {\bibinfo {author} {\bibfnamefont {C.}~\bibnamefont
  {Alexandrou}}, \bibinfo {author} {\bibfnamefont {K.}~\bibnamefont {Cichy}},
  \bibinfo {author} {\bibfnamefont {M.}~\bibnamefont {Constantinou}}, \bibinfo
  {author} {\bibfnamefont {K.}~\bibnamefont {Hadjiyiannakou}}, \bibinfo
  {author} {\bibfnamefont {K.}~\bibnamefont {Jansen}}, \bibinfo {author}
  {\bibfnamefont {H.}~\bibnamefont {Panagopoulos}}, \ and\ \bibinfo {author}
  {\bibfnamefont {F.}~\bibnamefont {Steffens}},\ }\href {\doibase
  10.1016/j.nuclphysb.2017.08.012} {\bibfield  {journal} {\bibinfo  {journal}
  {Nucl. Phys. B}\ }\textbf {\bibinfo {volume} {923}},\ \bibinfo {pages} {394}
  (\bibinfo {year} {2017})}\BibitemShut {NoStop}%
\bibitem [{\citenamefont {Chen}\ \emph
  {et~al.}(2018{\natexlab{a}})\citenamefont {Chen}, \citenamefont {Ishikawa},
  \citenamefont {Jin}, \citenamefont {Lin}, \citenamefont {Yang}, \citenamefont
  {Zhang},\ and\ \citenamefont {Zhao}}]{Chen:2017mzz}%
  \BibitemOpen
  \bibfield  {author} {\bibinfo {author} {\bibfnamefont {J.-W.}\ \bibnamefont
  {Chen}}, \bibinfo {author} {\bibfnamefont {T.}~\bibnamefont {Ishikawa}},
  \bibinfo {author} {\bibfnamefont {L.}~\bibnamefont {Jin}}, \bibinfo {author}
  {\bibfnamefont {H.-W.}\ \bibnamefont {Lin}}, \bibinfo {author} {\bibfnamefont
  {Y.-B.}\ \bibnamefont {Yang}}, \bibinfo {author} {\bibfnamefont {J.-H.}\
  \bibnamefont {Zhang}}, \ and\ \bibinfo {author} {\bibfnamefont
  {Y.}~\bibnamefont {Zhao}},\ }\href {\doibase 10.1103/PhysRevD.97.014505}
  {\bibfield  {journal} {\bibinfo  {journal} {Phys. Rev. D}\ }\textbf {\bibinfo
  {volume} {97}},\ \bibinfo {pages} {014505} (\bibinfo {year}
  {2018}{\natexlab{a}})}\BibitemShut {NoStop}%
\bibitem [{\citenamefont {Constantinou}\ and\ \citenamefont
  {Panagopoulos}(2017)}]{Constantinou:2017sej}%
  \BibitemOpen
  \bibfield  {author} {\bibinfo {author} {\bibfnamefont {M.}~\bibnamefont
  {Constantinou}}\ and\ \bibinfo {author} {\bibfnamefont {H.}~\bibnamefont
  {Panagopoulos}},\ }\href {\doibase 10.1103/PhysRevD.96.054506} {\bibfield
  {journal} {\bibinfo  {journal} {Phys. Rev. D}\ }\textbf {\bibinfo {volume}
  {96}},\ \bibinfo {pages} {054506} (\bibinfo {year} {2017})}\BibitemShut
  {NoStop}%
\bibitem [{\citenamefont {Stewart}\ and\ \citenamefont
  {Zhao}(2018)}]{Stewart:2017tvs}%
  \BibitemOpen
  \bibfield  {author} {\bibinfo {author} {\bibfnamefont {I.~W.}\ \bibnamefont
  {Stewart}}\ and\ \bibinfo {author} {\bibfnamefont {Y.}~\bibnamefont {Zhao}},\
  }\href {\doibase 10.1103/PhysRevD.97.054512} {\bibfield  {journal} {\bibinfo
  {journal} {Phys. Rev. D}\ }\textbf {\bibinfo {volume} {97}},\ \bibinfo
  {pages} {054512} (\bibinfo {year} {2018})}\BibitemShut {NoStop}%
\bibitem [{\citenamefont {Ji}\ \emph {et~al.}(2021{\natexlab{b}})\citenamefont
  {Ji}, \citenamefont {Liu}, \citenamefont {Sch\"afer}, \citenamefont {Wang},
  \citenamefont {Yang}, \citenamefont {Zhang},\ and\ \citenamefont
  {Zhao}}]{Ji:2020brr}%
  \BibitemOpen
  \bibfield  {author} {\bibinfo {author} {\bibfnamefont {X.}~\bibnamefont
  {Ji}}, \bibinfo {author} {\bibfnamefont {Y.}~\bibnamefont {Liu}}, \bibinfo
  {author} {\bibfnamefont {A.}~\bibnamefont {Sch\"afer}}, \bibinfo {author}
  {\bibfnamefont {W.}~\bibnamefont {Wang}}, \bibinfo {author} {\bibfnamefont
  {Y.-B.}\ \bibnamefont {Yang}}, \bibinfo {author} {\bibfnamefont {J.-H.}\
  \bibnamefont {Zhang}}, \ and\ \bibinfo {author} {\bibfnamefont
  {Y.}~\bibnamefont {Zhao}},\ }\href {\doibase 10.1016/j.nuclphysb.2021.115311}
  {\bibfield  {journal} {\bibinfo  {journal} {Nucl. Phys. B}\ }\textbf
  {\bibinfo {volume} {964}},\ \bibinfo {pages} {115311} (\bibinfo {year}
  {2021}{\natexlab{b}})}\BibitemShut {NoStop}%
\bibitem [{\citenamefont {Huo}\ \emph {et~al.}(2021)\citenamefont {Huo} \emph
  {et~al.}}]{LatticePartonLPC:2021gpi}%
  \BibitemOpen
  \bibfield  {author} {\bibinfo {author} {\bibfnamefont {Y.-K.}\ \bibnamefont
  {Huo}} \emph {et~al.} (\bibinfo {collaboration} {Lattice Parton (LPC)}),\
  }\href {\doibase 10.1016/j.nuclphysb.2021.115443} {\bibfield  {journal}
  {\bibinfo  {journal} {Nucl. Phys. B}\ }\textbf {\bibinfo {volume} {969}},\
  \bibinfo {pages} {115443} (\bibinfo {year} {2021})}\BibitemShut {NoStop}%
\bibitem [{\citenamefont {Izubuchi}\ \emph {et~al.}(2018)\citenamefont
  {Izubuchi}, \citenamefont {Ji}, \citenamefont {Jin}, \citenamefont
  {Stewart},\ and\ \citenamefont {Zhao}}]{Izubuchi:2018srq}%
  \BibitemOpen
  \bibfield  {author} {\bibinfo {author} {\bibfnamefont {T.}~\bibnamefont
  {Izubuchi}}, \bibinfo {author} {\bibfnamefont {X.}~\bibnamefont {Ji}},
  \bibinfo {author} {\bibfnamefont {L.}~\bibnamefont {Jin}}, \bibinfo {author}
  {\bibfnamefont {I.~W.}\ \bibnamefont {Stewart}}, \ and\ \bibinfo {author}
  {\bibfnamefont {Y.}~\bibnamefont {Zhao}},\ }\href {\doibase
  10.1103/PhysRevD.98.056004} {\bibfield  {journal} {\bibinfo  {journal} {Phys.
  Rev. D}\ }\textbf {\bibinfo {volume} {98}},\ \bibinfo {pages} {056004}
  (\bibinfo {year} {2018})}\BibitemShut {NoStop}%
\bibitem [{\citenamefont {Li}\ \emph {et~al.}(2021)\citenamefont {Li},
  \citenamefont {Ma},\ and\ \citenamefont {Qiu}}]{Li:2020xml}%
  \BibitemOpen
  \bibfield  {author} {\bibinfo {author} {\bibfnamefont {Z.-Y.}\ \bibnamefont
  {Li}}, \bibinfo {author} {\bibfnamefont {Y.-Q.}\ \bibnamefont {Ma}}, \ and\
  \bibinfo {author} {\bibfnamefont {J.-W.}\ \bibnamefont {Qiu}},\ }\href
  {\doibase 10.1103/PhysRevLett.126.072001} {\bibfield  {journal} {\bibinfo
  {journal} {Phys. Rev. Lett.}\ }\textbf {\bibinfo {volume} {126}},\ \bibinfo
  {pages} {072001} (\bibinfo {year} {2021})}\BibitemShut {NoStop}%
\bibitem [{\citenamefont {Chen}\ \emph {et~al.}(2021)\citenamefont {Chen},
  \citenamefont {Wang},\ and\ \citenamefont {Zhu}}]{Chen:2020ody}%
  \BibitemOpen
  \bibfield  {author} {\bibinfo {author} {\bibfnamefont {L.-B.}\ \bibnamefont
  {Chen}}, \bibinfo {author} {\bibfnamefont {W.}~\bibnamefont {Wang}}, \ and\
  \bibinfo {author} {\bibfnamefont {R.}~\bibnamefont {Zhu}},\ }\href {\doibase
  10.1103/PhysRevLett.126.072002} {\bibfield  {journal} {\bibinfo  {journal}
  {Phys. Rev. Lett.}\ }\textbf {\bibinfo {volume} {126}},\ \bibinfo {pages}
  {072002} (\bibinfo {year} {2021})}\BibitemShut {NoStop}%
\bibitem [{\citenamefont {Cheng}\ \emph {et~al.}(2025)\citenamefont {Cheng},
  \citenamefont {Huang}, \citenamefont {Li}, \citenamefont {Li},\ and\
  \citenamefont {Ma}}]{Cheng:2024wyu}%
  \BibitemOpen
  \bibfield  {author} {\bibinfo {author} {\bibfnamefont {C.}~\bibnamefont
  {Cheng}}, \bibinfo {author} {\bibfnamefont {L.-H.}\ \bibnamefont {Huang}},
  \bibinfo {author} {\bibfnamefont {X.}~\bibnamefont {Li}}, \bibinfo {author}
  {\bibfnamefont {Z.-Y.}\ \bibnamefont {Li}}, \ and\ \bibinfo {author}
  {\bibfnamefont {Y.-Q.}\ \bibnamefont {Ma}},\ }\href {\doibase
  10.1103/53ys-3n19} {\bibfield  {journal} {\bibinfo  {journal} {Phys. Rev.
  Lett.}\ }\textbf {\bibinfo {volume} {134}},\ \bibinfo {pages} {251902}
  (\bibinfo {year} {2025})}\BibitemShut {NoStop}%
\bibitem [{\citenamefont {Braun}\ \emph {et~al.}(2019)\citenamefont {Braun},
  \citenamefont {Vladimirov},\ and\ \citenamefont {Zhang}}]{Braun:2018brg}%
  \BibitemOpen
  \bibfield  {author} {\bibinfo {author} {\bibfnamefont {V.~M.}\ \bibnamefont
  {Braun}}, \bibinfo {author} {\bibfnamefont {A.}~\bibnamefont {Vladimirov}}, \
  and\ \bibinfo {author} {\bibfnamefont {J.-H.}\ \bibnamefont {Zhang}},\ }\href
  {\doibase 10.1103/PhysRevD.99.014013} {\bibfield  {journal} {\bibinfo
  {journal} {Phys. Rev. D}\ }\textbf {\bibinfo {volume} {99}},\ \bibinfo
  {pages} {014013} (\bibinfo {year} {2019})}\BibitemShut {NoStop}%
\bibitem [{\citenamefont {Zhang}\ \emph {et~al.}(2023)\citenamefont {Zhang},
  \citenamefont {Holligan}, \citenamefont {Ji},\ and\ \citenamefont
  {Su}}]{Zhang:2023bxs}%
  \BibitemOpen
  \bibfield  {author} {\bibinfo {author} {\bibfnamefont {R.}~\bibnamefont
  {Zhang}}, \bibinfo {author} {\bibfnamefont {J.}~\bibnamefont {Holligan}},
  \bibinfo {author} {\bibfnamefont {X.}~\bibnamefont {Ji}}, \ and\ \bibinfo
  {author} {\bibfnamefont {Y.}~\bibnamefont {Su}},\ }\href {\doibase
  10.1016/j.physletb.2023.138081} {\bibfield  {journal} {\bibinfo  {journal}
  {Phys. Lett. B}\ }\textbf {\bibinfo {volume} {844}},\ \bibinfo {pages}
  {138081} (\bibinfo {year} {2023})}\BibitemShut {NoStop}%
\bibitem [{\citenamefont {Holligan}\ \emph {et~al.}(2023)\citenamefont
  {Holligan}, \citenamefont {Ji}, \citenamefont {Lin}, \citenamefont {Su},\
  and\ \citenamefont {Zhang}}]{Holligan:2023rex}%
  \BibitemOpen
  \bibfield  {author} {\bibinfo {author} {\bibfnamefont {J.}~\bibnamefont
  {Holligan}}, \bibinfo {author} {\bibfnamefont {X.}~\bibnamefont {Ji}},
  \bibinfo {author} {\bibfnamefont {H.-W.}\ \bibnamefont {Lin}}, \bibinfo
  {author} {\bibfnamefont {Y.}~\bibnamefont {Su}}, \ and\ \bibinfo {author}
  {\bibfnamefont {R.}~\bibnamefont {Zhang}},\ }\href {\doibase
  10.1016/j.nuclphysb.2023.116282} {\bibfield  {journal} {\bibinfo  {journal}
  {Nucl. Phys. B}\ }\textbf {\bibinfo {volume} {993}},\ \bibinfo {pages}
  {116282} (\bibinfo {year} {2023})}\BibitemShut {NoStop}%
\bibitem [{\citenamefont {Gao}\ \emph {et~al.}(2021)\citenamefont {Gao},
  \citenamefont {Lee}, \citenamefont {Mukherjee}, \citenamefont {Shugert},\
  and\ \citenamefont {Zhao}}]{Gao:2021hxl}%
  \BibitemOpen
  \bibfield  {author} {\bibinfo {author} {\bibfnamefont {X.}~\bibnamefont
  {Gao}}, \bibinfo {author} {\bibfnamefont {K.}~\bibnamefont {Lee}}, \bibinfo
  {author} {\bibfnamefont {S.}~\bibnamefont {Mukherjee}}, \bibinfo {author}
  {\bibfnamefont {C.}~\bibnamefont {Shugert}}, \ and\ \bibinfo {author}
  {\bibfnamefont {Y.}~\bibnamefont {Zhao}},\ }\href {\doibase
  10.1103/PhysRevD.103.094504} {\bibfield  {journal} {\bibinfo  {journal}
  {Phys. Rev. D}\ }\textbf {\bibinfo {volume} {103}},\ \bibinfo {pages}
  {094504} (\bibinfo {year} {2021})}\BibitemShut {NoStop}%
\bibitem [{\citenamefont {Su}\ \emph {et~al.}(2023)\citenamefont {Su},
  \citenamefont {Holligan}, \citenamefont {Ji}, \citenamefont {Yao},
  \citenamefont {Zhang},\ and\ \citenamefont {Zhang}}]{Su:2022fiu}%
  \BibitemOpen
  \bibfield  {author} {\bibinfo {author} {\bibfnamefont {Y.}~\bibnamefont
  {Su}}, \bibinfo {author} {\bibfnamefont {J.}~\bibnamefont {Holligan}},
  \bibinfo {author} {\bibfnamefont {X.}~\bibnamefont {Ji}}, \bibinfo {author}
  {\bibfnamefont {F.}~\bibnamefont {Yao}}, \bibinfo {author} {\bibfnamefont
  {J.-H.}\ \bibnamefont {Zhang}}, \ and\ \bibinfo {author} {\bibfnamefont
  {R.}~\bibnamefont {Zhang}},\ }\href {\doibase
  10.1016/j.nuclphysb.2023.116201} {\bibfield  {journal} {\bibinfo  {journal}
  {Nucl. Phys. B}\ }\textbf {\bibinfo {volume} {991}},\ \bibinfo {pages}
  {116201} (\bibinfo {year} {2023})}\BibitemShut {NoStop}%
\bibitem [{\citenamefont {Ji}\ \emph {et~al.}(2023)\citenamefont {Ji},
  \citenamefont {Liu},\ and\ \citenamefont {Su}}]{Ji:2023pba}%
  \BibitemOpen
  \bibfield  {author} {\bibinfo {author} {\bibfnamefont {X.}~\bibnamefont
  {Ji}}, \bibinfo {author} {\bibfnamefont {Y.}~\bibnamefont {Liu}}, \ and\
  \bibinfo {author} {\bibfnamefont {Y.}~\bibnamefont {Su}},\ }\href {\doibase
  10.1007/JHEP08(2023)037} {\bibfield  {journal} {\bibinfo  {journal} {JHEP}\
  }\textbf {\bibinfo {volume} {08}},\ \bibinfo {pages} {037} (\bibinfo {year}
  {2023})}\BibitemShut {NoStop}%
\bibitem [{\citenamefont {Baker}\ \emph {et~al.}(2024)\citenamefont {Baker},
  \citenamefont {Bollweg}, \citenamefont {Boyle}, \citenamefont {Clo{\"e}t},
  \citenamefont {Gao}, \citenamefont {Mukherjee}, \citenamefont {Petreczky},
  \citenamefont {Zhang},\ and\ \citenamefont {Zhao}}]{Baker:2024zcd}%
  \BibitemOpen
  \bibfield  {author} {\bibinfo {author} {\bibfnamefont {E.}~\bibnamefont
  {Baker}}, \bibinfo {author} {\bibfnamefont {D.}~\bibnamefont {Bollweg}},
  \bibinfo {author} {\bibfnamefont {P.}~\bibnamefont {Boyle}}, \bibinfo
  {author} {\bibfnamefont {I.}~\bibnamefont {Clo{\"e}t}}, \bibinfo {author}
  {\bibfnamefont {X.}~\bibnamefont {Gao}}, \bibinfo {author} {\bibfnamefont
  {S.}~\bibnamefont {Mukherjee}}, \bibinfo {author} {\bibfnamefont
  {P.}~\bibnamefont {Petreczky}}, \bibinfo {author} {\bibfnamefont
  {R.}~\bibnamefont {Zhang}}, \ and\ \bibinfo {author} {\bibfnamefont
  {Y.}~\bibnamefont {Zhao}},\ }\href {\doibase 10.1007/JHEP07(2024)211}
  {\bibfield  {journal} {\bibinfo  {journal} {JHEP}\ }\textbf {\bibinfo
  {volume} {07}},\ \bibinfo {pages} {211} (\bibinfo {year} {2024})}\BibitemShut
  {NoStop}%
\bibitem [{\citenamefont {Ji}\ \emph {et~al.}(2025)\citenamefont {Ji},
  \citenamefont {Liu}, \citenamefont {Su},\ and\ \citenamefont
  {Zhang}}]{Ji:2024hit}%
  \BibitemOpen
  \bibfield  {author} {\bibinfo {author} {\bibfnamefont {X.}~\bibnamefont
  {Ji}}, \bibinfo {author} {\bibfnamefont {Y.}~\bibnamefont {Liu}}, \bibinfo
  {author} {\bibfnamefont {Y.}~\bibnamefont {Su}}, \ and\ \bibinfo {author}
  {\bibfnamefont {R.}~\bibnamefont {Zhang}},\ }\href {\doibase
  10.1007/JHEP03(2025)045} {\bibfield  {journal} {\bibinfo  {journal} {JHEP}\
  }\textbf {\bibinfo {volume} {03}},\ \bibinfo {pages} {045} (\bibinfo {year}
  {2025})}\BibitemShut {NoStop}%
\bibitem [{\citenamefont {Holligan}\ \emph {et~al.}(2025)\citenamefont
  {Holligan}, \citenamefont {Lin}, \citenamefont {Zhang},\ and\ \citenamefont
  {Zhao}}]{Holligan:2025baj}%
  \BibitemOpen
  \bibfield  {author} {\bibinfo {author} {\bibfnamefont {J.}~\bibnamefont
  {Holligan}}, \bibinfo {author} {\bibfnamefont {H.-W.}\ \bibnamefont {Lin}},
  \bibinfo {author} {\bibfnamefont {R.}~\bibnamefont {Zhang}}, \ and\ \bibinfo
  {author} {\bibfnamefont {Y.}~\bibnamefont {Zhao}},\ }\href {\doibase
  10.1007/JHEP07(2025)241} {\bibfield  {journal} {\bibinfo  {journal} {JHEP}\
  }\textbf {\bibinfo {volume} {207}},\ \bibinfo {pages} {241} (\bibinfo {year}
  {2025})}\BibitemShut {NoStop}%
\bibitem [{\citenamefont {Gao}\ \emph {et~al.}(2024{\natexlab{a}})\citenamefont
  {Gao}, \citenamefont {Liu},\ and\ \citenamefont {Zhao}}]{Gao:2023lny}%
  \BibitemOpen
  \bibfield  {author} {\bibinfo {author} {\bibfnamefont {X.}~\bibnamefont
  {Gao}}, \bibinfo {author} {\bibfnamefont {W.-Y.}\ \bibnamefont {Liu}}, \ and\
  \bibinfo {author} {\bibfnamefont {Y.}~\bibnamefont {Zhao}},\ }\href {\doibase
  10.1103/PhysRevD.109.094506} {\bibfield  {journal} {\bibinfo  {journal}
  {Phys. Rev. D}\ }\textbf {\bibinfo {volume} {109}},\ \bibinfo {pages}
  {094506} (\bibinfo {year} {2024}{\natexlab{a}})}\BibitemShut {NoStop}%
\bibitem [{\citenamefont {Zhao}(2024)}]{Zhao:2023ptv}%
  \BibitemOpen
  \bibfield  {author} {\bibinfo {author} {\bibfnamefont {Y.}~\bibnamefont
  {Zhao}},\ }\href {\doibase 10.1103/PhysRevLett.133.241904} {\bibfield
  {journal} {\bibinfo  {journal} {Phys. Rev. Lett.}\ }\textbf {\bibinfo
  {volume} {133}},\ \bibinfo {pages} {241904} (\bibinfo {year}
  {2024})}\BibitemShut {NoStop}%
\bibitem [{\citenamefont {Bollweg}\ \emph {et~al.}(2024)\citenamefont
  {Bollweg}, \citenamefont {Gao}, \citenamefont {Mukherjee},\ and\
  \citenamefont {Zhao}}]{Bollweg:2024zet}%
  \BibitemOpen
  \bibfield  {author} {\bibinfo {author} {\bibfnamefont {D.}~\bibnamefont
  {Bollweg}}, \bibinfo {author} {\bibfnamefont {X.}~\bibnamefont {Gao}},
  \bibinfo {author} {\bibfnamefont {S.}~\bibnamefont {Mukherjee}}, \ and\
  \bibinfo {author} {\bibfnamefont {Y.}~\bibnamefont {Zhao}},\ }\href {\doibase
  10.1016/j.physletb.2024.138617} {\bibfield  {journal} {\bibinfo  {journal}
  {Phys. Lett. B}\ }\textbf {\bibinfo {volume} {852}},\ \bibinfo {pages}
  {138617} (\bibinfo {year} {2024})}\BibitemShut {NoStop}%
\bibitem [{\citenamefont {Braun}\ and\ \citenamefont
  {M{\"u}ller}(2008)}]{Braun:2007wv}%
  \BibitemOpen
  \bibfield  {author} {\bibinfo {author} {\bibfnamefont {V.}~\bibnamefont
  {Braun}}\ and\ \bibinfo {author} {\bibfnamefont {D.}~\bibnamefont
  {M{\"u}ller}},\ }\href {\doibase 10.1140/epjc/s10052-008-0608-4} {\bibfield
  {journal} {\bibinfo  {journal} {Eur. Phys. J. C}\ }\textbf {\bibinfo {volume}
  {55}},\ \bibinfo {pages} {349} (\bibinfo {year} {2008})}\BibitemShut
  {NoStop}%
\bibitem [{\citenamefont {Radyushkin}(2017)}]{Radyushkin:2017cyf}%
  \BibitemOpen
  \bibfield  {author} {\bibinfo {author} {\bibfnamefont {A.~V.}\ \bibnamefont
  {Radyushkin}},\ }\href {\doibase 10.1103/PhysRevD.96.034025} {\bibfield
  {journal} {\bibinfo  {journal} {Phys. Rev. D}\ }\textbf {\bibinfo {volume}
  {96}},\ \bibinfo {pages} {034025} (\bibinfo {year} {2017})}\BibitemShut
  {NoStop}%
\bibitem [{\citenamefont {Orginos}\ \emph {et~al.}(2017)\citenamefont
  {Orginos}, \citenamefont {Radyushkin}, \citenamefont {Karpie},\ and\
  \citenamefont {Zafeiropoulos}}]{Orginos:2017kos}%
  \BibitemOpen
  \bibfield  {author} {\bibinfo {author} {\bibfnamefont {K.}~\bibnamefont
  {Orginos}}, \bibinfo {author} {\bibfnamefont {A.}~\bibnamefont {Radyushkin}},
  \bibinfo {author} {\bibfnamefont {J.}~\bibnamefont {Karpie}}, \ and\ \bibinfo
  {author} {\bibfnamefont {S.}~\bibnamefont {Zafeiropoulos}},\ }\href {\doibase
  10.1103/PhysRevD.96.094503} {\bibfield  {journal} {\bibinfo  {journal} {Phys.
  Rev. D}\ }\textbf {\bibinfo {volume} {96}},\ \bibinfo {pages} {094503}
  (\bibinfo {year} {2017})}\BibitemShut {NoStop}%
\bibitem [{\citenamefont {Ma}\ and\ \citenamefont {Qiu}(2018)}]{Ma:2017pxb}%
  \BibitemOpen
  \bibfield  {author} {\bibinfo {author} {\bibfnamefont {Y.-Q.}\ \bibnamefont
  {Ma}}\ and\ \bibinfo {author} {\bibfnamefont {J.-W.}\ \bibnamefont {Qiu}},\
  }\href {\doibase 10.1103/PhysRevLett.120.022003} {\bibfield  {journal}
  {\bibinfo  {journal} {Phys. Rev. Lett.}\ }\textbf {\bibinfo {volume} {120}},\
  \bibinfo {pages} {022003} (\bibinfo {year} {2018})}\BibitemShut {NoStop}%
\bibitem [{\citenamefont {Liu}\ and\ \citenamefont {Dong}(1994)}]{Liu:1993cv}%
  \BibitemOpen
  \bibfield  {author} {\bibinfo {author} {\bibfnamefont {K.-F.}\ \bibnamefont
  {Liu}}\ and\ \bibinfo {author} {\bibfnamefont {S.-J.}\ \bibnamefont {Dong}},\
  }\href {\doibase 10.1103/PhysRevLett.72.1790} {\bibfield  {journal} {\bibinfo
   {journal} {Phys. Rev. Lett.}\ }\textbf {\bibinfo {volume} {72}},\ \bibinfo
  {pages} {1790} (\bibinfo {year} {1994})}\BibitemShut {NoStop}%
\bibitem [{\citenamefont {Liang}\ \emph {et~al.}(2020)\citenamefont {Liang},
  \citenamefont {Draper}, \citenamefont {Liu}, \citenamefont {Rothkopf},\ and\
  \citenamefont {Yang}}]{Liang:2019frk}%
  \BibitemOpen
  \bibfield  {author} {\bibinfo {author} {\bibfnamefont {J.}~\bibnamefont
  {Liang}}, \bibinfo {author} {\bibfnamefont {T.}~\bibnamefont {Draper}},
  \bibinfo {author} {\bibfnamefont {K.-F.}\ \bibnamefont {Liu}}, \bibinfo
  {author} {\bibfnamefont {A.}~\bibnamefont {Rothkopf}}, \ and\ \bibinfo
  {author} {\bibfnamefont {Y.-B.}\ \bibnamefont {Yang}} (\bibinfo
  {collaboration} {XQCD}),\ }\href {\doibase 10.1103/PhysRevD.101.114503}
  {\bibfield  {journal} {\bibinfo  {journal} {Phys. Rev. D}\ }\textbf {\bibinfo
  {volume} {101}},\ \bibinfo {pages} {114503} (\bibinfo {year}
  {2020})}\BibitemShut {NoStop}%
\bibitem [{\citenamefont {Detmold}\ and\ \citenamefont
  {Lin}(2006)}]{Detmold:2005gg}%
  \BibitemOpen
  \bibfield  {author} {\bibinfo {author} {\bibfnamefont {W.}~\bibnamefont
  {Detmold}}\ and\ \bibinfo {author} {\bibfnamefont {C.~J.~D.}\ \bibnamefont
  {Lin}},\ }\href {\doibase 10.1103/PhysRevD.73.014501} {\bibfield  {journal}
  {\bibinfo  {journal} {Phys. Rev. D}\ }\textbf {\bibinfo {volume} {73}},\
  \bibinfo {pages} {014501} (\bibinfo {year} {2006})}\BibitemShut {NoStop}%
\bibitem [{\citenamefont {Detmold}\ \emph {et~al.}(2021)\citenamefont
  {Detmold}, \citenamefont {Grebe}, \citenamefont {Kanamori}, \citenamefont
  {Lin}, \citenamefont {Perry},\ and\ \citenamefont {Zhao}}]{Detmold:2021uru}%
  \BibitemOpen
  \bibfield  {author} {\bibinfo {author} {\bibfnamefont {W.}~\bibnamefont
  {Detmold}}, \bibinfo {author} {\bibfnamefont {A.~V.}\ \bibnamefont {Grebe}},
  \bibinfo {author} {\bibfnamefont {I.}~\bibnamefont {Kanamori}}, \bibinfo
  {author} {\bibfnamefont {C.~J.~D.}\ \bibnamefont {Lin}}, \bibinfo {author}
  {\bibfnamefont {R.~J.}\ \bibnamefont {Perry}}, \ and\ \bibinfo {author}
  {\bibfnamefont {Y.}~\bibnamefont {Zhao}} (\bibinfo {collaboration} {HOPE}),\
  }\href {\doibase 10.1103/PhysRevD.104.074511} {\bibfield  {journal} {\bibinfo
   {journal} {Phys. Rev. D}\ }\textbf {\bibinfo {volume} {104}},\ \bibinfo
  {pages} {074511} (\bibinfo {year} {2021})}\BibitemShut {NoStop}%
\bibitem [{\citenamefont {Chambers}\ \emph {et~al.}(2017)\citenamefont
  {Chambers}, \citenamefont {Horsley}, \citenamefont {Nakamura}, \citenamefont
  {Perlt}, \citenamefont {Rakow}, \citenamefont {Schierholz}, \citenamefont
  {Schiller}, \citenamefont {Somfleth}, \citenamefont {Young},\ and\
  \citenamefont {Zanotti}}]{Chambers:2017dov}%
  \BibitemOpen
  \bibfield  {author} {\bibinfo {author} {\bibfnamefont {A.~J.}\ \bibnamefont
  {Chambers}}, \bibinfo {author} {\bibfnamefont {R.}~\bibnamefont {Horsley}},
  \bibinfo {author} {\bibfnamefont {Y.}~\bibnamefont {Nakamura}}, \bibinfo
  {author} {\bibfnamefont {H.}~\bibnamefont {Perlt}}, \bibinfo {author}
  {\bibfnamefont {P.~E.~L.}\ \bibnamefont {Rakow}}, \bibinfo {author}
  {\bibfnamefont {G.}~\bibnamefont {Schierholz}}, \bibinfo {author}
  {\bibfnamefont {A.}~\bibnamefont {Schiller}}, \bibinfo {author}
  {\bibfnamefont {K.}~\bibnamefont {Somfleth}}, \bibinfo {author}
  {\bibfnamefont {R.~D.}\ \bibnamefont {Young}}, \ and\ \bibinfo {author}
  {\bibfnamefont {J.~M.}\ \bibnamefont {Zanotti}},\ }\href {\doibase
  10.1103/PhysRevLett.118.242001} {\bibfield  {journal} {\bibinfo  {journal}
  {Phys. Rev. Lett.}\ }\textbf {\bibinfo {volume} {118}},\ \bibinfo {pages}
  {242001} (\bibinfo {year} {2017})}\BibitemShut {NoStop}%
\bibitem [{\citenamefont {Davoudi}\ and\ \citenamefont
  {Savage}(2012)}]{Davoudi:2012ya}%
  \BibitemOpen
  \bibfield  {author} {\bibinfo {author} {\bibfnamefont {Z.}~\bibnamefont
  {Davoudi}}\ and\ \bibinfo {author} {\bibfnamefont {M.~J.}\ \bibnamefont
  {Savage}},\ }\href {\doibase 10.1103/PhysRevD.86.054505} {\bibfield
  {journal} {\bibinfo  {journal} {Phys. Rev. D}\ }\textbf {\bibinfo {volume}
  {86}},\ \bibinfo {pages} {054505} (\bibinfo {year} {2012})}\BibitemShut
  {NoStop}%
\bibitem [{\citenamefont {Shindler}(2024)}]{Shindler:2023xpd}%
  \BibitemOpen
  \bibfield  {author} {\bibinfo {author} {\bibfnamefont {A.}~\bibnamefont
  {Shindler}},\ }\href {\doibase 10.1103/PhysRevD.110.L051503} {\bibfield
  {journal} {\bibinfo  {journal} {Phys. Rev. D}\ }\textbf {\bibinfo {volume}
  {110}},\ \bibinfo {pages} {L051503} (\bibinfo {year} {2024})}\BibitemShut
  {NoStop}%
\bibitem [{\citenamefont {Francis}\ \emph {et~al.}(2026)\citenamefont {Francis}
  \emph {et~al.}}]{Francis:2025rya}%
  \BibitemOpen
  \bibfield  {author} {\bibinfo {author} {\bibfnamefont {A.}~\bibnamefont
  {Francis}} \emph {et~al.},\ }\href {\doibase 10.1103/z3wr-zk8n} {\bibfield
  {journal} {\bibinfo  {journal} {Phys. Rev. Lett.}\ }\textbf {\bibinfo
  {volume} {136}},\ \bibinfo {pages} {171903} (\bibinfo {year}
  {2026})}\BibitemShut {NoStop}%
\bibitem [{\citenamefont {Liu}\ \emph {et~al.}(2020)\citenamefont {Liu} \emph
  {et~al.}}]{LatticeParton:2018gjr}%
  \BibitemOpen
  \bibfield  {author} {\bibinfo {author} {\bibfnamefont {Y.-S.}\ \bibnamefont
  {Liu}} \emph {et~al.} (\bibinfo {collaboration} {Lattice Parton}),\ }\href
  {\doibase 10.1103/PhysRevD.101.034020} {\bibfield  {journal} {\bibinfo
  {journal} {Phys. Rev. D}\ }\textbf {\bibinfo {volume} {101}},\ \bibinfo
  {pages} {034020} (\bibinfo {year} {2020})}\BibitemShut {NoStop}%
\bibitem [{\citenamefont {Lin}\ \emph {et~al.}(2018{\natexlab{a}})\citenamefont
  {Lin}, \citenamefont {Chen}, \citenamefont {Ishikawa},\ and\ \citenamefont
  {Zhang}}]{Lin:2017ani}%
  \BibitemOpen
  \bibfield  {author} {\bibinfo {author} {\bibfnamefont {H.-W.}\ \bibnamefont
  {Lin}}, \bibinfo {author} {\bibfnamefont {J.-W.}\ \bibnamefont {Chen}},
  \bibinfo {author} {\bibfnamefont {T.}~\bibnamefont {Ishikawa}}, \ and\
  \bibinfo {author} {\bibfnamefont {J.-H.}\ \bibnamefont {Zhang}} (\bibinfo
  {collaboration} {LP3}),\ }\href {\doibase 10.1103/PhysRevD.98.054504}
  {\bibfield  {journal} {\bibinfo  {journal} {Phys. Rev. D}\ }\textbf {\bibinfo
  {volume} {98}},\ \bibinfo {pages} {054504} (\bibinfo {year}
  {2018}{\natexlab{a}})}\BibitemShut {NoStop}%
\bibitem [{\citenamefont {Alexandrou}\ \emph
  {et~al.}(2018{\natexlab{a}})\citenamefont {Alexandrou}, \citenamefont
  {Cichy}, \citenamefont {Constantinou}, \citenamefont {Jansen}, \citenamefont
  {Scapellato},\ and\ \citenamefont {Steffens}}]{Alexandrou:2018pbm}%
  \BibitemOpen
  \bibfield  {author} {\bibinfo {author} {\bibfnamefont {C.}~\bibnamefont
  {Alexandrou}}, \bibinfo {author} {\bibfnamefont {K.}~\bibnamefont {Cichy}},
  \bibinfo {author} {\bibfnamefont {M.}~\bibnamefont {Constantinou}}, \bibinfo
  {author} {\bibfnamefont {K.}~\bibnamefont {Jansen}}, \bibinfo {author}
  {\bibfnamefont {A.}~\bibnamefont {Scapellato}}, \ and\ \bibinfo {author}
  {\bibfnamefont {F.}~\bibnamefont {Steffens}},\ }\href {\doibase
  10.1103/PhysRevLett.121.112001} {\bibfield  {journal} {\bibinfo  {journal}
  {Phys. Rev. Lett.}\ }\textbf {\bibinfo {volume} {121}},\ \bibinfo {pages}
  {112001} (\bibinfo {year} {2018}{\natexlab{a}})}\BibitemShut {NoStop}%
\bibitem [{\citenamefont {Chen}\ \emph
  {et~al.}(2018{\natexlab{b}})\citenamefont {Chen}, \citenamefont {Jin},
  \citenamefont {Lin}, \citenamefont {Liu}, \citenamefont {Yang}, \citenamefont
  {Zhang},\ and\ \citenamefont {Zhao}}]{Chen:2018xof}%
  \BibitemOpen
  \bibfield  {author} {\bibinfo {author} {\bibfnamefont {J.-W.}\ \bibnamefont
  {Chen}}, \bibinfo {author} {\bibfnamefont {L.}~\bibnamefont {Jin}}, \bibinfo
  {author} {\bibfnamefont {H.-W.}\ \bibnamefont {Lin}}, \bibinfo {author}
  {\bibfnamefont {Y.-S.}\ \bibnamefont {Liu}}, \bibinfo {author} {\bibfnamefont
  {Y.-B.}\ \bibnamefont {Yang}}, \bibinfo {author} {\bibfnamefont {J.-H.}\
  \bibnamefont {Zhang}}, \ and\ \bibinfo {author} {\bibfnamefont
  {Y.}~\bibnamefont {Zhao}},\ }\href@noop {} {\  (\bibinfo {year}
  {2018}{\natexlab{b}})},\ \bibinfo {note} {arXiv:1803.04393}\BibitemShut
  {NoStop}%
\bibitem [{\citenamefont {Alexandrou}\ \emph {et~al.}(2019)\citenamefont
  {Alexandrou}, \citenamefont {Cichy}, \citenamefont {Constantinou},
  \citenamefont {Hadjiyiannakou}, \citenamefont {Jansen}, \citenamefont
  {Scapellato},\ and\ \citenamefont {Steffens}}]{Alexandrou:2019lfo}%
  \BibitemOpen
  \bibfield  {author} {\bibinfo {author} {\bibfnamefont {C.}~\bibnamefont
  {Alexandrou}}, \bibinfo {author} {\bibfnamefont {K.}~\bibnamefont {Cichy}},
  \bibinfo {author} {\bibfnamefont {M.}~\bibnamefont {Constantinou}}, \bibinfo
  {author} {\bibfnamefont {K.}~\bibnamefont {Hadjiyiannakou}}, \bibinfo
  {author} {\bibfnamefont {K.}~\bibnamefont {Jansen}}, \bibinfo {author}
  {\bibfnamefont {A.}~\bibnamefont {Scapellato}}, \ and\ \bibinfo {author}
  {\bibfnamefont {F.}~\bibnamefont {Steffens}},\ }\href {\doibase
  10.1103/PhysRevD.99.114504} {\bibfield  {journal} {\bibinfo  {journal} {Phys.
  Rev. D}\ }\textbf {\bibinfo {volume} {99}},\ \bibinfo {pages} {114504}
  (\bibinfo {year} {2019})}\BibitemShut {NoStop}%
\bibitem [{\citenamefont {Jo\'o}\ \emph {et~al.}(2019)\citenamefont {Jo\'o},
  \citenamefont {Karpie}, \citenamefont {Orginos}, \citenamefont {Radyushkin},
  \citenamefont {Richards},\ and\ \citenamefont {Zafeiropoulos}}]{Joo:2019jct}%
  \BibitemOpen
  \bibfield  {author} {\bibinfo {author} {\bibfnamefont {B.}~\bibnamefont
  {Jo\'o}}, \bibinfo {author} {\bibfnamefont {J.}~\bibnamefont {Karpie}},
  \bibinfo {author} {\bibfnamefont {K.}~\bibnamefont {Orginos}}, \bibinfo
  {author} {\bibfnamefont {A.}~\bibnamefont {Radyushkin}}, \bibinfo {author}
  {\bibfnamefont {D.}~\bibnamefont {Richards}}, \ and\ \bibinfo {author}
  {\bibfnamefont {S.}~\bibnamefont {Zafeiropoulos}},\ }\href {\doibase
  10.1007/JHEP12(2019)081} {\bibfield  {journal} {\bibinfo  {journal} {JHEP}\
  }\textbf {\bibinfo {volume} {12}},\ \bibinfo {pages} {081} (\bibinfo {year}
  {2019})}\BibitemShut {NoStop}%
\bibitem [{\citenamefont {Jo\'o}\ \emph {et~al.}(2020)\citenamefont {Jo\'o},
  \citenamefont {Karpie}, \citenamefont {Orginos}, \citenamefont {Radyushkin},
  \citenamefont {Richards},\ and\ \citenamefont {Zafeiropoulos}}]{Joo:2020spy}%
  \BibitemOpen
  \bibfield  {author} {\bibinfo {author} {\bibfnamefont {B.}~\bibnamefont
  {Jo\'o}}, \bibinfo {author} {\bibfnamefont {J.}~\bibnamefont {Karpie}},
  \bibinfo {author} {\bibfnamefont {K.}~\bibnamefont {Orginos}}, \bibinfo
  {author} {\bibfnamefont {A.~V.}\ \bibnamefont {Radyushkin}}, \bibinfo
  {author} {\bibfnamefont {D.~G.}\ \bibnamefont {Richards}}, \ and\ \bibinfo
  {author} {\bibfnamefont {S.}~\bibnamefont {Zafeiropoulos}},\ }\href {\doibase
  10.1103/PhysRevLett.125.232003} {\bibfield  {journal} {\bibinfo  {journal}
  {Phys. Rev. Lett.}\ }\textbf {\bibinfo {volume} {125}},\ \bibinfo {pages}
  {232003} (\bibinfo {year} {2020})}\BibitemShut {NoStop}%
\bibitem [{\citenamefont {Gao}\ \emph {et~al.}(2023)\citenamefont {Gao},
  \citenamefont {Hanlon}, \citenamefont {Holligan}, \citenamefont {Karthik},
  \citenamefont {Mukherjee}, \citenamefont {Petreczky}, \citenamefont
  {Syritsyn},\ and\ \citenamefont {Zhao}}]{Gao:2022uhg}%
  \BibitemOpen
  \bibfield  {author} {\bibinfo {author} {\bibfnamefont {X.}~\bibnamefont
  {Gao}}, \bibinfo {author} {\bibfnamefont {A.~D.}\ \bibnamefont {Hanlon}},
  \bibinfo {author} {\bibfnamefont {J.}~\bibnamefont {Holligan}}, \bibinfo
  {author} {\bibfnamefont {N.}~\bibnamefont {Karthik}}, \bibinfo {author}
  {\bibfnamefont {S.}~\bibnamefont {Mukherjee}}, \bibinfo {author}
  {\bibfnamefont {P.}~\bibnamefont {Petreczky}}, \bibinfo {author}
  {\bibfnamefont {S.}~\bibnamefont {Syritsyn}}, \ and\ \bibinfo {author}
  {\bibfnamefont {Y.}~\bibnamefont {Zhao}},\ }\href {\doibase
  10.1103/PhysRevD.107.074509} {\bibfield  {journal} {\bibinfo  {journal}
  {Phys. Rev. D}\ }\textbf {\bibinfo {volume} {107}},\ \bibinfo {pages}
  {074509} (\bibinfo {year} {2023})}\BibitemShut {NoStop}%
\bibitem [{\citenamefont {Liu}\ \emph {et~al.}(2018)\citenamefont {Liu},
  \citenamefont {Chen}, \citenamefont {Jin}, \citenamefont {Li}, \citenamefont
  {Lin}, \citenamefont {Yang}, \citenamefont {Zhang},\ and\ \citenamefont
  {Zhao}}]{Liu:2018hxv}%
  \BibitemOpen
  \bibfield  {author} {\bibinfo {author} {\bibfnamefont {Y.-S.}\ \bibnamefont
  {Liu}}, \bibinfo {author} {\bibfnamefont {J.-W.}\ \bibnamefont {Chen}},
  \bibinfo {author} {\bibfnamefont {L.}~\bibnamefont {Jin}}, \bibinfo {author}
  {\bibfnamefont {R.}~\bibnamefont {Li}}, \bibinfo {author} {\bibfnamefont
  {H.-W.}\ \bibnamefont {Lin}}, \bibinfo {author} {\bibfnamefont {Y.-B.}\
  \bibnamefont {Yang}}, \bibinfo {author} {\bibfnamefont {J.-H.}\ \bibnamefont
  {Zhang}}, \ and\ \bibinfo {author} {\bibfnamefont {Y.}~\bibnamefont {Zhao}},\
  }\href@noop {} {\  (\bibinfo {year} {2018})},\ \bibinfo {note}
  {arXiv:1810.05043}\BibitemShut {NoStop}%
\bibitem [{\citenamefont {Alexandrou}\ \emph
  {et~al.}(2018{\natexlab{b}})\citenamefont {Alexandrou}, \citenamefont
  {Cichy}, \citenamefont {Constantinou}, \citenamefont {Jansen}, \citenamefont
  {Scapellato},\ and\ \citenamefont {Steffens}}]{Alexandrou:2018eet}%
  \BibitemOpen
  \bibfield  {author} {\bibinfo {author} {\bibfnamefont {C.}~\bibnamefont
  {Alexandrou}}, \bibinfo {author} {\bibfnamefont {K.}~\bibnamefont {Cichy}},
  \bibinfo {author} {\bibfnamefont {M.}~\bibnamefont {Constantinou}}, \bibinfo
  {author} {\bibfnamefont {K.}~\bibnamefont {Jansen}}, \bibinfo {author}
  {\bibfnamefont {A.}~\bibnamefont {Scapellato}}, \ and\ \bibinfo {author}
  {\bibfnamefont {F.}~\bibnamefont {Steffens}},\ }\href {\doibase
  10.1103/PhysRevD.98.091503} {\bibfield  {journal} {\bibinfo  {journal} {Phys.
  Rev. D}\ }\textbf {\bibinfo {volume} {98}},\ \bibinfo {pages} {091503}
  (\bibinfo {year} {2018}{\natexlab{b}})}\BibitemShut {NoStop}%
\bibitem [{\citenamefont {Yao}\ \emph {et~al.}(2023)\citenamefont {Yao} \emph
  {et~al.}}]{LatticeParton:2022xsd}%
  \BibitemOpen
  \bibfield  {author} {\bibinfo {author} {\bibfnamefont {F.}~\bibnamefont
  {Yao}} \emph {et~al.} (\bibinfo {collaboration} {Lattice Parton}),\ }\href
  {\doibase 10.1103/PhysRevLett.131.261901} {\bibfield  {journal} {\bibinfo
  {journal} {Phys. Rev. Lett.}\ }\textbf {\bibinfo {volume} {131}},\ \bibinfo
  {pages} {261901} (\bibinfo {year} {2023})}\BibitemShut {NoStop}%
\bibitem [{\citenamefont {Gao}\ \emph {et~al.}(2024{\natexlab{b}})\citenamefont
  {Gao}, \citenamefont {Hanlon}, \citenamefont {Mukherjee}, \citenamefont
  {Petreczky}, \citenamefont {Shi}, \citenamefont {Syritsyn},\ and\
  \citenamefont {Zhao}}]{Gao:2023ktu}%
  \BibitemOpen
  \bibfield  {author} {\bibinfo {author} {\bibfnamefont {X.}~\bibnamefont
  {Gao}}, \bibinfo {author} {\bibfnamefont {A.~D.}\ \bibnamefont {Hanlon}},
  \bibinfo {author} {\bibfnamefont {S.}~\bibnamefont {Mukherjee}}, \bibinfo
  {author} {\bibfnamefont {P.}~\bibnamefont {Petreczky}}, \bibinfo {author}
  {\bibfnamefont {Q.}~\bibnamefont {Shi}}, \bibinfo {author} {\bibfnamefont
  {S.}~\bibnamefont {Syritsyn}}, \ and\ \bibinfo {author} {\bibfnamefont
  {Y.}~\bibnamefont {Zhao}},\ }\href {\doibase 10.1103/PhysRevD.109.054506}
  {\bibfield  {journal} {\bibinfo  {journal} {Phys. Rev. D}\ }\textbf {\bibinfo
  {volume} {109}},\ \bibinfo {pages} {054506} (\bibinfo {year}
  {2024}{\natexlab{b}})}\BibitemShut {NoStop}%
\bibitem [{\citenamefont {Chen}\ \emph {et~al.}(2016)\citenamefont {Chen},
  \citenamefont {Cohen}, \citenamefont {Ji}, \citenamefont {Lin},\ and\
  \citenamefont {Zhang}}]{Chen:2016utp}%
  \BibitemOpen
  \bibfield  {author} {\bibinfo {author} {\bibfnamefont {J.-W.}\ \bibnamefont
  {Chen}}, \bibinfo {author} {\bibfnamefont {S.~D.}\ \bibnamefont {Cohen}},
  \bibinfo {author} {\bibfnamefont {X.}~\bibnamefont {Ji}}, \bibinfo {author}
  {\bibfnamefont {H.-W.}\ \bibnamefont {Lin}}, \ and\ \bibinfo {author}
  {\bibfnamefont {J.-H.}\ \bibnamefont {Zhang}},\ }\href {\doibase
  10.1016/j.nuclphysb.2016.07.033} {\bibfield  {journal} {\bibinfo  {journal}
  {Nucl. Phys. B}\ }\textbf {\bibinfo {volume} {911}},\ \bibinfo {pages} {246}
  (\bibinfo {year} {2016})}\BibitemShut {NoStop}%
\bibitem [{\citenamefont {Lin}\ \emph {et~al.}(2018{\natexlab{b}})\citenamefont
  {Lin}, \citenamefont {Chen}, \citenamefont {Ji}, \citenamefont {Jin},
  \citenamefont {Li}, \citenamefont {Liu}, \citenamefont {Yang}, \citenamefont
  {Zhang},\ and\ \citenamefont {Zhao}}]{Lin:2018pvv}%
  \BibitemOpen
  \bibfield  {author} {\bibinfo {author} {\bibfnamefont {H.-W.}\ \bibnamefont
  {Lin}}, \bibinfo {author} {\bibfnamefont {J.-W.}\ \bibnamefont {Chen}},
  \bibinfo {author} {\bibfnamefont {X.}~\bibnamefont {Ji}}, \bibinfo {author}
  {\bibfnamefont {L.}~\bibnamefont {Jin}}, \bibinfo {author} {\bibfnamefont
  {R.}~\bibnamefont {Li}}, \bibinfo {author} {\bibfnamefont {Y.-S.}\
  \bibnamefont {Liu}}, \bibinfo {author} {\bibfnamefont {Y.-B.}\ \bibnamefont
  {Yang}}, \bibinfo {author} {\bibfnamefont {J.-H.}\ \bibnamefont {Zhang}}, \
  and\ \bibinfo {author} {\bibfnamefont {Y.}~\bibnamefont {Zhao}},\ }\href
  {\doibase 10.1103/PhysRevLett.121.242003} {\bibfield  {journal} {\bibinfo
  {journal} {Phys. Rev. Lett.}\ }\textbf {\bibinfo {volume} {121}},\ \bibinfo
  {pages} {242003} (\bibinfo {year} {2018}{\natexlab{b}})}\BibitemShut
  {NoStop}%
\bibitem [{\citenamefont {Lin}\ and\ \citenamefont
  {Zhang}(2019)}]{Lin:2019ocg}%
  \BibitemOpen
  \bibfield  {author} {\bibinfo {author} {\bibfnamefont {H.-W.}\ \bibnamefont
  {Lin}}\ and\ \bibinfo {author} {\bibfnamefont {R.}~\bibnamefont {Zhang}},\
  }\href {\doibase 10.1103/PhysRevD.100.074502} {\bibfield  {journal} {\bibinfo
   {journal} {Phys. Rev. D}\ }\textbf {\bibinfo {volume} {100}},\ \bibinfo
  {pages} {074502} (\bibinfo {year} {2019})}\BibitemShut {NoStop}%
\bibitem [{\citenamefont {Alexandrou}\ \emph
  {et~al.}(2021{\natexlab{a}})\citenamefont {Alexandrou}, \citenamefont
  {Cichy}, \citenamefont {Constantinou}, \citenamefont {Green}, \citenamefont
  {Hadjiyiannakou}, \citenamefont {Jansen}, \citenamefont {Manigrasso},
  \citenamefont {Scapellato},\ and\ \citenamefont
  {Steffens}}]{Alexandrou:2020qtt}%
  \BibitemOpen
  \bibfield  {author} {\bibinfo {author} {\bibfnamefont {C.}~\bibnamefont
  {Alexandrou}}, \bibinfo {author} {\bibfnamefont {K.}~\bibnamefont {Cichy}},
  \bibinfo {author} {\bibfnamefont {M.}~\bibnamefont {Constantinou}}, \bibinfo
  {author} {\bibfnamefont {J.~R.}\ \bibnamefont {Green}}, \bibinfo {author}
  {\bibfnamefont {K.}~\bibnamefont {Hadjiyiannakou}}, \bibinfo {author}
  {\bibfnamefont {K.}~\bibnamefont {Jansen}}, \bibinfo {author} {\bibfnamefont
  {F.}~\bibnamefont {Manigrasso}}, \bibinfo {author} {\bibfnamefont
  {A.}~\bibnamefont {Scapellato}}, \ and\ \bibinfo {author} {\bibfnamefont
  {F.}~\bibnamefont {Steffens}},\ }\href {\doibase 10.1103/PhysRevD.103.094512}
  {\bibfield  {journal} {\bibinfo  {journal} {Phys. Rev. D}\ }\textbf {\bibinfo
  {volume} {103}},\ \bibinfo {pages} {094512} (\bibinfo {year}
  {2021}{\natexlab{a}})}\BibitemShut {NoStop}%
\bibitem [{\citenamefont {Alexandrou}\ \emph
  {et~al.}(2021{\natexlab{b}})\citenamefont {Alexandrou}, \citenamefont
  {Constantinou}, \citenamefont {Hadjiyiannakou}, \citenamefont {Jansen},\ and\
  \citenamefont {Manigrasso}}]{Alexandrou:2021oih}%
  \BibitemOpen
  \bibfield  {author} {\bibinfo {author} {\bibfnamefont {C.}~\bibnamefont
  {Alexandrou}}, \bibinfo {author} {\bibfnamefont {M.}~\bibnamefont
  {Constantinou}}, \bibinfo {author} {\bibfnamefont {K.}~\bibnamefont
  {Hadjiyiannakou}}, \bibinfo {author} {\bibfnamefont {K.}~\bibnamefont
  {Jansen}}, \ and\ \bibinfo {author} {\bibfnamefont {F.}~\bibnamefont
  {Manigrasso}},\ }\href {\doibase 10.1103/PhysRevD.104.054503} {\bibfield
  {journal} {\bibinfo  {journal} {Phys. Rev. D}\ }\textbf {\bibinfo {volume}
  {104}},\ \bibinfo {pages} {054503} (\bibinfo {year}
  {2021}{\natexlab{b}})}\BibitemShut {NoStop}%
\bibitem [{\citenamefont {Alexandrou}\ \emph
  {et~al.}(2021{\natexlab{c}})\citenamefont {Alexandrou}, \citenamefont
  {Constantinou}, \citenamefont {Hadjiyiannakou}, \citenamefont {Jansen},\ and\
  \citenamefont {Manigrasso}}]{Alexandrou:2020uyt}%
  \BibitemOpen
  \bibfield  {author} {\bibinfo {author} {\bibfnamefont {C.}~\bibnamefont
  {Alexandrou}}, \bibinfo {author} {\bibfnamefont {M.}~\bibnamefont
  {Constantinou}}, \bibinfo {author} {\bibfnamefont {K.}~\bibnamefont
  {Hadjiyiannakou}}, \bibinfo {author} {\bibfnamefont {K.}~\bibnamefont
  {Jansen}}, \ and\ \bibinfo {author} {\bibfnamefont {F.}~\bibnamefont
  {Manigrasso}},\ }\href {\doibase 10.1103/PhysRevLett.126.102003} {\bibfield
  {journal} {\bibinfo  {journal} {Phys. Rev. Lett.}\ }\textbf {\bibinfo
  {volume} {126}},\ \bibinfo {pages} {102003} (\bibinfo {year}
  {2021}{\natexlab{c}})}\BibitemShut {NoStop}%
\bibitem [{\citenamefont {Fan}\ \emph {et~al.}(2020)\citenamefont {Fan},
  \citenamefont {Gao}, \citenamefont {Li}, \citenamefont {Lin}, \citenamefont
  {Karthik}, \citenamefont {Mukherjee}, \citenamefont {Petreczky},
  \citenamefont {Syritsyn}, \citenamefont {Yang},\ and\ \citenamefont
  {Zhang}}]{Fan:2020nzz}%
  \BibitemOpen
  \bibfield  {author} {\bibinfo {author} {\bibfnamefont {Z.}~\bibnamefont
  {Fan}}, \bibinfo {author} {\bibfnamefont {X.}~\bibnamefont {Gao}}, \bibinfo
  {author} {\bibfnamefont {R.}~\bibnamefont {Li}}, \bibinfo {author}
  {\bibfnamefont {H.-W.}\ \bibnamefont {Lin}}, \bibinfo {author} {\bibfnamefont
  {N.}~\bibnamefont {Karthik}}, \bibinfo {author} {\bibfnamefont
  {S.}~\bibnamefont {Mukherjee}}, \bibinfo {author} {\bibfnamefont
  {P.}~\bibnamefont {Petreczky}}, \bibinfo {author} {\bibfnamefont
  {S.}~\bibnamefont {Syritsyn}}, \bibinfo {author} {\bibfnamefont {Y.-B.}\
  \bibnamefont {Yang}}, \ and\ \bibinfo {author} {\bibfnamefont
  {R.}~\bibnamefont {Zhang}},\ }\href {\doibase 10.1103/PhysRevD.102.074504}
  {\bibfield  {journal} {\bibinfo  {journal} {Phys. Rev. D}\ }\textbf {\bibinfo
  {volume} {102}},\ \bibinfo {pages} {074504} (\bibinfo {year}
  {2020})}\BibitemShut {NoStop}%
\bibitem [{\citenamefont {Edwards}\ \emph {et~al.}(2023)\citenamefont {Edwards}
  \emph {et~al.}}]{HadStruc:2022nay}%
  \BibitemOpen
  \bibfield  {author} {\bibinfo {author} {\bibfnamefont {R.~G.}\ \bibnamefont
  {Edwards}} \emph {et~al.} (\bibinfo {collaboration} {HadStruc}),\ }\href
  {\doibase 10.1007/JHEP03(2023)086} {\bibfield  {journal} {\bibinfo  {journal}
  {JHEP}\ }\textbf {\bibinfo {volume} {03}},\ \bibinfo {pages} {086} (\bibinfo
  {year} {2023})}\BibitemShut {NoStop}%
\bibitem [{\citenamefont {Holligan}\ and\ \citenamefont
  {Lin}(2024)}]{Holligan:2024wpv}%
  \BibitemOpen
  \bibfield  {author} {\bibinfo {author} {\bibfnamefont {J.}~\bibnamefont
  {Holligan}}\ and\ \bibinfo {author} {\bibfnamefont {H.-W.}\ \bibnamefont
  {Lin}},\ }\href {\doibase 10.1016/j.physletb.2024.138731} {\bibfield
  {journal} {\bibinfo  {journal} {Phys. Lett. B}\ }\textbf {\bibinfo {volume}
  {854}},\ \bibinfo {pages} {138731} (\bibinfo {year} {2024})}\BibitemShut
  {NoStop}%
\bibitem [{\citenamefont {Zhao}(2019)}]{Zhao:2018fyu}%
  \BibitemOpen
  \bibfield  {author} {\bibinfo {author} {\bibfnamefont {Y.}~\bibnamefont
  {Zhao}},\ }\href {\doibase 10.1142/S0217751X18300338} {\bibfield  {journal}
  {\bibinfo  {journal} {Int. J. Mod. Phys. A}\ }\textbf {\bibinfo {volume}
  {33}},\ \bibinfo {pages} {1830033} (\bibinfo {year} {2019})}\BibitemShut
  {NoStop}%
\bibitem [{\citenamefont {Cichy}\ and\ \citenamefont
  {Constantinou}(2019)}]{Cichy:2018mum}%
  \BibitemOpen
  \bibfield  {author} {\bibinfo {author} {\bibfnamefont {K.}~\bibnamefont
  {Cichy}}\ and\ \bibinfo {author} {\bibfnamefont {M.}~\bibnamefont
  {Constantinou}},\ }\href {\doibase 10.1155/2019/3036904} {\bibfield
  {journal} {\bibinfo  {journal} {Adv. High Energy Phys.}\ }\textbf {\bibinfo
  {volume} {2019}},\ \bibinfo {pages} {3036904} (\bibinfo {year}
  {2019})}\BibitemShut {NoStop}%
\bibitem [{\citenamefont {Monahan}(2018)}]{Monahan:2018euv}%
  \BibitemOpen
  \bibfield  {author} {\bibinfo {author} {\bibfnamefont {C.}~\bibnamefont
  {Monahan}},\ }\href {\doibase 10.22323/1.334.0018} {\bibfield  {journal}
  {\bibinfo  {journal} {PoS}\ }\textbf {\bibinfo {volume} {LATTICE2018}},\
  \bibinfo {pages} {018} (\bibinfo {year} {2018})}\BibitemShut {NoStop}%
\bibitem [{\citenamefont {Constantinou}(2021)}]{Constantinou:2020pek}%
  \BibitemOpen
  \bibfield  {author} {\bibinfo {author} {\bibfnamefont {M.}~\bibnamefont
  {Constantinou}},\ }\href {\doibase 10.1140/epja/s10050-021-00353-7}
  {\bibfield  {journal} {\bibinfo  {journal} {Eur. Phys. J. A}\ }\textbf
  {\bibinfo {volume} {57}},\ \bibinfo {pages} {77} (\bibinfo {year}
  {2021})}\BibitemShut {NoStop}%
\bibitem [{\citenamefont {Burkardt}(2013)}]{Burkardt:2008ps}%
  \BibitemOpen
  \bibfield  {author} {\bibinfo {author} {\bibfnamefont {M.}~\bibnamefont
  {Burkardt}},\ }\href {\doibase 10.1103/PhysRevD.88.114502} {\bibfield
  {journal} {\bibinfo  {journal} {Phys. Rev. D}\ }\textbf {\bibinfo {volume}
  {88}},\ \bibinfo {pages} {114502} (\bibinfo {year} {2013})}\BibitemShut
  {NoStop}%
\bibitem [{\citenamefont {Armstrong}\ \emph {et~al.}(2019)\citenamefont
  {Armstrong} \emph {et~al.}}]{SANE:2018pwx}%
  \BibitemOpen
  \bibfield  {author} {\bibinfo {author} {\bibfnamefont {W.}~\bibnamefont
  {Armstrong}} \emph {et~al.} (\bibinfo {collaboration} {SANE}),\ }\href
  {\doibase 10.1103/PhysRevLett.122.022002} {\bibfield  {journal} {\bibinfo
  {journal} {Phys. Rev. Lett.}\ }\textbf {\bibinfo {volume} {122}},\ \bibinfo
  {pages} {022002} (\bibinfo {year} {2019})}\BibitemShut {NoStop}%
\bibitem [{\citenamefont {Deur}\ \emph {et~al.}(2019)\citenamefont {Deur},
  \citenamefont {Brodsky},\ and\ \citenamefont
  {De~T{\'e}ramond}}]{Deur:2018roz}%
  \BibitemOpen
  \bibfield  {author} {\bibinfo {author} {\bibfnamefont {A.}~\bibnamefont
  {Deur}}, \bibinfo {author} {\bibfnamefont {S.~J.}\ \bibnamefont {Brodsky}}, \
  and\ \bibinfo {author} {\bibfnamefont {G.~F.}\ \bibnamefont
  {De~T{\'e}ramond}},\ }\href {\doibase 10.1088/1361-6633/ab0b8f} {\bibfield
  {journal} {\bibinfo  {journal} {Rept. Prog. Phys.}\ }\textbf {\bibinfo
  {volume} {82}},\ \bibinfo {pages} {076201} (\bibinfo {year}
  {2019})}\BibitemShut {NoStop}%
\bibitem [{\citenamefont {Airapetian}\ \emph {et~al.}(2012)\citenamefont
  {Airapetian} \emph {et~al.}}]{HERMES:2011xgd}%
  \BibitemOpen
  \bibfield  {author} {\bibinfo {author} {\bibfnamefont {A.}~\bibnamefont
  {Airapetian}} \emph {et~al.} (\bibinfo {collaboration} {HERMES}),\ }\href
  {\doibase 10.1140/epjc/s10052-012-1921-5} {\bibfield  {journal} {\bibinfo
  {journal} {Eur. Phys. J. C}\ }\textbf {\bibinfo {volume} {72}},\ \bibinfo
  {pages} {1921} (\bibinfo {year} {2012})}\BibitemShut {NoStop}%
\bibitem [{\citenamefont {Flay}\ \emph {et~al.}(2016)\citenamefont {Flay} \emph
  {et~al.}}]{JeffersonLabHallA:2016neg}%
  \BibitemOpen
  \bibfield  {author} {\bibinfo {author} {\bibfnamefont {D.}~\bibnamefont
  {Flay}} \emph {et~al.} (\bibinfo {collaboration} {Jefferson Lab Hall A}),\
  }\href {\doibase 10.1103/PhysRevD.94.052003} {\bibfield  {journal} {\bibinfo
  {journal} {Phys. Rev. D}\ }\textbf {\bibinfo {volume} {94}},\ \bibinfo
  {pages} {052003} (\bibinfo {year} {2016})}\BibitemShut {NoStop}%
\bibitem [{\citenamefont {Cocuzza}\ \emph {et~al.}(2025)\citenamefont
  {Cocuzza}, \citenamefont {Hunt-Smith}, \citenamefont {Melnitchouk},
  \citenamefont {Sato},\ and\ \citenamefont {Thomas}}]{Cocuzza:2025qvf}%
  \BibitemOpen
  \bibfield  {author} {\bibinfo {author} {\bibfnamefont {C.}~\bibnamefont
  {Cocuzza}}, \bibinfo {author} {\bibfnamefont {N.~T.}\ \bibnamefont
  {Hunt-Smith}}, \bibinfo {author} {\bibfnamefont {W.}~\bibnamefont
  {Melnitchouk}}, \bibinfo {author} {\bibfnamefont {N.}~\bibnamefont {Sato}}, \
  and\ \bibinfo {author} {\bibfnamefont {A.~W.}\ \bibnamefont {Thomas}}
  (\bibinfo {collaboration} {JAM Collaboration (Spin PDF Analysis Group)}),\
  }\href {\doibase 10.1103/6fn9-1wqb} {\bibfield  {journal} {\bibinfo
  {journal} {Phys. Rev. D}\ }\textbf {\bibinfo {volume} {112}},\ \bibinfo
  {pages} {114017} (\bibinfo {year} {2025})}\BibitemShut {NoStop}%
\bibitem [{\citenamefont {Bhattacharya}\ \emph
  {et~al.}(2020{\natexlab{a}})\citenamefont {Bhattacharya}, \citenamefont
  {Cichy}, \citenamefont {Constantinou}, \citenamefont {Metz}, \citenamefont
  {Scapellato},\ and\ \citenamefont {Steffens}}]{Bhattacharya:2020cen}%
  \BibitemOpen
  \bibfield  {author} {\bibinfo {author} {\bibfnamefont {S.}~\bibnamefont
  {Bhattacharya}}, \bibinfo {author} {\bibfnamefont {K.}~\bibnamefont {Cichy}},
  \bibinfo {author} {\bibfnamefont {M.}~\bibnamefont {Constantinou}}, \bibinfo
  {author} {\bibfnamefont {A.}~\bibnamefont {Metz}}, \bibinfo {author}
  {\bibfnamefont {A.}~\bibnamefont {Scapellato}}, \ and\ \bibinfo {author}
  {\bibfnamefont {F.}~\bibnamefont {Steffens}},\ }\href {\doibase
  10.1103/PhysRevD.102.111501} {\bibfield  {journal} {\bibinfo  {journal}
  {Phys. Rev. D}\ }\textbf {\bibinfo {volume} {102}},\ \bibinfo {pages}
  {111501} (\bibinfo {year} {2020}{\natexlab{a}})}\BibitemShut {NoStop}%
\bibitem [{\citenamefont {Bhattacharya}\ \emph
  {et~al.}(2020{\natexlab{b}})\citenamefont {Bhattacharya}, \citenamefont
  {Cichy}, \citenamefont {Constantinou}, \citenamefont {Metz}, \citenamefont
  {Scapellato},\ and\ \citenamefont {Steffens}}]{Bhattacharya:2020xlt}%
  \BibitemOpen
  \bibfield  {author} {\bibinfo {author} {\bibfnamefont {S.}~\bibnamefont
  {Bhattacharya}}, \bibinfo {author} {\bibfnamefont {K.}~\bibnamefont {Cichy}},
  \bibinfo {author} {\bibfnamefont {M.}~\bibnamefont {Constantinou}}, \bibinfo
  {author} {\bibfnamefont {A.}~\bibnamefont {Metz}}, \bibinfo {author}
  {\bibfnamefont {A.}~\bibnamefont {Scapellato}}, \ and\ \bibinfo {author}
  {\bibfnamefont {F.}~\bibnamefont {Steffens}},\ }\href {\doibase
  10.1103/PhysRevD.102.034005} {\bibfield  {journal} {\bibinfo  {journal}
  {Phys. Rev. D}\ }\textbf {\bibinfo {volume} {102}},\ \bibinfo {pages}
  {034005} (\bibinfo {year} {2020}{\natexlab{b}})},\ \bibinfo {note} {[Erratum:
  Phys.Rev.D 108, 039901 (2023)]}\BibitemShut {NoStop}%
\bibitem [{\citenamefont {Bhattacharya}\ \emph {et~al.}(2021)\citenamefont
  {Bhattacharya}, \citenamefont {Cichy}, \citenamefont {Constantinou},
  \citenamefont {Metz}, \citenamefont {Scapellato},\ and\ \citenamefont
  {Steffens}}]{Bhattacharya:2021moj}%
  \BibitemOpen
  \bibfield  {author} {\bibinfo {author} {\bibfnamefont {S.}~\bibnamefont
  {Bhattacharya}}, \bibinfo {author} {\bibfnamefont {K.}~\bibnamefont {Cichy}},
  \bibinfo {author} {\bibfnamefont {M.}~\bibnamefont {Constantinou}}, \bibinfo
  {author} {\bibfnamefont {A.}~\bibnamefont {Metz}}, \bibinfo {author}
  {\bibfnamefont {A.}~\bibnamefont {Scapellato}}, \ and\ \bibinfo {author}
  {\bibfnamefont {F.}~\bibnamefont {Steffens}},\ }\href {\doibase
  10.1103/PhysRevD.104.114510} {\bibfield  {journal} {\bibinfo  {journal}
  {Phys. Rev. D}\ }\textbf {\bibinfo {volume} {104}},\ \bibinfo {pages}
  {114510} (\bibinfo {year} {2021})}\BibitemShut {NoStop}%
\bibitem [{\citenamefont {Wandzura}\ and\ \citenamefont
  {Wilczek}(1977)}]{Wandzura:1977qf}%
  \BibitemOpen
  \bibfield  {author} {\bibinfo {author} {\bibfnamefont {S.}~\bibnamefont
  {Wandzura}}\ and\ \bibinfo {author} {\bibfnamefont {F.}~\bibnamefont
  {Wilczek}},\ }\href {\doibase 10.1016/0370-2693(77)90700-6} {\bibfield
  {journal} {\bibinfo  {journal} {Phys. Lett. B}\ }\textbf {\bibinfo {volume}
  {72}},\ \bibinfo {pages} {195} (\bibinfo {year} {1977})}\BibitemShut
  {NoStop}%
\bibitem [{\citenamefont {Braun}\ \emph {et~al.}(2021)\citenamefont {Braun},
  \citenamefont {Ji},\ and\ \citenamefont {Vladimirov}}]{Braun:2021aon}%
  \BibitemOpen
  \bibfield  {author} {\bibinfo {author} {\bibfnamefont {V.~M.}\ \bibnamefont
  {Braun}}, \bibinfo {author} {\bibfnamefont {Y.}~\bibnamefont {Ji}}, \ and\
  \bibinfo {author} {\bibfnamefont {A.}~\bibnamefont {Vladimirov}},\ }\href
  {\doibase 10.1007/JHEP05(2021)086} {\bibfield  {journal} {\bibinfo  {journal}
  {JHEP}\ }\textbf {\bibinfo {volume} {05}},\ \bibinfo {pages} {086} (\bibinfo
  {year} {2021})}\BibitemShut {NoStop}%
\bibitem [{\citenamefont {Gockeler}\ \emph {et~al.}(2005)\citenamefont
  {Gockeler}, \citenamefont {Horsley}, \citenamefont {Pleiter}, \citenamefont
  {Rakow}, \citenamefont {Schafer}, \citenamefont {Schierholz}, \citenamefont
  {Stuben},\ and\ \citenamefont {Zanotti}}]{Gockeler:2005vw}%
  \BibitemOpen
  \bibfield  {author} {\bibinfo {author} {\bibfnamefont {M.}~\bibnamefont
  {Gockeler}}, \bibinfo {author} {\bibfnamefont {R.}~\bibnamefont {Horsley}},
  \bibinfo {author} {\bibfnamefont {D.}~\bibnamefont {Pleiter}}, \bibinfo
  {author} {\bibfnamefont {P.~E.~L.}\ \bibnamefont {Rakow}}, \bibinfo {author}
  {\bibfnamefont {A.}~\bibnamefont {Schafer}}, \bibinfo {author} {\bibfnamefont
  {G.}~\bibnamefont {Schierholz}}, \bibinfo {author} {\bibfnamefont
  {H.}~\bibnamefont {Stuben}}, \ and\ \bibinfo {author} {\bibfnamefont {J.~M.}\
  \bibnamefont {Zanotti}},\ }\href {\doibase 10.1103/PhysRevD.72.054507}
  {\bibfield  {journal} {\bibinfo  {journal} {Phys. Rev. D}\ }\textbf {\bibinfo
  {volume} {72}},\ \bibinfo {pages} {054507} (\bibinfo {year}
  {2005})}\BibitemShut {NoStop}%
\bibitem [{\citenamefont {B{\"u}rger}\ \emph {et~al.}(2022)\citenamefont
  {B{\"u}rger}, \citenamefont {Wurm}, \citenamefont {L{\"o}ffler},
  \citenamefont {G{\"o}ckeler}, \citenamefont {Bali}, \citenamefont {Collins},
  \citenamefont {Sch{\"a}fer},\ and\ \citenamefont
  {Sternbeck}}]{Burger:2021knd}%
  \BibitemOpen
  \bibfield  {author} {\bibinfo {author} {\bibfnamefont {S.}~\bibnamefont
  {B{\"u}rger}}, \bibinfo {author} {\bibfnamefont {T.}~\bibnamefont {Wurm}},
  \bibinfo {author} {\bibfnamefont {M.}~\bibnamefont {L{\"o}ffler}}, \bibinfo
  {author} {\bibfnamefont {M.}~\bibnamefont {G{\"o}ckeler}}, \bibinfo {author}
  {\bibfnamefont {G.}~\bibnamefont {Bali}}, \bibinfo {author} {\bibfnamefont
  {S.}~\bibnamefont {Collins}}, \bibinfo {author} {\bibfnamefont
  {A.}~\bibnamefont {Sch{\"a}fer}}, \ and\ \bibinfo {author} {\bibfnamefont
  {A.}~\bibnamefont {Sternbeck}} (\bibinfo {collaboration} {RQCD}),\ }\href
  {\doibase 10.1103/PhysRevD.105.054504} {\bibfield  {journal} {\bibinfo
  {journal} {Phys. Rev. D}\ }\textbf {\bibinfo {volume} {105}},\ \bibinfo
  {pages} {054504} (\bibinfo {year} {2022})}\BibitemShut {NoStop}%
\bibitem [{\citenamefont {Crawford}\ \emph {et~al.}(2025)\citenamefont
  {Crawford}, \citenamefont {Can}, \citenamefont {Horsley}, \citenamefont
  {Rakow}, \citenamefont {Schierholz}, \citenamefont {St{\"u}ben},
  \citenamefont {Young},\ and\ \citenamefont {Zanotti}}]{Crawford:2024wzx}%
  \BibitemOpen
  \bibfield  {author} {\bibinfo {author} {\bibfnamefont {J.~A.}\ \bibnamefont
  {Crawford}}, \bibinfo {author} {\bibfnamefont {K.~U.}\ \bibnamefont {Can}},
  \bibinfo {author} {\bibfnamefont {R.}~\bibnamefont {Horsley}}, \bibinfo
  {author} {\bibfnamefont {P.~E.~L.}\ \bibnamefont {Rakow}}, \bibinfo {author}
  {\bibfnamefont {G.}~\bibnamefont {Schierholz}}, \bibinfo {author}
  {\bibfnamefont {H.}~\bibnamefont {St{\"u}ben}}, \bibinfo {author}
  {\bibfnamefont {R.~D.}\ \bibnamefont {Young}}, \ and\ \bibinfo {author}
  {\bibfnamefont {J.~M.}\ \bibnamefont {Zanotti}} (\bibinfo {collaboration}
  {QCDSF}),\ }\href {\doibase 10.1103/PhysRevLett.134.071901} {\bibfield
  {journal} {\bibinfo  {journal} {Phys. Rev. Lett.}\ }\textbf {\bibinfo
  {volume} {134}},\ \bibinfo {pages} {071901} (\bibinfo {year}
  {2025})}\BibitemShut {NoStop}%
\bibitem [{\citenamefont {Follana}\ \emph {et~al.}(2007)\citenamefont
  {Follana}, \citenamefont {Mason}, \citenamefont {Davies}, \citenamefont
  {Hornbostel}, \citenamefont {Lepage}, \citenamefont {Shigemitsu},
  \citenamefont {Trottier},\ and\ \citenamefont {Wong}}]{Follana:2006rc}%
  \BibitemOpen
  \bibfield  {author} {\bibinfo {author} {\bibfnamefont {E.}~\bibnamefont
  {Follana}}, \bibinfo {author} {\bibfnamefont {Q.}~\bibnamefont {Mason}},
  \bibinfo {author} {\bibfnamefont {C.}~\bibnamefont {Davies}}, \bibinfo
  {author} {\bibfnamefont {K.}~\bibnamefont {Hornbostel}}, \bibinfo {author}
  {\bibfnamefont {G.~P.}\ \bibnamefont {Lepage}}, \bibinfo {author}
  {\bibfnamefont {J.}~\bibnamefont {Shigemitsu}}, \bibinfo {author}
  {\bibfnamefont {H.}~\bibnamefont {Trottier}}, \ and\ \bibinfo {author}
  {\bibfnamefont {K.}~\bibnamefont {Wong}} (\bibinfo {collaboration} {HPQCD,
  UKQCD}),\ }\href {\doibase 10.1103/PhysRevD.75.054502} {\bibfield  {journal}
  {\bibinfo  {journal} {Phys. Rev. D}\ }\textbf {\bibinfo {volume} {75}},\
  \bibinfo {pages} {054502} (\bibinfo {year} {2007})}\BibitemShut {NoStop}%
\bibitem [{\citenamefont {Bazavov}\ \emph {et~al.}(2019)\citenamefont {Bazavov}
  \emph {et~al.}}]{Bazavov:2019www}%
  \BibitemOpen
  \bibfield  {author} {\bibinfo {author} {\bibfnamefont {A.}~\bibnamefont
  {Bazavov}} \emph {et~al.},\ }\href {\doibase 10.1103/PhysRevD.100.094510}
  {\bibfield  {journal} {\bibinfo  {journal} {Phys. Rev. D}\ }\textbf {\bibinfo
  {volume} {100}},\ \bibinfo {pages} {094510} (\bibinfo {year}
  {2019})}\BibitemShut {NoStop}%
\bibitem [{\citenamefont {Bazavov}\ \emph {et~al.}(2014)\citenamefont {Bazavov}
  \emph {et~al.}}]{HotQCD:2014kol}%
  \BibitemOpen
  \bibfield  {author} {\bibinfo {author} {\bibfnamefont {A.}~\bibnamefont
  {Bazavov}} \emph {et~al.} (\bibinfo {collaboration} {HotQCD}),\ }\href
  {\doibase 10.1103/PhysRevD.90.094503} {\bibfield  {journal} {\bibinfo
  {journal} {Phys. Rev. D}\ }\textbf {\bibinfo {volume} {90}},\ \bibinfo
  {pages} {094503} (\bibinfo {year} {2014})}\BibitemShut {NoStop}%
\bibitem [{\citenamefont {Sheikholeslami}\ and\ \citenamefont
  {Wohlert}(1985)}]{Sheikholeslami:1985ij}%
  \BibitemOpen
  \bibfield  {author} {\bibinfo {author} {\bibfnamefont {B.}~\bibnamefont
  {Sheikholeslami}}\ and\ \bibinfo {author} {\bibfnamefont {R.}~\bibnamefont
  {Wohlert}},\ }\href {\doibase 10.1016/0550-3213(85)90002-1} {\bibfield
  {journal} {\bibinfo  {journal} {Nucl. Phys. B}\ }\textbf {\bibinfo {volume}
  {259}},\ \bibinfo {pages} {572} (\bibinfo {year} {1985})}\BibitemShut
  {NoStop}%
\bibitem [{\citenamefont {Hasenfratz}\ and\ \citenamefont
  {Knechtli}(2001)}]{Hasenfratz:2001hp}%
  \BibitemOpen
  \bibfield  {author} {\bibinfo {author} {\bibfnamefont {A.}~\bibnamefont
  {Hasenfratz}}\ and\ \bibinfo {author} {\bibfnamefont {F.}~\bibnamefont
  {Knechtli}},\ }\href {\doibase 10.1103/PhysRevD.64.034504} {\bibfield
  {journal} {\bibinfo  {journal} {Phys. Rev. D}\ }\textbf {\bibinfo {volume}
  {64}},\ \bibinfo {pages} {034504} (\bibinfo {year} {2001})}\BibitemShut
  {NoStop}%
\bibitem [{\citenamefont {Clark}\ \emph {et~al.}(2010)\citenamefont {Clark},
  \citenamefont {Babich}, \citenamefont {Barros}, \citenamefont {Brower},\ and\
  \citenamefont {Rebbi}}]{Clark:2009wm}%
  \BibitemOpen
  \bibfield  {author} {\bibinfo {author} {\bibfnamefont {M.~A.}\ \bibnamefont
  {Clark}}, \bibinfo {author} {\bibfnamefont {R.}~\bibnamefont {Babich}},
  \bibinfo {author} {\bibfnamefont {K.}~\bibnamefont {Barros}}, \bibinfo
  {author} {\bibfnamefont {R.~C.}\ \bibnamefont {Brower}}, \ and\ \bibinfo
  {author} {\bibfnamefont {C.}~\bibnamefont {Rebbi}} (\bibinfo {collaboration}
  {QUDA}),\ }\href {\doibase 10.1016/j.cpc.2010.05.002} {\bibfield  {journal}
  {\bibinfo  {journal} {Comput. Phys. Commun.}\ }\textbf {\bibinfo {volume}
  {181}},\ \bibinfo {pages} {1517} (\bibinfo {year} {2010})}\BibitemShut
  {NoStop}%
\bibitem [{\citenamefont {Babich}\ \emph {et~al.}(2011)\citenamefont {Babich},
  \citenamefont {Clark}, \citenamefont {Joo}, \citenamefont {Shi},
  \citenamefont {Brower},\ and\ \citenamefont {Gottlieb}}]{Babich:2011np}%
  \BibitemOpen
  \bibfield  {author} {\bibinfo {author} {\bibfnamefont {R.}~\bibnamefont
  {Babich}}, \bibinfo {author} {\bibfnamefont {M.~A.}\ \bibnamefont {Clark}},
  \bibinfo {author} {\bibfnamefont {B.}~\bibnamefont {Joo}}, \bibinfo {author}
  {\bibfnamefont {G.}~\bibnamefont {Shi}}, \bibinfo {author} {\bibfnamefont
  {R.~C.}\ \bibnamefont {Brower}}, \ and\ \bibinfo {author} {\bibfnamefont
  {S.}~\bibnamefont {Gottlieb}} (\bibinfo {collaboration} {QUDA}),\ }in\ \href
  {\doibase 10.1145/2063384.2063478} {\emph {\bibinfo {booktitle}
  {{International Conference for High Performance Computing, Networking,
  Storage and Analysis}}}}\ (\bibinfo {year} {2011})\BibitemShut {NoStop}%
\bibitem [{\citenamefont {Shintani}\ \emph {et~al.}(2015)\citenamefont
  {Shintani}, \citenamefont {Arthur}, \citenamefont {Blum}, \citenamefont
  {Izubuchi}, \citenamefont {Jung},\ and\ \citenamefont
  {Lehner}}]{Shintani:2014vja}%
  \BibitemOpen
  \bibfield  {author} {\bibinfo {author} {\bibfnamefont {E.}~\bibnamefont
  {Shintani}}, \bibinfo {author} {\bibfnamefont {R.}~\bibnamefont {Arthur}},
  \bibinfo {author} {\bibfnamefont {T.}~\bibnamefont {Blum}}, \bibinfo {author}
  {\bibfnamefont {T.}~\bibnamefont {Izubuchi}}, \bibinfo {author}
  {\bibfnamefont {C.}~\bibnamefont {Jung}}, \ and\ \bibinfo {author}
  {\bibfnamefont {C.}~\bibnamefont {Lehner}},\ }\href {\doibase
  10.1103/PhysRevD.91.114511} {\bibfield  {journal} {\bibinfo  {journal} {Phys.
  Rev. D}\ }\textbf {\bibinfo {volume} {91}},\ \bibinfo {pages} {114511}
  (\bibinfo {year} {2015})}\BibitemShut {NoStop}%
\bibitem [{\citenamefont {Bali}\ \emph {et~al.}(2016)\citenamefont {Bali},
  \citenamefont {Lang}, \citenamefont {Musch},\ and\ \citenamefont
  {Sch\"afer}}]{Bali:2016lva}%
  \BibitemOpen
  \bibfield  {author} {\bibinfo {author} {\bibfnamefont {G.~S.}\ \bibnamefont
  {Bali}}, \bibinfo {author} {\bibfnamefont {B.}~\bibnamefont {Lang}}, \bibinfo
  {author} {\bibfnamefont {B.~U.}\ \bibnamefont {Musch}}, \ and\ \bibinfo
  {author} {\bibfnamefont {A.}~\bibnamefont {Sch\"afer}},\ }\href {\doibase
  10.1103/PhysRevD.93.094515} {\bibfield  {journal} {\bibinfo  {journal} {Phys.
  Rev. D}\ }\textbf {\bibinfo {volume} {93}},\ \bibinfo {pages} {094515}
  (\bibinfo {year} {2016})}\BibitemShut {NoStop}%
\bibitem [{\citenamefont {Izubuchi}\ \emph {et~al.}(2019)\citenamefont
  {Izubuchi}, \citenamefont {Jin}, \citenamefont {Kallidonis}, \citenamefont
  {Karthik}, \citenamefont {Mukherjee}, \citenamefont {Petreczky},
  \citenamefont {Shugert},\ and\ \citenamefont {Syritsyn}}]{Izubuchi:2019lyk}%
  \BibitemOpen
  \bibfield  {author} {\bibinfo {author} {\bibfnamefont {T.}~\bibnamefont
  {Izubuchi}}, \bibinfo {author} {\bibfnamefont {L.}~\bibnamefont {Jin}},
  \bibinfo {author} {\bibfnamefont {C.}~\bibnamefont {Kallidonis}}, \bibinfo
  {author} {\bibfnamefont {N.}~\bibnamefont {Karthik}}, \bibinfo {author}
  {\bibfnamefont {S.}~\bibnamefont {Mukherjee}}, \bibinfo {author}
  {\bibfnamefont {P.}~\bibnamefont {Petreczky}}, \bibinfo {author}
  {\bibfnamefont {C.}~\bibnamefont {Shugert}}, \ and\ \bibinfo {author}
  {\bibfnamefont {S.}~\bibnamefont {Syritsyn}},\ }\href {\doibase
  10.1103/PhysRevD.100.034516} {\bibfield  {journal} {\bibinfo  {journal}
  {Phys. Rev. D}\ }\textbf {\bibinfo {volume} {100}},\ \bibinfo {pages}
  {034516} (\bibinfo {year} {2019})}\BibitemShut {NoStop}%
\bibitem [{\citenamefont {Buckley}\ \emph {et~al.}(2015)\citenamefont
  {Buckley}, \citenamefont {Ferrando}, \citenamefont {Lloyd}, \citenamefont
  {Nordstr\"om}, \citenamefont {Page}, \citenamefont {R\"ufenacht},
  \citenamefont {Sch\"onherr},\ and\ \citenamefont {Watt}}]{Buckley:2014ana}%
  \BibitemOpen
  \bibfield  {author} {\bibinfo {author} {\bibfnamefont {A.}~\bibnamefont
  {Buckley}}, \bibinfo {author} {\bibfnamefont {J.}~\bibnamefont {Ferrando}},
  \bibinfo {author} {\bibfnamefont {S.}~\bibnamefont {Lloyd}}, \bibinfo
  {author} {\bibfnamefont {K.}~\bibnamefont {Nordstr\"om}}, \bibinfo {author}
  {\bibfnamefont {B.}~\bibnamefont {Page}}, \bibinfo {author} {\bibfnamefont
  {M.}~\bibnamefont {R\"ufenacht}}, \bibinfo {author} {\bibfnamefont
  {M.}~\bibnamefont {Sch\"onherr}}, \ and\ \bibinfo {author} {\bibfnamefont
  {G.}~\bibnamefont {Watt}},\ }\href {\doibase 10.1140/epjc/s10052-015-3318-8}
  {\bibfield  {journal} {\bibinfo  {journal} {Eur. Phys. J. C}\ }\textbf
  {\bibinfo {volume} {75}},\ \bibinfo {pages} {132} (\bibinfo {year}
  {2015})}\BibitemShut {NoStop}%
\bibitem [{\citenamefont {Moch}\ \emph {et~al.}(2014)\citenamefont {Moch},
  \citenamefont {Vermaseren},\ and\ \citenamefont {Vogt}}]{Moch:2014sna}%
  \BibitemOpen
  \bibfield  {author} {\bibinfo {author} {\bibfnamefont {S.}~\bibnamefont
  {Moch}}, \bibinfo {author} {\bibfnamefont {J.~A.~M.}\ \bibnamefont
  {Vermaseren}}, \ and\ \bibinfo {author} {\bibfnamefont {A.}~\bibnamefont
  {Vogt}},\ }\href {\doibase 10.1016/j.nuclphysb.2014.10.016} {\bibfield
  {journal} {\bibinfo  {journal} {Nucl. Phys. B}\ }\textbf {\bibinfo {volume}
  {889}},\ \bibinfo {pages} {351} (\bibinfo {year} {2014})}\BibitemShut
  {NoStop}%
\bibitem [{\citenamefont {Kniehl}\ \emph {et~al.}(2025)\citenamefont {Kniehl},
  \citenamefont {Moch}, \citenamefont {Velizhanin},\ and\ \citenamefont
  {Vogt}}]{Kniehl:2025ttz}%
  \BibitemOpen
  \bibfield  {author} {\bibinfo {author} {\bibfnamefont {B.~A.}\ \bibnamefont
  {Kniehl}}, \bibinfo {author} {\bibfnamefont {S.}~\bibnamefont {Moch}},
  \bibinfo {author} {\bibfnamefont {V.~N.}\ \bibnamefont {Velizhanin}}, \ and\
  \bibinfo {author} {\bibfnamefont {A.}~\bibnamefont {Vogt}},\ }\href {\doibase
  10.1103/hkg5-88hr} {\bibfield  {journal} {\bibinfo  {journal} {Phys. Rev.
  Lett.}\ }\textbf {\bibinfo {volume} {135}},\ \bibinfo {pages} {071902}
  (\bibinfo {year} {2025})}\BibitemShut {NoStop}%
\bibitem [{\citenamefont {Herzog}\ \emph {et~al.}(2017)\citenamefont {Herzog},
  \citenamefont {Ruijl}, \citenamefont {Ueda}, \citenamefont {Vermaseren},\
  and\ \citenamefont {Vogt}}]{Herzog:2017ohr}%
  \BibitemOpen
  \bibfield  {author} {\bibinfo {author} {\bibfnamefont {F.}~\bibnamefont
  {Herzog}}, \bibinfo {author} {\bibfnamefont {B.}~\bibnamefont {Ruijl}},
  \bibinfo {author} {\bibfnamefont {T.}~\bibnamefont {Ueda}}, \bibinfo {author}
  {\bibfnamefont {J.~A.~M.}\ \bibnamefont {Vermaseren}}, \ and\ \bibinfo
  {author} {\bibfnamefont {A.}~\bibnamefont {Vogt}},\ }\href {\doibase
  10.1007/JHEP02(2017)090} {\bibfield  {journal} {\bibinfo  {journal} {JHEP}\
  }\textbf {\bibinfo {volume} {02}},\ \bibinfo {pages} {090} (\bibinfo {year}
  {2017})}\BibitemShut {NoStop}%
\bibitem [{\citenamefont {Ji}(2024)}]{Ji:2024oka}%
  \BibitemOpen
  \bibfield  {author} {\bibinfo {author} {\bibfnamefont {X.}~\bibnamefont
  {Ji}},\ }\href {\doibase 10.1016/j.nuclphysb.2024.116670} {\bibfield
  {journal} {\bibinfo  {journal} {Nucl. Phys. B}\ }\textbf {\bibinfo {volume}
  {1007}},\ \bibinfo {pages} {116670} (\bibinfo {year} {2024})}\BibitemShut
  {NoStop}%
\bibitem [{\citenamefont {Chen}\ \emph {et~al.}(2017)\citenamefont {Chen},
  \citenamefont {Ji},\ and\ \citenamefont {Zhang}}]{Chen:2016fxx}%
  \BibitemOpen
  \bibfield  {author} {\bibinfo {author} {\bibfnamefont {J.-W.}\ \bibnamefont
  {Chen}}, \bibinfo {author} {\bibfnamefont {X.}~\bibnamefont {Ji}}, \ and\
  \bibinfo {author} {\bibfnamefont {J.-H.}\ \bibnamefont {Zhang}},\ }\href
  {\doibase 10.1016/j.nuclphysb.2016.12.004} {\bibfield  {journal} {\bibinfo
  {journal} {Nucl. Phys. B}\ }\textbf {\bibinfo {volume} {915}},\ \bibinfo
  {pages} {1} (\bibinfo {year} {2017})}\BibitemShut {NoStop}%
\bibitem [{\citenamefont {Zhang}\ \emph {et~al.}(2019)\citenamefont {Zhang},
  \citenamefont {Ji}, \citenamefont {Sch{\"a}fer}, \citenamefont {Wang},\ and\
  \citenamefont {Zhao}}]{Zhang:2018diq}%
  \BibitemOpen
  \bibfield  {author} {\bibinfo {author} {\bibfnamefont {J.-H.}\ \bibnamefont
  {Zhang}}, \bibinfo {author} {\bibfnamefont {X.}~\bibnamefont {Ji}}, \bibinfo
  {author} {\bibfnamefont {A.}~\bibnamefont {Sch{\"a}fer}}, \bibinfo {author}
  {\bibfnamefont {W.}~\bibnamefont {Wang}}, \ and\ \bibinfo {author}
  {\bibfnamefont {S.}~\bibnamefont {Zhao}},\ }\href {\doibase
  10.1103/PhysRevLett.122.142001} {\bibfield  {journal} {\bibinfo  {journal}
  {Phys. Rev. Lett.}\ }\textbf {\bibinfo {volume} {122}},\ \bibinfo {pages}
  {142001} (\bibinfo {year} {2019})}\BibitemShut {NoStop}%
\bibitem [{\citenamefont {Li}\ \emph {et~al.}(2019)\citenamefont {Li},
  \citenamefont {Ma},\ and\ \citenamefont {Qiu}}]{Li:2018tpe}%
  \BibitemOpen
  \bibfield  {author} {\bibinfo {author} {\bibfnamefont {Z.-Y.}\ \bibnamefont
  {Li}}, \bibinfo {author} {\bibfnamefont {Y.-Q.}\ \bibnamefont {Ma}}, \ and\
  \bibinfo {author} {\bibfnamefont {J.-W.}\ \bibnamefont {Qiu}},\ }\href
  {\doibase 10.1103/PhysRevLett.122.062002} {\bibfield  {journal} {\bibinfo
  {journal} {Phys. Rev. Lett.}\ }\textbf {\bibinfo {volume} {122}},\ \bibinfo
  {pages} {062002} (\bibinfo {year} {2019})}\BibitemShut {NoStop}%
\bibitem [{\citenamefont {Chou}\ and\ \citenamefont
  {Chen}(2022)}]{Chou:2022drv}%
  \BibitemOpen
  \bibfield  {author} {\bibinfo {author} {\bibfnamefont {C.-Y.}\ \bibnamefont
  {Chou}}\ and\ \bibinfo {author} {\bibfnamefont {J.-W.}\ \bibnamefont
  {Chen}},\ }\href {\doibase 10.1103/PhysRevD.106.014507} {\bibfield  {journal}
  {\bibinfo  {journal} {Phys. Rev. D}\ }\textbf {\bibinfo {volume} {106}},\
  \bibinfo {pages} {014507} (\bibinfo {year} {2022})}\BibitemShut {NoStop}%
\bibitem [{\citenamefont {Bazavov}\ \emph {et~al.}(2016)\citenamefont
  {Bazavov}, \citenamefont {Brambilla}, \citenamefont {Ding}, \citenamefont
  {Petreczky}, \citenamefont {Schadler}, \citenamefont {Vairo},\ and\
  \citenamefont {Weber}}]{Bazavov:2016uvm}%
  \BibitemOpen
  \bibfield  {author} {\bibinfo {author} {\bibfnamefont {A.}~\bibnamefont
  {Bazavov}}, \bibinfo {author} {\bibfnamefont {N.}~\bibnamefont {Brambilla}},
  \bibinfo {author} {\bibfnamefont {H.~T.}\ \bibnamefont {Ding}}, \bibinfo
  {author} {\bibfnamefont {P.}~\bibnamefont {Petreczky}}, \bibinfo {author}
  {\bibfnamefont {H.~P.}\ \bibnamefont {Schadler}}, \bibinfo {author}
  {\bibfnamefont {A.}~\bibnamefont {Vairo}}, \ and\ \bibinfo {author}
  {\bibfnamefont {J.~H.}\ \bibnamefont {Weber}},\ }\href {\doibase
  10.1103/PhysRevD.93.114502} {\bibfield  {journal} {\bibinfo  {journal} {Phys.
  Rev. D}\ }\textbf {\bibinfo {volume} {93}},\ \bibinfo {pages} {114502}
  (\bibinfo {year} {2016})}\BibitemShut {NoStop}%
\bibitem [{\citenamefont {Bazavov}\ \emph
  {et~al.}(2018{\natexlab{a}})\citenamefont {Bazavov}, \citenamefont
  {Petreczky},\ and\ \citenamefont {Weber}}]{Bazavov:2017dsy}%
  \BibitemOpen
  \bibfield  {author} {\bibinfo {author} {\bibfnamefont {A.}~\bibnamefont
  {Bazavov}}, \bibinfo {author} {\bibfnamefont {P.}~\bibnamefont {Petreczky}},
  \ and\ \bibinfo {author} {\bibfnamefont {J.~H.}\ \bibnamefont {Weber}},\
  }\href {\doibase 10.1103/PhysRevD.97.014510} {\bibfield  {journal} {\bibinfo
  {journal} {Phys. Rev. D}\ }\textbf {\bibinfo {volume} {97}},\ \bibinfo
  {pages} {014510} (\bibinfo {year} {2018}{\natexlab{a}})}\BibitemShut
  {NoStop}%
\bibitem [{\citenamefont {Bazavov}\ \emph
  {et~al.}(2018{\natexlab{b}})\citenamefont {Bazavov}, \citenamefont
  {Brambilla}, \citenamefont {Petreczky}, \citenamefont {Vairo},\ and\
  \citenamefont {Weber}}]{Bazavov:2018wmo}%
  \BibitemOpen
  \bibfield  {author} {\bibinfo {author} {\bibfnamefont {A.}~\bibnamefont
  {Bazavov}}, \bibinfo {author} {\bibfnamefont {N.}~\bibnamefont {Brambilla}},
  \bibinfo {author} {\bibfnamefont {P.}~\bibnamefont {Petreczky}}, \bibinfo
  {author} {\bibfnamefont {A.}~\bibnamefont {Vairo}}, \ and\ \bibinfo {author}
  {\bibfnamefont {J.~H.}\ \bibnamefont {Weber}} (\bibinfo {collaboration}
  {TUMQCD}),\ }\href {\doibase 10.1103/PhysRevD.98.054511} {\bibfield
  {journal} {\bibinfo  {journal} {Phys. Rev. D}\ }\textbf {\bibinfo {volume}
  {98}},\ \bibinfo {pages} {054511} (\bibinfo {year}
  {2018}{\natexlab{b}})}\BibitemShut {NoStop}%
\bibitem [{\citenamefont {Pineda}(2001)}]{Pineda:2001zq}%
  \BibitemOpen
  \bibfield  {author} {\bibinfo {author} {\bibfnamefont {A.}~\bibnamefont
  {Pineda}},\ }\href {\doibase 10.1088/1126-6708/2001/06/022} {\bibfield
  {journal} {\bibinfo  {journal} {JHEP}\ }\textbf {\bibinfo {volume} {06}},\
  \bibinfo {pages} {022} (\bibinfo {year} {2001})}\BibitemShut {NoStop}%
\bibitem [{\citenamefont {Ding}\ \emph {et~al.}(2025)\citenamefont {Ding},
  \citenamefont {Gao}, \citenamefont {Mukherjee}, \citenamefont {Petreczky},
  \citenamefont {Shi}, \citenamefont {Syritsyn},\ and\ \citenamefont
  {Zhao}}]{Ding:2024saz}%
  \BibitemOpen
  \bibfield  {author} {\bibinfo {author} {\bibfnamefont {H.-T.}\ \bibnamefont
  {Ding}}, \bibinfo {author} {\bibfnamefont {X.}~\bibnamefont {Gao}}, \bibinfo
  {author} {\bibfnamefont {S.}~\bibnamefont {Mukherjee}}, \bibinfo {author}
  {\bibfnamefont {P.}~\bibnamefont {Petreczky}}, \bibinfo {author}
  {\bibfnamefont {Q.}~\bibnamefont {Shi}}, \bibinfo {author} {\bibfnamefont
  {S.}~\bibnamefont {Syritsyn}}, \ and\ \bibinfo {author} {\bibfnamefont
  {Y.}~\bibnamefont {Zhao}},\ }\href {\doibase 10.1007/JHEP02(2025)056}
  {\bibfield  {journal} {\bibinfo  {journal} {JHEP}\ }\textbf {\bibinfo
  {volume} {02}},\ \bibinfo {pages} {056} (\bibinfo {year} {2025})}\BibitemShut
  {NoStop}%
\bibitem [{\citenamefont {Dutrieux}\ \emph {et~al.}(2026)\citenamefont
  {Dutrieux}, \citenamefont {Karpie}, \citenamefont {Monahan}, \citenamefont
  {Orginos}, \citenamefont {Radyushkin}, \citenamefont {Richards},\ and\
  \citenamefont {Zafeiropoulos}}]{Dutrieux:2025jed}%
  \BibitemOpen
  \bibfield  {author} {\bibinfo {author} {\bibfnamefont {H.}~\bibnamefont
  {Dutrieux}}, \bibinfo {author} {\bibfnamefont {J.}~\bibnamefont {Karpie}},
  \bibinfo {author} {\bibfnamefont {C.~J.}\ \bibnamefont {Monahan}}, \bibinfo
  {author} {\bibfnamefont {K.}~\bibnamefont {Orginos}}, \bibinfo {author}
  {\bibfnamefont {A.}~\bibnamefont {Radyushkin}}, \bibinfo {author}
  {\bibfnamefont {D.}~\bibnamefont {Richards}}, \ and\ \bibinfo {author}
  {\bibfnamefont {S.}~\bibnamefont {Zafeiropoulos}},\ }\href {\doibase
  10.1103/4f6n-vy13} {\bibfield  {journal} {\bibinfo  {journal} {Phys. Rev. D}\
  }\textbf {\bibinfo {volume} {113}},\ \bibinfo {pages} {074524} (\bibinfo
  {year} {2026})}\BibitemShut {NoStop}%
\bibitem [{\citenamefont {Gao}\ \emph {et~al.}(2022{\natexlab{a}})\citenamefont
  {Gao}, \citenamefont {Hanlon}, \citenamefont {Mukherjee}, \citenamefont
  {Petreczky}, \citenamefont {Scior}, \citenamefont {Syritsyn},\ and\
  \citenamefont {Zhao}}]{Gao:2021dbh}%
  \BibitemOpen
  \bibfield  {author} {\bibinfo {author} {\bibfnamefont {X.}~\bibnamefont
  {Gao}}, \bibinfo {author} {\bibfnamefont {A.~D.}\ \bibnamefont {Hanlon}},
  \bibinfo {author} {\bibfnamefont {S.}~\bibnamefont {Mukherjee}}, \bibinfo
  {author} {\bibfnamefont {P.}~\bibnamefont {Petreczky}}, \bibinfo {author}
  {\bibfnamefont {P.}~\bibnamefont {Scior}}, \bibinfo {author} {\bibfnamefont
  {S.}~\bibnamefont {Syritsyn}}, \ and\ \bibinfo {author} {\bibfnamefont
  {Y.}~\bibnamefont {Zhao}},\ }\href {\doibase 10.1103/PhysRevLett.128.142003}
  {\bibfield  {journal} {\bibinfo  {journal} {Phys. Rev. Lett.}\ }\textbf
  {\bibinfo {volume} {128}},\ \bibinfo {pages} {142003} (\bibinfo {year}
  {2022}{\natexlab{a}})}\BibitemShut {NoStop}%
\bibitem [{\citenamefont {Chen}\ \emph {et~al.}(2026)\citenamefont {Chen} \emph
  {et~al.}}]{Chen:2025cxr}%
  \BibitemOpen
  \bibfield  {author} {\bibinfo {author} {\bibfnamefont {J.-W.}\ \bibnamefont
  {Chen}} \emph {et~al.},\ }\href {\doibase 10.1103/fflw-qpcc} {\bibfield
  {journal} {\bibinfo  {journal} {Phys. Rev. D}\ }\textbf {\bibinfo {volume}
  {113}},\ \bibinfo {pages} {014509} (\bibinfo {year} {2026})}\BibitemShut
  {NoStop}%
\bibitem [{\citenamefont {Ji}\ \emph {et~al.}(2026)\citenamefont {Ji},
  \citenamefont {Liu},\ and\ \citenamefont {Su}}]{Ji:2026vir}%
  \BibitemOpen
  \bibfield  {author} {\bibinfo {author} {\bibfnamefont {X.}~\bibnamefont
  {Ji}}, \bibinfo {author} {\bibfnamefont {Y.}~\bibnamefont {Liu}}, \ and\
  \bibinfo {author} {\bibfnamefont {Y.}~\bibnamefont {Su}},\ }\href@noop {} {\
  (\bibinfo {year} {2026})},\ \bibinfo {note} {arXiv:2601.12189}\BibitemShut
  {NoStop}%
\bibitem [{\citenamefont {Becher}\ \emph {et~al.}(2007)\citenamefont {Becher},
  \citenamefont {Neubert},\ and\ \citenamefont {Pecjak}}]{Becher:2006mr}%
  \BibitemOpen
  \bibfield  {author} {\bibinfo {author} {\bibfnamefont {T.}~\bibnamefont
  {Becher}}, \bibinfo {author} {\bibfnamefont {M.}~\bibnamefont {Neubert}}, \
  and\ \bibinfo {author} {\bibfnamefont {B.~D.}\ \bibnamefont {Pecjak}},\
  }\href {\doibase 10.1088/1126-6708/2007/01/076} {\bibfield  {journal}
  {\bibinfo  {journal} {JHEP}\ }\textbf {\bibinfo {volume} {01}},\ \bibinfo
  {pages} {076} (\bibinfo {year} {2007})}\BibitemShut {NoStop}%
\bibitem [{\citenamefont {Gao}\ \emph {et~al.}(2026)\citenamefont {Gao},
  \citenamefont {He}, \citenamefont {Lin}, \citenamefont {Mukherjee},
  \citenamefont {Petreczky}, \citenamefont {Zhang},\ and\ \citenamefont
  {Zhao}}]{Gao:2026hix}%
  \BibitemOpen
  \bibfield  {author} {\bibinfo {author} {\bibfnamefont {X.}~\bibnamefont
  {Gao}}, \bibinfo {author} {\bibfnamefont {J.}~\bibnamefont {He}}, \bibinfo
  {author} {\bibfnamefont {J.}~\bibnamefont {Lin}}, \bibinfo {author}
  {\bibfnamefont {S.}~\bibnamefont {Mukherjee}}, \bibinfo {author}
  {\bibfnamefont {P.}~\bibnamefont {Petreczky}}, \bibinfo {author}
  {\bibfnamefont {R.}~\bibnamefont {Zhang}}, \ and\ \bibinfo {author}
  {\bibfnamefont {Y.}~\bibnamefont {Zhao}},\ }\href@noop {} {\  (\bibinfo
  {year} {2026})},\ \bibinfo {note} {arXiv:2602.11283}\BibitemShut {NoStop}%
\bibitem [{\citenamefont {Zhang}\ \emph {et~al.}(2025)\citenamefont {Zhang},
  \citenamefont {Grebe}, \citenamefont {Hackett}, \citenamefont {Wagman},\ and\
  \citenamefont {Zhao}}]{Zhang:2025hyo}%
  \BibitemOpen
  \bibfield  {author} {\bibinfo {author} {\bibfnamefont {R.}~\bibnamefont
  {Zhang}}, \bibinfo {author} {\bibfnamefont {A.~V.}\ \bibnamefont {Grebe}},
  \bibinfo {author} {\bibfnamefont {D.~C.}\ \bibnamefont {Hackett}}, \bibinfo
  {author} {\bibfnamefont {M.~L.}\ \bibnamefont {Wagman}}, \ and\ \bibinfo
  {author} {\bibfnamefont {Y.}~\bibnamefont {Zhao}},\ }\href {\doibase
  10.1103/6dh4-6k4t} {\bibfield  {journal} {\bibinfo  {journal} {Phys. Rev. D}\
  }\textbf {\bibinfo {volume} {112}},\ \bibinfo {pages} {L051502} (\bibinfo
  {year} {2025})}\BibitemShut {NoStop}%
\bibitem [{\citenamefont {Ji}(2025)}]{Ji:2022ezo}%
  \BibitemOpen
  \bibfield  {author} {\bibinfo {author} {\bibfnamefont {X.}~\bibnamefont
  {Ji}},\ }\href {\doibase 10.34133/research.0695} {\bibfield  {journal}
  {\bibinfo  {journal} {Research}\ }\textbf {\bibinfo {volume} {8}},\ \bibinfo
  {pages} {0695} (\bibinfo {year} {2025})}\BibitemShut {NoStop}%
\bibitem [{\citenamefont {Ji}\ \emph {et~al.}(2017)\citenamefont {Ji},
  \citenamefont {Zhang},\ and\ \citenamefont {Zhao}}]{Ji:2017rah}%
  \BibitemOpen
  \bibfield  {author} {\bibinfo {author} {\bibfnamefont {X.}~\bibnamefont
  {Ji}}, \bibinfo {author} {\bibfnamefont {J.-H.}\ \bibnamefont {Zhang}}, \
  and\ \bibinfo {author} {\bibfnamefont {Y.}~\bibnamefont {Zhao}},\ }\href
  {\doibase 10.1016/j.nuclphysb.2017.09.001} {\bibfield  {journal} {\bibinfo
  {journal} {Nucl. Phys. B}\ }\textbf {\bibinfo {volume} {924}},\ \bibinfo
  {pages} {366} (\bibinfo {year} {2017})}\BibitemShut {NoStop}%
\bibitem [{\citenamefont {Zhang}\ \emph {et~al.}(2026)\citenamefont {Zhang},
  \citenamefont {Mondal}, \citenamefont {Xu}, \citenamefont {Zhao},\ and\
  \citenamefont {Vary}}]{zhang2026transverseforcetomographyinside}%
  \BibitemOpen
  \bibfield  {author} {\bibinfo {author} {\bibfnamefont {Z.}~\bibnamefont
  {Zhang}}, \bibinfo {author} {\bibfnamefont {C.}~\bibnamefont {Mondal}},
  \bibinfo {author} {\bibfnamefont {S.}~\bibnamefont {Xu}}, \bibinfo {author}
  {\bibfnamefont {X.}~\bibnamefont {Zhao}}, \ and\ \bibinfo {author}
  {\bibfnamefont {J.~P.}\ \bibnamefont {Vary}},\ }\href
  {https://arxiv.org/abs/2603.25548} {\enquote {\bibinfo {title} {Transverse
  force tomography inside a proton from basis light-front quantization},}\ }
  (\bibinfo {year} {2026})\BibitemShut {NoStop}%
\bibitem [{\citenamefont {Pochinsky}\ \emph {et~al.}(sent)\citenamefont
  {Pochinsky} \emph {et~al.}}]{qlua}%
  \BibitemOpen
  \bibfield  {author} {\bibinfo {author} {\bibfnamefont {A.}~\bibnamefont
  {Pochinsky}} \emph {et~al.},\ }\href {https://usqcd.lns.mit.edu/qlua} {\
  (\bibinfo {year} {2008–present})}\BibitemShut {NoStop}%
\bibitem [{\citenamefont {Gao}\ \emph {et~al.}(2022{\natexlab{b}})\citenamefont
  {Gao}, \citenamefont {Hanlon}, \citenamefont {Karthik}, \citenamefont
  {Mukherjee}, \citenamefont {Petreczky}, \citenamefont {Scior}, \citenamefont
  {Syritsyn},\ and\ \citenamefont {Zhao}}]{Gao:2022vyh}%
  \BibitemOpen
  \bibfield  {author} {\bibinfo {author} {\bibfnamefont {X.}~\bibnamefont
  {Gao}}, \bibinfo {author} {\bibfnamefont {A.~D.}\ \bibnamefont {Hanlon}},
  \bibinfo {author} {\bibfnamefont {N.}~\bibnamefont {Karthik}}, \bibinfo
  {author} {\bibfnamefont {S.}~\bibnamefont {Mukherjee}}, \bibinfo {author}
  {\bibfnamefont {P.}~\bibnamefont {Petreczky}}, \bibinfo {author}
  {\bibfnamefont {P.}~\bibnamefont {Scior}}, \bibinfo {author} {\bibfnamefont
  {S.}~\bibnamefont {Syritsyn}}, \ and\ \bibinfo {author} {\bibfnamefont
  {Y.}~\bibnamefont {Zhao}},\ }\href {\doibase 10.1103/PhysRevD.106.074505}
  {\bibfield  {journal} {\bibinfo  {journal} {Phys. Rev. D}\ }\textbf {\bibinfo
  {volume} {106}},\ \bibinfo {pages} {074505} (\bibinfo {year}
  {2022}{\natexlab{b}})}\BibitemShut {NoStop}%
\end{thebibliography}
%

\end{document}